\documentclass[11pt]{article}
\usepackage{amsfonts}
\usepackage{amssymb}
\usepackage{graphicx}
\usepackage{amsmath}
\usepackage{makeidx}
\usepackage{indentfirst}
\usepackage[T1]{fontenc}
\usepackage[utf8]{inputenc}

\newcounter{resultnum}[section]
\newcounter{conclusionnum}[section]
\newcounter{conditionnum}[section]
\newcounter{conjecturenum}[section]
\newcounter{examplenum}[section]
\newcounter{exercisenum}[section]
\newcounter{lemmanum}[section]
\newcounter{notationnum}[section]
\newcounter{theoremnum}[section]
\newcounter{definitionnum}[section]
\newcounter{corollarynum}[section]
\newcounter{remarknum}[section]
\newcounter{propositionnum}[section]
\newcounter{acknowledgementnum}[section]
\newcounter{algorithmnum}[section]
\newcounter{axiomnum}[section]
\newcounter{casenum}[section]
\newcounter{claimnum}[section]
\newcounter{summarynum}[section]
\newcounter{problemnum}[section]
\begin{document}

\title{Quantum geometric information flows and relativistic generalizations
of G. Perelman thermodynamics for nonholonomic Einstein systems with
black holes and stationary solitonic hierarchies}
\date{Jan 19, 2022\ (published in QINP)}
\author{ {\ \textbf{Iuliana Bubuianu} \vspace{.1 in} }\thanks{%
email: iulia.bubu@gmail.com }  \\  {\small \textit{Radio Ia\c{s}i, \ 44
Lasc\v{a}r Catargi street, Ia\c{s}i, \ 700107, Romania}}
 \vspace{.1 in} \\
  \vspace{.1 in}
 \textbf{Sergiu I. Vacaru}
\thanks{{\bf corresponding author};\
emails: sergiu.vacaru@gmail.com ;  \newline
\textit{Address for post correspondence in 2021-2022 as a visitor senior researcher at YF CNU Ukraine is:\ }
37 Yu. Gagarin street, ap. 3, Chernivtsi, Ukraine, 58008}  \\
{\small \textit{Physics Department, California State University at Fresno,
Fresno, CA 93740, USA; and }}\\
 {\small \textit{Dep. Theoretical Physics and Computer Modelling, 101  Storozhynetska street;}}\\
 {\small \textit{ Yuriy Fedkovych Chernivtsi National University,  Chernivtsi, 58029, Ukraine}}
 \vspace{.1 in}
  \\
{\ \textbf{El\c{s}en Veli Veliev} \vspace{.1 in} }\thanks{%
email: elsen@kocaeli.edu.tr and elsenveli@hotmail.com } \\
{\small \textit{Department of Physics,\ Kocaeli University, 41380, Izmit,
Turkey}} }
\maketitle

\begin{abstract}
We investigate classical and quantum geometric information flow theories (GIFs and QGIFs) when the geometric flow evolution and field equations for nonholonomic Einstein systems, NES, are derived from Perelman-Lyapunov type entropic type functionals. The term NES encodes models when the fundamental physical equations are subjected to nonholonomic (equivalently, non-integrable, anholonomic) constraints. There are used canonical geometric variables that allow a general decoupling and integration of systems of nonlinear partial differential equations describing GIFs and QGIFs and  Ricci soliton type configurations. Our approach is different from the constructions elaborated for special classes of solutions characterized by area-hypersurface entropy, related holographic, and dual gauge-gravity models involving generalizations of the Bekenstein-Hawking entropy. We formulate the theory of QGIFs which in certain quasi-classical limits encodes GIFs and models with flow evolution of NES. There are computed respectively the von Neumann, relative and conditional entropy; mutual information, entanglement, and R\'{e}nyi entropy. We construct explicit examples of generic off-diagonal exact and parametric solutions describing stationary solitonic gravitational hierarchies and deformations of black hole configurations. Finally, we show how Perelman's thermodynamic values and extensions to QGIF models can be computed for various new classes of exact solutions which can not be described following the Bekenstein-Hawking approach.

\vskip5pt


\vskip3pt

PACS2010:\ 02.40.-k, 02.90.+p, 03.67.-a, 04.50.-h, 04.50.Kd, 04.90.+e,
05.90.+m

\end{abstract}

\newpage

\tableofcontents


\section{Introduction}

This work provides an exposition of the subject of modelling modified
gravity theories, MGTs, and general relativity, GR, in the framework of
classical and quantum geometric information flow (respectively, GIF and
QGIF) theories. Although a great amount of research has been devoted
recently to quantum information theory, gravity and entanglement (for
reviews of results and some important directions of research, see \cite%
{preskill,witten18,ryu06,
raamsdonk10,bub21,faulkner14,swingle12,jacobson15,taylorm18,pastawski15,casini11, aolita14,nishioka18}%
), almost all constructions and applications related to gravity theories
involve area--hypersurface, holography and dual gauge-gravity models for
various types of entropic and thermodynamic type MGTs. Such approaches are
elaborated with respective generalizations of the concepts and formulas for
the Bekenstein--Hawking entropy and black hole, BH, thermodynamics \cite%
{bekenstein72,bekenstein73,bardeen73,hawking75}. 

In a series of papers \cite%
{vacaru09,ruchin13,gheorghiu16,rajpoot17,bubuianu19,vacaru19,vacaru19a}, we
deal with the following question: \textit{Does a more general concept of
gravitational entropy can be elaborated in such a form that would allow us
to characterize more general classes of generic off-diagonal solutions (not
only for BHs and cosmological solutions with horizons and holographic
configurations) depending on all spacetime/ phase space coordinates in
various types of MGTs, GR and Finsler like generalizations ?} The answer was
found positive for relativistic and various type generalizations of the
geometric flow theorye and by constructing in explicit form and providing
corresponding physical interpretations of a series of new classes of exact
and parametric solutions in GR and MGTs (supersymmetric/ noncommutative
string and brane gravity models, entropic gravity...) and nonholonomc
geometric flow models. The main idea exploited and developed in our works
was to characterize and derive such gravity theories and their solutions
using the concepts of F- and W-functional due to G. Perelman \cite{perelman1}
used in order to prove the Thurston-Poincar\'{e} conjecture in the theory of
Ricci flows of Riemannian metrics. In this article, we do not address issues
on modifications of the theory of Ricci flows from the viewpoint of
formulating (and possible proofs) of a generalized geometric conjectures for
pseudo-Riemannian (or non-commutative/supersymmetric etc.) metrics or/and
non-Riemannian connections, which is a very difficult mathematical task
requesting studies on thousands of pages with rigorous topological and
geometric analysis methods as in monographs \cite%
{hamilt1,monogrrf1,monogrrf2,monogrrf3} and related applications in physics
\cite{friedan2,vacaru09,ruchin13,gheorghiu16,rajpoot17}. We develop
corresponding geometric methods and find new classes of exact solutions when
the W-functional (it is also called the W-entropy because it is like a
"minus-entropy") allows us to elaborate on associated geometric and quantum
thermodynamic and information models which are more general than those
involving the Bekenstein--Hawking entropy and allow to characterize general
classes of solutions in geometric flow and MGTs. 

The G. Perelman functionals and related geometric evolution and
thermodynamic models can not be applied directly for deriving and
investigating realistic classical and quantum gravity and mater field
theories. To ensure the relativistic invariance and compatibility with GR
and/or other MGTs is necessary to consider certain types of nonholonomic
deformations of the F- and W-functionals. Here we note that in literature on
mathematics, mechanics and physics the term "anholonomic", i.e. with
non-integrable distributions/ structures is used equivalently with
"nonholonomic", see \cite{odintsov1,stavr1,bubuianu19,vacaru19,vacaru19a}
and references therein for reviews and applications of such methods in
modern gravity theories. Respective geometric flow theories can be
elaborated with different types of evolution parameters\footnote{%
for a temperature like one, which is used in this work, or as a time like
parameter treated as a "complex temperature"; such theories have, for
instance, very different topological and global classical and quantum
properties but, for additional assumptions, preserve much similarities which
are exploited for local and perturbative models, for instance, in quantum
statistics, particle and condensed matter physics etc.}. Nevertheless, Ricci
soliton configurations (determined by self-similar geometric flows) can be
defined for all types of such models which are described equivalently by
certain types of (modified) Einstein equations. Such equations can be solved
in very general forms using geometric and analytic methods elaborated in our
works, see details and reviews in \cite%
{bubuianu19,vacaru19,vacaru19a,vacaru11,vacaru18tc,bubuianu18,gheorghiu16,rajpoot17}%
. 

Even it is not clear at present if relativistic/ noncommutative /
supersymmetric analogues of the Poincar\'{e} hypothesis can be formulated
and proven, we can always construct various classes of exact and parametric
solutions for systems of nonlinear partial differential equations, PDEs,
describing a diversity of evolution/ dynamical/ diffusion/ kinetic /
thermodynamic / information classical and quantum systems. Such physically
important nonlinear PDEs can be decoupled and integrated in general forms
using the so-called anholonomic frame deformation method, AFDM (reviews of
results and applications to modern gravity and geometric flow theories can
be found in \cite{vacaru11,vacaru18tc,bubuianu18,bubuianu19,vacaru19a} and
references therein). To generate exact generic off-diagonal solutions is
important to write down the equations in so-called canonical nonholonomic
geometric variables (in our works, we use "hats" on symbols in order to
emphasize that respective geometric objects are constructed in canonical
forms adapted to a respective nonlinear connection structure, see definition
in next section).

In this article, we perform all classical and quantum models constructions
using canonical geometric data for conventional nonholonomic Einstein
systems, NES, which can be under (quantum) geometric flow evolution on a
temperature like parameter, or described by certain self-similar Ricci type
soliton configurations (with a fixed value of such a parameter). Here we
note that only certain classes of black hole and cosmological solutions in
MGTs and GR (for very special cases with respective conditions on higher
symmetry, hypersurface horizons and asymptotic/ initial behaviour) are
characterized by Bekenstein--Hawing type entropies. In another turn,
generalized G. Perelman entropic/ thermodynamic values can be defined and
computed in all cases at least for any finite spacetime region with
well-defined causality and regularizing singularities conditions. It is
possible to write equivalently such geometric (flow) and statistical
thermodynamic like values in canonical nonholonomic variables, or in certain
analogous Finsler-Lagrange-Hamilton, or almost K\"{a}hler, variables. This
is very important for elaborating analogous, entropic, deformation
quantization and brane/ gauge like gravity models (see, for instance, \cite%
{vacaru07,vacaru09b,vacaru09,bubuianu19,vacaru19,vacaru19a}). To apply such
methods to GR and standard particle physics we have to impose at the end
(when corresponding classes of exact solutions were found in exact form)
certain additional nonholonomic constraints in order to extract Levi-Civita
(torsionless), LC, configurations. If we work from the very beginning only
with the LC-connection, it is not possible to decouple/ integrate in general
forms nonlinear classical and quantum fundamental evolution/ dynamical
equations. This is a general property of the systems of nonlinear PDEs on
curved spacetimes. 

The formalism developed in this and partner works \cite%
{vacaru19b,vacaru19c,vacaru19e} allows us to address issues related to
various types of quasi-classical and quantum equations (nonlinear Schr\"{o}%
dinger, Liouville, and Dirac ones, and noncommutative quantum deformations
of the Einstein equations). We are able to deal with the basic elements
fundamental concepts of quantum information theory such as entanglement and
multipartite states, teleportation, and quantum interference for NES under
geometric flow evolution. There are three main goals in this article: 1) To
prove a general decoupling and integrability of stationary (on spacetime
coordinates) of geometric flow equations with a temperature like evolution
parameter. 2) To provide an introduction to the theory of GIFs and QGIFs of
NES and show how generalized Perelman entropies and associated quantum
mechanical, statistical and thermodynamic geometric theories are used for
elaborating such theories. 3) To consider possible applications of the AFDM
and study properties of stationary solutions (in particular, of BHs and
their solitonic off-diagonal deformations) under nonlinear geometric flows,
interactions and for solitonic gravitational and matter field configurations
when the W-entropy can be defined and computed but the concept of
Bekenstein-Hawking entropy is not applicable, or not enough, for
characterizing such classes of exact solutions and QGIFs. In this paper, we
use canonical nonholonomic variables which allow us to provide analogous
thermodynamic and GIF and QGIF descriptions of any solution found following
the AFDM, or other type methods, for MGTs and GR, and/or can be derived as
some emergent/ entropic gravity models. 

Here we consider also this important issue: One of the basic concepts in
quantum information theory is that of qubit. Given a Hilbert space
associated to a physical system is possible to realize a qubit as any
two-dimensional subspace of that Hilbert space. Such a physical realization
is not, in general, localized in a physical space. In \cite%
{vacaru19b,vacaru19c}, we study the concept of quibit for QGIFs when certain
analogous Hamilton and/or thermodynamic models are determined by W-entropy.
For such theories, we can formulate physical realizations that are
well-localized in certain effective phase spaces and associate a qubit as a
two-dimensional quantum state attached to a point in a base space. In a
curved spacetime context, we can represent a qubit as a sequence of
two-dimensional quantum states evolving along a spacetime trajectory and/or
with sets of world lines of qubits covering a spacetime region. Other type
ambiguities are related to the fact that there are no finite-dimensional (in
particular, two dimensional ones) faithful unitary representations of the
Lorentz group. Naively, it would appear that it is impossible to elaborate a
mathematical formalism which would describe localized qubits in a unique
both relativistic and unitary form. The problem (and an explicit method how
to solve it using WKB approximations) is analyzed in \cite{palmer12}. Here
we note that for QGIF theories there are used analogous statistical
thermodynamic models and respective quantum generalizations of GIF
entropies. Such constructions are well-defined for geometric flows in
spacetimes and phase spaces with conventional 3+1 splitting of dimensions
\cite{ruchin13} and there are rigorous proofs for noncommutative geometric
models (in the A. Connes approach and for theories with almost K\"{a}hler
structures and/or Siberg-Witten transforms) and deformation quantization
\cite{vacaru07,vacaru09b,vacaru09}. 

Let us explain the structure of our work: This paper aims to be
self-contained and that why we include in section \ref{s2} (and some
footnotes in other sections) a necessary background material on the geometry
of nonholonomic Lorentz manifolds, their relativistic geometric flow
evolution equations and nonholonomic Ricci solitons described by modified
Einstein equations. The geometric constructions are performed in canonical
nonholonomic variables which allow a straightforward decoupling and
integration of geometric flow and gravitational field equations. We define
the nonholonomic canonical version of G. Perelman F-functional and
W-functional (equivalently, W-entropy) and show how associated thermodynamic
models can be elaborated. 

Section \ref{s3} is devoted to the theory of (classical) geometric
information flows, GIFs, and quantum information geometric flows, QGIFs, and
respective entanglement of nonholonomic Einstein systems, NES. Using
statistical distribution functions determined by the W-entropy and
thermodynamic values constructed for GIFs, we define and show how to compute
the Shanon, conditional and relative entropy of such systems. Quantum
mechanical models, QMs, are elaborated for the canonical density matrix
 and von Neumann entropy as respective quantum analogues of the
statistical density matrix for NES GIFs. We investigate properties of
entanglement and gravity for QGIFs using (and showing how to compute)
respective inequalities for relative entropy, mutual information and the R%
\'{e}nyi entropy for classical and quantum geometric thermodynamics and
information flows. 

In section \ref{s4}, we apply the AFDM and prove the existence of an important
 decoupling properties and explicit integrability of nonholonomic
geometric flows and Ricci soliton equations for stationary (i.e. depending
on certain space coordinates) NES determined by solitonic configurations of
gravitational and matter fields. Such new classes of solutions are described
by generic off-diagonal metrics and generalized connections depending on all
spacetime coordinates and temperature like parameters via general classes of
generating functions and (effective) sources of solitonic configurations of
gravity and matter fields. Possible stationary parameterizations of
main geometric objects and matter sources are stated in Table 1. 

In section \ref{s5}, we show how exact and parametric solutions can be
constructed for flow evolution of stationary configurations and NES.  There are reviewed and cited \cite%
{ruchin13,gheorghiu16,rajpoot17,vacaru19,vacaru11,vacaru18tc,bubuianu18,
bubuianu19,vacaru19a} for details, methods and examples of other type
solutions. We provide Table 2 summarizing the AFDM for generating stationary
solutions for geometric flows of NES and nonholonomic Ricci solitons and
off-diagonal solitonic configurations. The formulas are re-defined for new
classes of solutions describing for instance, black hole (BH) nonholonomic/
ellipsoid deformations of horizons by a nontrivial vacuum with a solitonic
configuration, or embedding of BH in Ricci solitonic vacuum and nonvacuum
backgrounds with encoding solitonic gravitational and matter field
structures. In explicit form, the AFDM formalism is considered for computing
small parametric and generic off--diagonal deformations of the Kerr solution
under solitonic geometric flows and nontrivial solitonic gravitational and
matter field configurations. We emphasize that the thermodynamic properties
of such generalized classes of stationary solitonic solutions can not be
described by a Bekenstein-Hawking entropy and respective classical or
quantum generalizations.

G. Perelman W-entropy and main geometric thermodynamic values can be always
defined and computed in explicit form for various classes of solutions in
geometric flow, MGTs and GR. In section \ref{s6} we consider key steps and
provide explicit examples how to compute such values for three classes of
generic off-diagonal solutions stationary solitonic configurations and
respective models GIF and QGIF theories. To avoid cumbersome formulas we
consider special classes of normalization, generating and integration
functions. The fist example is for stationary solitonic generating
functions which can be of solitonic or non-solitonic character. Then, in the second example, we compute G. Perelman's
thermodynamic values for BHs deformed by general (not introducing small parameters) stationary solitonic generating
functions and sources. Finally, the formulas for such values are considered
for small parametric stationary deformations of BH solutions and respective
GIF and QGIF models.

An outlook, conclusions and discussions are presented in section \ref{sconcl}%
.

\section{Geometric flow equations for nonholonomic Einstein systems}

\label{s2} In this section, we provide an introduction to the theory
relativistic geometric evolution flows and nonholonomic Einstein systems,
NES. The constructions are performed in canonical nonholonomic variables
with the so-called "hat" connections (all correspondingly adapted to
non--integrable distributions of geometric/physical objects) which will
allow us to decouple and integrate physically important systems of nonlinear
PDEs, see section \ref{s4}. For geometric details, proofs, examples and
constructions with other type variables for geometric flows and MGTs, we
cite \cite{ruchin13,gheorghiu16,rajpoot17,bubuianu18,bubuianu19} and
references therein.

\subsection{Preliminaries: nonholonomic Lorentz manifolds with $2+2$
splitting}

We consider a pseudo-Riemannian manifold $\mathbf{V}$ determined by a metric
field $\mathbf{g}=\{g_{\alpha \beta }(u)\}$ with local pseudo-Euclidean
signature $(+++-)$ and enabled with a conventional 2+2 splitting into
horizontal (h) and vertical (v) components defined by a Whitney sum
\begin{equation}
\mathbf{N}:\ T\mathbf{V}=h\mathbf{\mathbf{V\oplus }}v\mathbf{V},
\label{ncon}
\end{equation}%
where $T\mathbf{V}$ is the tangent bundle.\footnote{%
On abstract and coordinate indices (we shall underline such indices if order
to emphasize that they are for a coordinate base) we adopt such conventions:
The local coordinates are labelled $u^{\mu}=(x^{i},y^{a}),$ (in brief, we
shall write $u=(x,y)$), where $i,j,...=1,2$ and $a,b,...=3,4.$ Cumulative
small Greek indices run values $\alpha ,\beta ,...=1,2,3,4,$ where $%
u^{4}=y^{4}=t$ is a time like coordinate. An arbitrary local basis is
denoted $e^{\alpha }=(e^{i},e^{a})$ and the corresponding dual one,
co-basis, is written in the form $e_{\beta }=(e_{j},e_{b}).$ There are
always some nontrivial frame transforms to corresponding coordinate bases, $%
\partial _{\alpha ^{\prime }}=(\partial _{i^{\prime }},\partial _{a^{\prime
}})$ [for instance, $\partial _{i^{\prime }}=\partial /\partial x^{i^{\prime
}}],$ and cobasis, when $e_{\beta }=A_{\beta }^{\ \beta ^{\prime
}}(u)\partial _{\beta ^{\prime }}$ and $e^{\alpha }=A_{\ \alpha ^{\prime
}}^{\alpha }(u)du^{\alpha ^{\prime }},$ for $du^{\alpha ^{\prime
}}=(dx^{i^{\prime }},dy^{a^{\prime }}),$ are considered as frame (vierbein)
transforms. There are used primed, underlined indices etc. for other type
frame/coordinate systems related via respective classes of nonlinear
transforms. It is applied the Einstein summation rule on repeating up--low
indices if the contrary will be not stated.} A N--connection structure (\ref%
{ncon}) is determined locally by a corresponding set of coefficients $%
N_{i}^{a},$ when $\mathbf{N}=N_{i}^{a}(u)dx^{i}\otimes \partial _{a}.$ \ For
any h-v--splitting, we can define N--adapted local bases, $\mathbf{e}_{\nu
}=(\mathbf{e}_{i},e_{a}),$ and cobases, $\mathbf{e}^{\mu }=(e^{i},\mathbf{e}%
^{a}),$ when
\begin{eqnarray}
\mathbf{e}_{\nu } &=&(\mathbf{e}_{i}=\partial /\partial x^{i}-\
N_{i}^{a}\partial /\partial y^{a},\ e_{a}=\partial _{a}=\partial /\partial
y^{a}),  \label{nader} \\
\mathbf{e}^{\mu } &=&(e^{i}=dx^{i},\mathbf{e}^{a}=dy^{a}+\ N_{i}^{a}dx^{i}).
\label{nadif}
\end{eqnarray}%
In general, such N--adapted bases are nonholonomic because there are
satisfied relations of type
\begin{equation}
\lbrack \mathbf{e}_{\alpha },\mathbf{e}_{\beta }]=\mathbf{e}_{\alpha }%
\mathbf{e}_{\beta }-\mathbf{e}_{\beta }\mathbf{e}_{\alpha }=W_{\alpha \beta
}^{\gamma }\mathbf{e}_{\gamma },  \label{anhcoef}
\end{equation}%
with nontrivial anholonomy coefficients $W_{ia}^{b}=\partial
_{a}N_{i}^{b},W_{ji}^{a}=\Omega _{ij}^{a}=\mathbf{e}_{j}\left(
N_{i}^{a}\right) -\mathbf{e}_{i}(N_{j}^{a}).$ There are encoded holonomic
(integrable) bases if and only if $W_{\alpha \beta }^{\gamma }=0.$ We can
elaborate on $\mathbf{V}$ and respective (co) tangent bundles a N--adapted
differential and integral calculus and a corresponding variational formalism
there are used N--elongated operators (\ref{nader}) and (\ref{nadif}) and
introduced additionally a covariant derivative $D$ defined as a
metric-affine connection. All geometric constructions and physical models
can be re-defined in terms of distinguished objects (in brief, d--objects)
when the coefficients are determined with respect to N--adapted (co) frames
and their tensor products.\footnote{%
For instance, any vector $Y(u)\in T\mathbf{V}$ can be parameterized as a
d--vector, $\mathbf{Y}=$ $\mathbf{Y}^{\alpha }\mathbf{e}_{\alpha }=\mathbf{Y}%
^{i}\mathbf{e}_{i}+\mathbf{Y}^{a}e_{a},$ or $\mathbf{Y}=(hY,vY),$ with $hY=\{%
\mathbf{Y}^{i}\}$ and $vY=\{\mathbf{Y}^{a}\}.$ Similarly, we can determine
and compute the coefficients of d--tensors, N--adapted differential forms,
d--connections, d--spinors etc.}

A manifold $(\mathbf{V,N})$ endowed with a nontrivial structure $W_{\alpha
\beta }^{\gamma }$ (\ref{anhcoef}) \ is called nonholonomic (equivalently,
anholonomic) if it is defined as a union of non-integrable distributions in
any point $u\in \mathbf{V.}$ We call it as a \textbf{nonholonomic Lorentz
manifold} if at least locally such a curved spacetime possess a causal
structure like that for the special relativity theory determined by a local
Minkowski metric.

For nonholonomic manifolds, there is a subclass of linear connections which
are adapted to the N--connection structure and called \textbf{distinguished
connections} (in brief, \textbf{d--connections}). Such a
d-connection $\mathbf{D}=(h\mathbf{D},v\mathbf{D})$ on $\mathbf{V}$
preserves under parallel transport the N--connection splitting (\ref{ncon}).
A general linear connection $D$ is not adapted to a chosen $h$-$v$%
--decomposition, i.e. it is not a d--connection.\footnote{%
We do not use boldface symbols for not N--adapted geometric objects.} For
instance, the Levi--Civita, LC, connection in GR is not a d--connection. To
a d--connection $\mathbf{D,}$ we can associate an operator of covariant
derivative, $\mathbf{D}_{\mathbf{X}}\mathbf{Y}$ (for a d--vector $\mathbf{Y}$
in the direction of a d--vector $\mathbf{X)}.$ We can compute N--adapted
coefficients for $\mathbf{D}=\{\mathbf{\Gamma }_{\ \alpha \beta }^{\gamma
}=(L_{jk}^{i},L_{bk}^{a},C_{jc}^{i},C_{bc}^{a})\}$ defined with respect to
tensor products of N--adapted frames (\ref{nader}) and (\ref{nadif}).

For any d--connection $\mathbf{D}$ and d--vectors $\mathbf{X,Y\in }T\mathbf{%
V,}$ the d--torsion, $\mathbf{T,}$ the nonmetricity, $\mathbf{Q},$ and the
d--curvature, $\mathbf{R},$ tensors (in N--adapted forms, they are
d--tensors) are defined and computed in standard form,
\begin{equation*}
\mathbf{T}(\mathbf{X,Y}):=\mathbf{D}_{\mathbf{X}}\mathbf{Y}-\mathbf{D}_{%
\mathbf{Y}}\mathbf{X}-[\mathbf{X,Y}],\ \mathbf{Q}(\mathbf{X}):=\mathbf{D}_{%
\mathbf{X}}\mathbf{g}, \mathbf{R}(\mathbf{X,Y}:= \mathbf{D}_{\mathbf{X}}%
\mathbf{D}_{\mathbf{Y}}-\mathbf{D}_{\mathbf{Y}}\mathbf{D}_{\mathbf{X}}-%
\mathbf{D}_{\mathbf{[X,Y]}}.
\end{equation*}%
Such values are defined locally by N--adapted coefficients are
correspondingly labeled using $h$- and $v$--indices,
\begin{eqnarray}
\mathbf{T} &=&\{\mathbf{T}_{\ \alpha \beta }^{\gamma }=\left( T_{\
jk}^{i},T_{\ ja}^{i},T_{\ ji}^{a},T_{\ bi}^{a},T_{\ bc}^{a}\right) \},%
\mathbf{Q}=\mathbf{\{Q}_{\ \alpha \beta }^{\gamma }\},  \notag \\
\mathbf{R} &=& \mathbf{\{R}_{\ \beta \gamma \delta }^{\alpha }=\left( R_{\
hjk}^{i}\mathbf{,}R_{\ bjk}^{a}\mathbf{,}R_{\ hja}^{i},R_{\ bja}^{c},R_{\
hba}^{i},R_{\ bea}^{c}\right) \},  \label{rnmc}
\end{eqnarray}%
see explicit formulas in \cite%
{ruchin13,gheorghiu16,rajpoot17,bubuianu18,bubuianu19}.

With respect to a dual local coordinate basis $du^{\alpha },$ any metric
tensor $\mathbf{g}$ on $\left( \mathbf{V,N}\right) $ can be parameterized in
a (general) off-diagonal form,
\begin{equation}
\mathbf{g}=\underline{g}_{\alpha \beta }(u)du^{\alpha }\otimes du^{\beta },%
\mbox{\ where \ }\underline{g}_{\alpha \beta }(u)=\left[
\begin{array}{cc}
g_{ij}+N_{i}^{a}N_{j}^{b}g_{ab} & N_{j}^{e}g_{ae} \\
N_{i}^{e}g_{be} & g_{ab}%
\end{array}%
\right] ,  \label{ofdans}
\end{equation}%
with 6 independent coefficients.\footnote{%
Any symmetric second rank tensor on a 4-d $\mathbf{V}$ has, in general, 10
independent coefficients but 4 components of a metric tensor can be always
transformed into zero by locall coordinate transform. We can not transform
arbitrary geometric data $\left( \mathbf{g,N}\right) $ into a diagonal
metric in a finite region $U\subset \mathbf{V}$ even a Minkowski metric can
be assigned (using frame/coordinate transforms) in a point $u\in \mathbf{V.}$%
} Any metric $\mathbf{g}$ can be written equivalently as a d--tensor
(d--metric)
\begin{equation}
\mathbf{g}=g_{\alpha }(u)\mathbf{e}^{\alpha }\otimes \mathbf{e}^{\beta
}=g_{i}(x)dx^{i}\otimes dx^{i}+g_{a}(x,y)\mathbf{e}^{a}\otimes \mathbf{e}%
^{a},  \label{dm1}
\end{equation}%
or, in brief, $\mathbf{g}=(h\mathbf{g},v\mathbf{g}).$ A metric $\mathbf{g}$ (%
\ref{ofdans}) with N--coefficients $N_{j}^{e}$ is generic off--diagonal if
the anholonomy coefficients $W_{\alpha \beta }^{\gamma }$ (\ref{anhcoef})
are not zero.

For any metric field / d--metric $\mathbf{g,}$ we can define two important
linear connection structures following such geometric conditions: {%
\begin{equation}
\mathbf{g}\rightarrow \left\{
\begin{array}{ccccc}
\nabla : &  & \nabla \mathbf{g}=0;\ ^{\nabla }\mathbf{T}=0, &  &
\mbox{ the
Levi--Civita connection;} \\
\widehat{\mathbf{D}}: &  & \widehat{\mathbf{D}}\ \mathbf{g}=0;\ h\widehat{%
\mathbf{T}}=0,\ v\widehat{\mathbf{T}}=0, &  &
\mbox{ the canonical
d--connection.}%
\end{array}%
\right.  \label{lcconcdcon}
\end{equation}%
The LC--connection }$\nabla $ can be introduced without any N--connection
structure but the canonical d--connection $\widehat{\mathbf{D}}$ (in a
series of our works, it is called also the "hat"-connection) depends
generically on a prescribed nonholonomic $h$- and $v$-splitting. In above
formulas, $h\widehat{\mathbf{T}}$ and $\ v\widehat{\mathbf{T}}$ are
respective torsion components of $\widehat{\mathbf{D}}$ which vanish on
conventional h- and v--subspaces. Nevertheless, there are nonzero torsion
components, $hv\widehat{\mathbf{T}}.$ Here we note that $\widehat{\mathbf{T}}
$ is defined completely by the coefficients of a d-metric and N-connection
(i.e. of respective off-diagonal metric) structures which is different from
the well-known Cartan connection in Riemann-Cartan geometry.

It should be emphasized that all geometric constructions on a $\mathbf{V}$
can be performed equivalently with $\nabla $ and/or $\widehat{\mathbf{D}}$
and related via a canonical distorting relation
\begin{equation}
\widehat{\mathbf{D}}\mathbf{[g,N]}=\nabla \lbrack \mathbf{g}]+\widehat{%
\mathbf{Z}}\mathbf{[g,N]}.  \label{distr}
\end{equation}%
In this formula, both linear connections and the distorting tensor $\widehat{%
\mathbf{Z}}$ are uniquely determined by data $(\mathbf{g,N)}$ as an
algebraic combination of coefficients of $\widehat{\mathbf{T}}_{\ \alpha
\beta }^{\gamma }.$ The N--adapted coefficients for $\widehat{\mathbf{D}}$
and corresponding torsion, $\widehat{\mathbf{T}}_{\ \alpha \beta }^{\gamma }$%
, Ricci d--tensor, $\widehat{\mathbf{R}}_{\ \beta \gamma }$, and Einstein
d--tensor, $\widehat{\mathbf{E}}_{\ \beta \gamma }$, can be computed in
standard form as in the case of LC-connection $\nabla $, or any
metric-affine connection $D$ (abstract formulas are similar, but the
N-adapted coefficients are different and can be defined only for
d-connections). The canonical distortion relation (\ref{distr}) defines
respective distortion relations of the Riemannian, Ricci and Einstein
tensors and respective curvature scalars which are uniquely determined by
data $(\mathbf{g,N).}$ Any (pseudo) Riemannian geometry can be equivalently
formulated using $(\mathbf{g,\nabla )}$ or $(\mathbf{g},\widehat{\mathbf{D}}%
).$ In our partner works \cite{vacaru19b,vacaru19c} (see references
therein), we use also generalized Finsler-Lagrange-Hamilton variables with
respective canonical d-connection structures (they can be introduced even in
GR but also in other type MGTs) which is important for elaborating QGIF
theories of Lagrange-Hamilton nonlinear mechanical systems and analogous/
emergent gravity theories.

The canonical d--connection $\widehat{\mathbf{D}}$ has a very important role
in elaborating the AFDM for constructing exact and parametric solutions in
geometric flow and MGTs. It allows us to decouple the geometric flow
evolution and gravitational and matter field equations with respect to
N--adapted frames of reference. This is not possible if we work only with $%
\nabla $ and there are very limited possibilities for "pure" Finsler
connections (like the Cartan, Chern, or Berwald d-connections)$.$
Constructing certain general classes of solutions for $\widehat{\mathbf{D}}$%
, we can impose at the end the condition $\widehat{\mathbf{T}}=0$ and
extract LC--configurations%
\begin{equation}
\widehat{\mathbf{D}}_{\mid \widehat{\mathbf{T}}=0}=\nabla ,  \label{lccond}
\end{equation}
or to re-define the solutions and respective physical models in certain
Lagrange-Hamilton, or almost symplectic, variables which are more convenient
for elaborating and study quantum field theories, quantum mechanical models
and related quantum information theories.

\subsection{Hypersurface and (non) relativistic nonholonomic geometric flows}

In this work, we consider families of 4--d Lorentz nonholonomic manifolds $%
\mathbf{V}(\tau )$ determined by metrics $\mathbf{g}(\tau )=\mathbf{g}(\tau
,u)$ of signature $(+++-)$ and N--connections $\mathbf{N}(\tau )$
parameterized by a positive parameter $\tau ,0\leq \tau \leq \tau _{0}$. Any
$\mathbf{V}\subset \mathbf{V}(\tau )$ can be enabled with a double
nonholonomic 2+2 and 3+1 splitting, see \cite%
{ruchin13,gheorghiu16,rajpoot17,bubuianu18,bubuianu19} for details on the
geometry of spacetimes enabled with such double distributions. The local
coordinates are labeled as on nonholonomic Lorentz manifold when $%
u^{\alpha}=(x^{i},y^{a})=(x^{\grave{\imath}},u^{4}=t)$ for $i,j,k,...=1,2;$ $%
a,b,c,...=3,4;$ and $\grave{\imath},\grave{j},\grave{k}=1,2,3,$ but
dependencies on a temperature like parameter $\tau $ will be emphasized for
geometric flow evolution models. The 3+1 splitting can be chosen in such a
form that any open region $U\subset $ $\mathbf{V}$ is covered by a family of
3-d spacelike hypersurfaces $\widehat{\Xi }_{t}$ parameterized by a time
like parameter $t.$\footnote{%
There are two generic different types of geometric flow theories when 1) $%
\tau (\chi )$ is a re-parametrization of a temperature like parameter used
for labeling 4-d Lorentz spacetime configurations and 2) $\xi (t)$ is a time
like parameter when 3-d spacelike configurations evolve relativistically on
a "redefined" time like coordinate. In next sections, we shall study in
details only theories of type 1 even a number of formulas and geometric
constructions will be presented with a relativistic time/ parameters which
can be used for theories of type 2.}

The (non-relativistic) Ricci flow evolution equations were postulated
heuristically by R. Hamilton \cite{hamilt1} (in physics, similar equations
had been considered by D. Friedan \cite{friedan2} a few years earlier). We
write such equations in the form
\begin{equation}
\frac{\partial g_{\grave{\imath}\grave{j}}}{\partial \xi }=-2\ R_{\grave{%
\imath}\grave{j}}.  \label{heq1a}
\end{equation}
The equations (\ref{heq1a}) describe a nonlinear diffusion process for
geometric flow evolution of 3-d Riemannian metrics.\footnote{%
This can be found for small deformations of a 3--d Euclidean metric $g_{%
\grave{\imath}\grave{j}}\approx \delta _{\grave{\imath}\grave{j}}+$ $h_{%
\grave{\imath}\grave{j}},$ with $\delta _{\grave{\imath}\grave{j}%
}=diag[1,1,1]$ and $h_{\grave{\imath}\grave{j}}|\ll 1$, when the Ricci
tensor approximates the Laplace operator $\Delta =\frac{\partial ^{2}}{%
(\partial u^{1})^{2}}+\frac{\partial ^{2}}{(\partial u^{2})^{2}}+\frac{%
\partial ^{2}}{(\partial u^{3})^{2}}$ and we obtain a linear diffusion
equation if $R_{\grave{\imath}\grave{j}}\sim \Delta h_{\grave{\imath}\grave{j%
}}.$} In modified and normalized forms (see below some more details related
to nonholonomic generalizations and formulas (\ref{heq1c}) and (\ref{heq1d}%
)), the Hamilton equations (\ref{heq1a}) and various nonholonomic
deformations can be proven following a corresponding variational calculus
for Perelman's W- and F--functionals \cite{monogrrf1,monogrrf2,monogrrf3}.

For self-similar configurations in fixed points, the geometric flows (\ref%
{heq1a}) are described by the so-called Ricci soliton equations
\begin{equation}
R_{\grave{\imath}\grave{j}}-\lambda g_{\grave{\imath}\grave{j}}=\nabla _{%
\grave{\imath}}v_{\grave{j}}+\nabla _{\grave{j}}v_{\grave{\imath}},
\label{friedsoliton}
\end{equation}%
for $\lambda =\pm 1,0$ and a vector field $v_{\grave{j}}.$ The formulas (\ref%
{friedsoliton}) consist a variant of Einstein equations with cosmological
constant and a special source determined by $v_{\grave{j}}$ if the 3-d
Riemannian metrics are transformed into pseudo-Riemannian ones.

\subsubsection{Geometric evolution of d-metrics and the Laplace d-operator
for 3+1 splitting}

For a region $U\subset $ $\mathbf{V}$ with 2+2 splitting defined by data $(%
\mathbf{N,g}),$ we consider an additional structure of 3-d hypersurfaces $%
\Xi _{t}$ parameterized by time like coordinate $y^{4}=t$ for coordinates $%
u^{\alpha }=(x^{i},y^{a})=(x^{\grave{\imath}},t).$ The metric structure can
be represented in a d-metric form \ and/or with 3+1 splitting,%
\begin{eqnarray}
\mathbf{g} &=&\mathbf{g}_{\alpha ^{\prime }\beta ^{\prime }}(\tau ,\ u)d\
\mathbf{e}^{\alpha ^{\prime }}(\tau )\otimes d\mathbf{e}^{\beta ^{\prime
}}(\tau )  \label{decomp31} \\
&=&q_{i}(\tau ,x^{k})dx^{i}\otimes dx^{i}+\mathbf{q}_{3}(\tau ,x^{k},y^{a})%
\mathbf{e}^{3}(\tau )\otimes \mathbf{e}^{3}(\tau )-[\ _{q}N(\tau
,x^{k},y^{a})]^{2}\mathbf{e}^{4}(\tau )\otimes \mathbf{e}^{4}(\tau ).  \notag
\end{eqnarray}%
In (\ref{decomp31}), there are considered "shift" coefficients $\mathbf{q}_{%
\grave{\imath}}=(q_{i},\mathbf{q}_{3})$ related to a 3-d metric $\mathbf{q}%
_{ij}=diag(\mathbf{q}_{\grave{\imath}})=(q_{i},\mathbf{q}_{3})$ on a
hypersurface $\Xi _{t}$ if $\mathbf{q}_{3}=\mathbf{g}_{3}$ and $[\
_{q}N]^{2}=-\mathbf{g}_{4},$ where $\ _{q}N$ is the lapse function (our
notations are different from those in \cite{misner} when we use a left label
$q$ in order to avoid ambiguities with the coefficients $N_{i}^{a}$ for the
N-connection). The parameter $\tau $ can be of temperature type like in
thermodynamic theories or considered as a time like one when $\tau $ is
identified with $y^{4}=ct$ (depending on the type of geometric evolution
flow model we study).

To elaborate on relativistic geometric flows and thermodynamical models we
can use a N--adapted 3+1 decomposition for the canonical d--connection, $%
\mathbf{D}=(\ _{\shortmid }\mathbf{D},\ ^{t}D)$ and d--metric $\mathbf{g}:=(%
\mathbf{q,}\ _{q}N)$ of a 4--d nonholonomic Lorentz manifold $\mathbf{V.}$
On closed 3-d spacelike hypersurfaces, both the geometric flow and MGTs can
be formulated in two equivalent forms using the connections $\ _{\shortmid
}\nabla $ and/or $\ _{\shortmid }\mathbf{D}$ when the evolution of geometric
objects are determined by the evolution of the hypersurface metric $\mathbf{q%
}$ and an extension to $\mathbf{g}$. We introduce the canonical Laplacian
d-operator, $\ _{\shortmid }\widehat{\Delta }:=$ $\ _{\shortmid }\mathbf{D}$
$\ _{\shortmid }\mathbf{D}$ and consider the canonical distortion tensor $\
_{\shortmid }\mathbf{Z.}$ Using distortions $\ _{\shortmid }\nabla =\
_{\shortmid }\mathbf{D}-\ _{\shortmid }\mathbf{Z}$ (which is a 3-d version
of the canonical distortion relation (\ref{distr})), we compute
\begin{equation*}
\mathbf{\ }_{\shortmid }\widehat{\Delta }=\mathbf{\ }_{\shortmid }\mathbf{D}%
_{\alpha }\mathbf{\ }_{\shortmid }\mathbf{D}^{\alpha }=\mathbf{\ }%
_{\shortmid }\Delta +\ _{\shortmid }^{Z}\widehat{\Delta }
\end{equation*}%
\begin{eqnarray}
\mbox{ where }\ _{\shortmid }\Delta &=&\ _{\shortmid }\nabla _{\grave{\imath}%
}\ \mathbf{\ }_{\shortmid }\nabla ^{\grave{\imath}}=\mathbf{\ }_{\shortmid
}\nabla _{\alpha }\ \mathbf{\ }_{\shortmid }\nabla ^{\alpha },
\label{distrel2} \\
\mathbf{\ }_{\shortmid }^{Z}\widehat{\Delta } &=&\ _{\shortmid }\mathbf{Z}_{%
\grave{\imath}}\ \mathbf{\ }_{\shortmid }\mathbf{Z}^{\grave{\imath}}-[\
_{\shortmid }\mathbf{D}_{\grave{\imath}}(\mathbf{\ }_{\shortmid }\mathbf{Z}^{%
\grave{\imath}})+\mathbf{\ }_{\shortmid }\mathbf{Z}_{\grave{\imath}}(\
_{\shortmid }\mathbf{D}^{\grave{\imath}})]=\mathbf{\ }_{\shortmid }\mathbf{Z}%
_{\alpha }\mathbf{\ }_{\shortmid }\mathbf{Z}^{\alpha }-[\ _{\shortmid }%
\mathbf{D}_{\alpha }(\mathbf{\ }_{\shortmid }\mathbf{Z}^{\alpha })+\mathbf{\
}_{\shortmid }\mathbf{Z}_{\alpha }(\ _{\shortmid }\mathbf{D}^{\alpha })];
\notag \\
\mathbf{\ }_{\shortmid }\mathbf{R}_{\grave{\imath}\grave{j}} &=&\mathbf{\ }%
_{\shortmid }R_{\grave{\imath}\grave{j}}-\mathbf{\ }_{\shortmid }\mathbf{Z}%
ic_{\grave{\imath}\grave{j}},\mathbf{\ }_{\shortmid }\mathbf{R}_{\ \beta
\gamma }=\mathbf{\ }_{\shortmid }R_{\ \beta \gamma }-\mathbf{\ }_{\shortmid }%
\mathbf{Z}ic_{\beta \gamma },\   \notag
\end{eqnarray}%
\begin{eqnarray*}
\ _{\shortmid }^{s}R &=&\ _{\shortmid }R-\mathbf{g}^{\beta \gamma }\
_{\shortmid }\mathbf{Z}ic_{\beta \gamma }=\mathbf{\ }_{\shortmid }R-\mathbf{q%
}^{\grave{\imath}\grave{j}}\mathbf{\ }_{\shortmid }\mathbf{Z}ic_{\grave{%
\imath}\grave{j}}=\mathbf{\ }_{\shortmid }R-\mathbf{\ }_{\shortmid }\mathbf{Z%
}, \\
\mathbf{\ }_{\shortmid }\mathbf{Z} &=&\mathbf{g}^{\beta \gamma }\mathbf{\ }%
_{\shortmid }\mathbf{Z}ic_{\beta \gamma }=\mathbf{q}^{\grave{\imath}\grave{j}%
}\mathbf{\ }_{\shortmid }\mathbf{Z}ic_{\grave{\imath}\grave{j}}=\ _{h}%
\widehat{Z}+\ _{v}\widehat{Z},\ _{h}\widehat{Z}=g^{ij}\ \mathbf{Z}ic_{ij},\
_{v}\widehat{Z}=h^{ab}\ \mathbf{Z}ic_{ab}; \\
R &=&\ _{h}R+\ _{v}R,\ \ _{h}R:=g^{ij}\ R_{ij},\ _{v}R=h^{ab}\ R_{ab}.
\end{eqnarray*}%
Such values can be computed in explicit form for any class of exact
solutions of modified Einstein equations when a double 2+2 and 3+1 splitting
is prescribed and the LC--conditions can be imposed additionally. The 3-d
distortion formulas (\ref{distrel2}) are important for defining and
computing gravitational thermodynamic values on space like hypersurfaces.

\subsubsection{Nonholonomc 3--d hypersurface Perelman's functionals}

On a normalized 3-d spacelike closed hypersurface $\ ^{c}\widehat{\Xi }%
\subset \mathbf{V},$ the normalized version of R. Hamilton equations can be
written in a coordinate basis,
\begin{equation}
\partial _{\xi }q_{\grave{\imath}\grave{j}} = -2\ _{\shortmid }R_{\grave{%
\imath}\grave{j}}+\frac{2\grave{r}}{5}q_{\grave{\imath}\grave{j}},\ q_{%
\grave{\imath}\grave{j}\mid \xi =0} =q_{\grave{\imath}\grave{j}}^{[0]}[x^{%
\grave{\imath}}].  \label{heq1b}
\end{equation}%
For these formulas, the left label "c" is used for "compact and closed"
regions. We shall prefer to write explicitly only the dependence on
parameter variable $\xi $ (writing in brief $q_{\grave{\imath}\grave{j}}(\xi
)=q_{\grave{\imath}\grave{j}}(x^{\grave{\imath}},\xi ))$ if that will do not
result in ambiguities. The Ricci tensor $\ _{\shortmid }R_{\grave{\imath}%
\grave{j}}$ is computed for the Levi--Civita connection $\ _{\shortmid
}\nabla $ of $q_{\grave{\imath}\grave{j}}(\xi )$ parameterized by a real
variable $\xi ,$ $0\leq \xi <\xi _{0},$ for a differentiable function $\xi
(t)$ and can be distorted to 3-d nonholonomic canonical variables. In the
standard Riemannian approach, the boundary conditions in (\ref{heq1b}) are
stated for $\xi =0$ when a normalizing factor $\grave{r}=\int_{\ ^{c}%
\widehat{\Xi }}\ _{\shortmid }R\sqrt{|q_{\grave{\imath}\grave{j}}|}d\grave{x}%
^{3}/\int_{\ ^{c}\widehat{\Xi }}\sqrt{|q_{\grave{\imath}\grave{j}}|}d\grave{x%
}^{3}$ is introduced in order to preserve the volume of $\ ^{c}\widehat{\Xi }%
,$ i.e. $\int_{\ ^{c}\Xi }\sqrt{|q_{\grave{\imath}\grave{j}}|}d\grave{x}%
^{3}. $

To generate solutions of (\ref{heq1b}) for $q_{\grave{\imath}\grave{j}%
}\subset g_{\alpha \beta }$ with $g_{\alpha \beta }$ considered as a
solution of the 4-d (modified) Einstein equations we have to relate a
nontrivial normalizing factor $\grave{r}$ and a respective cosmological
constant. The equation (\ref{heq1a}) can be written in any nonholonomic
basis using respective formulas for hypersurface geometric evolution of
frame fields, $\ \partial _{\xi }e_{\grave{\imath}}^{\ \underline{\grave{%
\imath}}}=q^{\underline{\grave{\imath}}\underline{\grave{j}}}\ _{\shortmid
}R_{\underline{\grave{j}}\underline{\grave{k}}}e_{\grave{\imath}}^{\
\underline{\grave{k}}},$ when $q_{\grave{\imath}\grave{j}}(\xi )=q_{%
\underline{\grave{\imath}}\underline{\grave{j}}}(\xi )e_{\grave{\imath}}^{\
\underline{\grave{\imath}}}(\xi )e_{\grave{j}}^{\ \underline{\grave{j}}}(\xi
)$ for $e_{\grave{\imath}}(\xi )=e_{\grave{\imath}}^{\ \underline{\grave{%
\imath}}}(\xi )\partial _{\underline{\grave{\imath}}}$ and $e^{j}(\xi )=e_{\
\underline{\grave{j}}}^{j}(\xi )dx^{\underline{\grave{j}}}.$ As in standard
Hamilton--Perelman theory, there is a unique solution for such systems of
linear ordinary differential equations, ODEs, for any $\xi \in \lbrack 0,\xi
_{0})$ and such a solution can be extended for a family of 3-d hypersurfaces.

In nonholonomic variables and for the hypersurface canonical d--connection $%
\ _{\shortmid }\widehat{\mathbf{D}},$ the Perelman's functionals can be
written in terms of integrals on families of 3-d hypersurfaces%
\begin{equation}
\ _{\shortmid }\widehat{\mathcal{F}} = \int_{\widehat{\Xi }_{t}}e^{-f}\sqrt{%
|q_{\grave{\imath}\grave{j}}|}d\grave{x}^{3}(\ _{\shortmid }^{s}\widehat{R}%
+|\ _{\shortmid }\widehat{\mathbf{D}}f|^{2}), \mbox{ and } \ _{\shortmid }%
\widehat{\mathcal{W}} = \int_{\widehat{\Xi }_{t}}M\sqrt{|q_{\grave{\imath}%
\grave{j}}|}d\grave{x}^{3}[\xi (\ _{\shortmid }^{s}\widehat{R}+|\ \
_{\shortmid }^{h}\widehat{\mathbf{D}}f|+|\ \ _{\shortmid }^{v}\widehat{%
\mathbf{D}}f|)^{2}+f-6],  \label{3dwp}
\end{equation}%
where the scaling function $f$ satisfies $\int_{\widehat{\Xi }_{t}}M\sqrt{%
|q_{\grave{\imath}\grave{j}}|}d\grave{x}^{3}=1$ for $M=\left( 4\pi \xi
\right) ^{-3}e^{-f}$ $.$ The functionals $\ _{\shortmid }\widehat{\mathcal{F}%
}$ and $\ _{\shortmid }\widehat{\mathcal{W}}$ \ transform into standard
Perelman functionals on a hypersurface $\widehat{\Xi }_{t}$ if $\
_{\shortmid }\widehat{\mathbf{D}}\rightarrow \ _{\shortmid }\nabla .$ The
W--entropy $\ _{\shortmid }\widehat{\mathcal{W}}$ \ is a Lyapunov type
non--decreasing functional and can be considered as an alternative to the
Hawking-Bekenstein entropy for the case of hypersufaces for BHs and various
holographic generalizations. Such entropy type functionals can be used for
elaborating hypersurface thermodynamic models and computing respective
statistical distribution and energy functionals (we shall consider this in
next sections for 4-d configurations).

\subsubsection{Nonholonomic Ricci flow equations for 3--d hypersurface
metrics}

We can consider another type dependencies of geometric objects in formulas (%
\ref{3dwp}) on a smooth parameter $\upsilon (\xi )$ for which $\partial
\upsilon /\partial \xi =-1.$ For simplicity, \ we can omit the normalization
terms. Elaborating on a variational N-adapted calculus or using geometric
abstract symbolic methods for global constructions on manifolds, we prove
the nonholonomic geometric evolution (modified by nonholonomic distortion)
Hamilton equations for any induced 3--d metric $\mathbf{q}$ and canonical
d--connection $\ _{\shortmid }\widehat{\mathbf{D}}. $ We obtain a flow
evolution system of PDEs,
\begin{eqnarray}
\partial _{\upsilon }\mathbf{q}_{\grave{\imath}\grave{j}} &=&-2(\mathbf{\ }%
_{\shortmid }\widehat{\mathbf{R}}_{\grave{\imath}\grave{j}}+\mathbf{\ }%
_{\shortmid }\widehat{\mathbf{Z}}ic_{\grave{\imath}\grave{j}}),
\label{heq1c} \\
\mathbf{\ }_{\shortmid }\widehat{\mathbf{R}}_{i\grave{a}} &=&-\mathbf{\ }%
_{\shortmid }\widehat{\mathbf{Z}}ic_{i\grave{a}},  \label{heq1d} \\
\partial _{\upsilon }f &=&-(\mathbf{\ }_{\shortmid }\widehat{\Delta }-%
\mathbf{\ }_{\shortmid }^{Z}\widehat{\Delta })f+|(\ _{\shortmid }\widehat{%
\mathbf{D}}-\mathbf{\ }_{\shortmid }\widehat{\mathbf{Z}})f|^{2}-\mathbf{\ }%
_{\shortmid }^{s}\widehat{R}-\mathbf{\ }_{\shortmid }\widehat{\mathbf{Z}}.
\notag
\end{eqnarray}%
The distortion tensors in these equations are completely determined by $%
\mathbf{q}_{\grave{\imath}\grave{j}},$ see formulas (\ref{distrel2}), when
\begin{equation*}
\partial _{\upsilon }\ _{\shortmid }\widehat{\mathcal{F}}(q,\ _{\shortmid }%
\widehat{\mathbf{D}},f)=2\int_{\ \ ^{c}\widehat{\Xi }_{t}}e^{-f}\sqrt{|q_{%
\grave{\imath}\grave{j}}|}d\grave{x}^{3}[|\ _{\shortmid }\widehat{\mathbf{D}}%
_{\grave{\imath}\grave{j}}+\ _{\shortmid }\widehat{\mathbf{Z}}ic_{\grave{%
\imath}\grave{j}}+(\ _{\shortmid }\widehat{\mathbf{D}}_{\grave{\imath}}-\
_{\shortmid }\widehat{\mathbf{Z}}_{\grave{\imath}})(\ _{\shortmid }\widehat{%
\mathbf{D}}_{\grave{j}}-\ _{\shortmid }\widehat{\mathbf{Z}}_{\grave{j}%
})f|^{2}],
\end{equation*}%
when the normalizing function $f$ is subjected to the condition that $%
\int_{\ ^{c}\widehat{\Xi }_{t}}e^{-f}\sqrt{|q_{\grave{\imath}\grave{j}}|}d%
\grave{x}^{3}$ is constant for a fixed $\xi $ and $f(\upsilon (\xi ))=f(\xi
).$

The system of nonlinear PDEs (\ref{heq1c}) and (\ref{heq1d}) is equivalent
to (\ref{heq1b}) (for self-similar conditions when $\partial _{\upsilon }%
\mathbf{q}_{\grave{\imath}\grave{j}}=0,$ we obtain 3-d Ricci soliton
equations) up to certain re--definition of nonholonomic frames and
variables. We have to consider (\ref{heq1d}) as additional constraints
because in nonholonomic variables the Ricci d--tensor is (in general)
nonsymmetric.

\subsubsection{Geometric evolution to nonholonomic 4--d Lorentz
configurations}

The geometric evolution of 3-d d-metrics embedded into 4-d d-metrics $%
\mathbf{q}(\xi (t))\subset \mathbf{g}(\xi (t)):=$\newline
$\left( \mathbf{q}(\xi (t))\mathbf{,}\ _{q}N(\xi (t))\right) $ is described
by respective generalizations of functionals (\ref{3dwp}) resulting in
nonholonomic deformations of the Hamilton equations.

For foliations $\widehat{\Xi }_{t}$ adapted to a N-connection structures and
parameterized by a spacetime coordinate $t,$ we can introduce such 4--d
functionals
\begin{equation*}
\widehat{\mathcal{F}}(\mathbf{q},\ _{\shortmid }\widehat{\mathbf{D}}%
,f)=\int_{t_{1}}^{t_{2}}dt\ \ _{q}N(\xi )\ _{\shortmid }\widehat{\mathcal{F}}%
(\mathbf{q},\ _{\shortmid }\widehat{\mathbf{D}},f),\mbox{ and } \widehat{%
\mathcal{W}}(\mathbf{q},\ _{\shortmid }\widehat{\mathbf{D}}%
,f)=\int_{t_{1}}^{t_{2}}dt\ \ _{q}N(\xi )\ \ _{\shortmid }\widehat{\mathcal{W%
}}(\mathbf{q},\ _{\shortmid }\widehat{\mathbf{D}},f(\xi )).
\end{equation*}%
In these relativistic flow formulas, the parameter $\xi $ for 3-d Ricci
flows is extended as a time like coordinate on a open region in a 4-d $%
\mathbf{V}$ determined by a Lorentzian d-metric $\mathbf{g}(\xi ),$ For
elaborating alternative models (which consists the main goals of this work),
we can consider additional dependencies on a temperature parameter $\tau $
both for the 3-d and 4-d configurations. Using frame transforms on 4-d
nonholonomic Lorentz manifolds, such values can be re--defined respectively
in terms of data $(\mathbf{g}(\tau ),\widehat{\mathbf{D}}(\tau )).$ We
obtain
\begin{equation}
\widehat{\mathcal{F}}(\tau ) =\int_{t_{1}}^{t_{2}}\int_{\widehat{\Xi }%
_{t}}e^{-\widehat{f}}\sqrt{|\mathbf{g}_{\alpha \beta }|}d^{4}u(\ ^{s}R+|%
\widehat{\mathbf{D}}\widehat{f}|^{2}),\ \widehat{\mathcal{W}}(\tau ) =
\int_{t_{1}}^{t_{2}}\int_{\widehat{\Xi }_{t}}\widehat{M}\sqrt{|\mathbf{g}%
_{\alpha \beta }|}d^{4}u[\tau (\ ^{s}R+|\ ^{h}\widehat{\mathbf{D}}\widehat{f}%
|+|\ ^{v}\widehat{\mathbf{D}}\widehat{f}|)^{2}+\widehat{f}-8],
\label{wfperelm4}
\end{equation}%
where the scaling function $\widehat{f}$ satisfies $\int_{t_{1}}^{t_{2}}%
\int_{\widehat{\Xi }_{t}}\widehat{M}\sqrt{|\mathbf{g}_{\alpha \beta }|}%
d^{4}u=1$ for $\widehat{M}=\left( 4\pi \tau \right) ^{-3}e^{-\widehat{f}}$.
In these formulas $\tau $ is a temperature like parameter describing
relativistic evolution of geometric objects and associated thermodynamic
values.

The functionals $\widehat{\mathcal{F}}(\tau )$ and $\widehat{\mathcal{W}}%
(\tau )$ \ for 4-d pseudo-Riemannian metrics are not just static
thermodynamic entropies like in the 3--d Riemannian case. They determine
certain relativistic thermo field models with flows of entropic values on
respective temperature like parameter $\tau $ and a time like coordinate $%
\xi .$ Nevertheless, they describe nonlinear general relativistic diffusion
type processes if $\mathbf{q}\subset \mathbf{g}\ $\ and $\ _{\shortmid }%
\widehat{\mathbf{D}}$ $\subset \widehat{\mathbf{D}}$ are determined by
certain lapse and shift functions as certain solutions of 4--d gravitational
equations. The canonical values (\ref{wfperelm4}) can be geometrically and
physically motivated for any exact/parametric solution in GR or MGT, when
the AFDM is applicable. It is not clear if such functionals are well defined
for arbitrary geometric flows of pseudo-Riemannian metrics.

The systems of nonliner PDEs (\ref{heq1c}) and (\ref{heq1d}) can be
generalized to 4--d configurations when the coefficients are determined by
the Ricci d--tensors and distortions contain the left label "$\shortmid ".$
Using the hypersurface d-metric $\mathbf{q}_{\alpha \beta }=\mathbf{g}%
_{\alpha \beta }+\mathbf{n}_{\alpha }\mathbf{n}_{\beta },$ we write%
\begin{eqnarray}
\partial _{\upsilon }\mathbf{g}_{\alpha \beta } &=&-2(\mathbf{\ }_{\shortmid
}\widehat{\mathbf{R}}_{\alpha \beta }+\ _{\shortmid }\widehat{\mathbf{Z}}%
ic_{\alpha \beta })-\partial _{\upsilon }(\mathbf{n}_{\alpha }\mathbf{n}%
_{\beta }),  \label{heq1cc} \\
\ _{\shortmid }\widehat{\mathbf{R}}_{ia} &=&-\ _{\shortmid }\widehat{\mathbf{%
Z}}ic_{ia},\mbox{
for }\ _{\shortmid }\widehat{\mathbf{R}}_{\alpha \beta }\mbox{ with }\alpha
\neq \beta .  \notag
\end{eqnarray}%
The term $\partial _{\upsilon }(\mathbf{n}_{\alpha }\mathbf{n}_{\beta })$
can be computed in explicit form using formulas for the geometric
relativistic evolution of N--adapted frames.\footnote{%
There are some important formulas on flow evolution on a parameter $\upsilon
\in \lbrack 0,\upsilon _{0})$ of N--adapted frames in a 4--d nonholonomic
Lorentz manifold computed as $\mathbf{e}_{\alpha }(\upsilon )=\ \mathbf{e}%
_{\alpha }^{\ \underline{\alpha }}(\upsilon ,u)\partial _{\underline{\alpha }%
}$. For N-adapted frame/coordinate transforms, the frame coefficients are
\begin{equation*}
\ \mathbf{e}_{\alpha }^{\ \underline{\alpha }}(\upsilon ,u)=\left[
\begin{array}{cc}
\ e_{i}^{\ \underline{i}}(\upsilon ,u) & -~N_{i}^{b}(\upsilon ,u)\ e_{b}^{\
\underline{a}}(\upsilon ,u) \\
0 & \ e_{a}^{\ \underline{a}}(\upsilon ,u)%
\end{array}%
\right] ,\ \mathbf{e}_{\ \underline{\alpha }}^{\alpha }(\upsilon ,u)\ =\left[
\begin{array}{cc}
e_{\ \underline{i}}^{i}=\delta _{\underline{i}}^{i} & e_{\ \underline{i}%
}^{b}=N_{k}^{b}(\upsilon ,u)\ \ \delta _{\underline{i}}^{k} \\
e_{\ \underline{a}}^{i}=0 & e_{\ \underline{a}}^{a}=\delta _{\underline{a}%
}^{a}%
\end{array}%
\right] ,
\end{equation*}%
when the h- and v- components of a d-metric evolve as $\tilde{g}%
_{ij}(\upsilon )=\ e_{i}^{\ \underline{i}}(\upsilon ,u)\ e_{j}^{\ \underline{%
j}}(\upsilon ,u)\eta _{\underline{i}\underline{j}}$ and $\tilde{g}%
_{ab}(\upsilon )=\ e_{a}^{\ \underline{a}}(\upsilon ,u)\ e_{b}^{\ \underline{%
b}}(\upsilon ,u)\eta _{\underline{a}\underline{b}}.$ For local
parameterizations of Minkowski type with $\eta _{\underline{i}\underline{j}%
}=diag[+,+]$ and $\eta _{\underline{a}\underline{b}}=diag[+,-]$
corresponding to the chosen signature of $\ \mathbf{\tilde{g}}_{\alpha \beta
}^{[0]}(u),$ we have the evolution equations $\frac{\partial }{\partial
\upsilon }\mathbf{e}_{\ \underline{\alpha }}^{\alpha }\ =\ \mathbf{g}%
^{\alpha \beta }~\widehat{\mathbf{R}}_{\beta \gamma }~\ \mathbf{e}_{\
\underline{\alpha }}^{\gamma }.$}

For 4-d configurations with a corresponding re--definition of the scaling
function, $f\rightarrow \widehat{f},$ and using necessary type N--adapted
distributions, we can construct models of geometric evolution of vacuum
gravitational fields with $h$-- and $v$--splitting for $\widehat{\mathbf{D}}%
, $
\begin{eqnarray}
\partial _{\upsilon }\mathbf{g}_{ij} &=&-2\widehat{\mathbf{R}}_{ij},\
\partial _{\upsilon }\mathbf{g}_{ab}=-2\widehat{\mathbf{R}}_{ab},
\label{rfcandc} \\
\partial _{\upsilon }\widehat{f} &=&-\widehat{\square }\widehat{f}%
+\left\vert \widehat{\mathbf{D}}\widehat{f}\right\vert ^{2}-\ ^{s}\widehat{R}%
,  \notag
\end{eqnarray}%
Proofs are possible if we follow a similar calculus to that presented in the
proof of Proposition 1.5.3 of \cite{monogrrf1} but in N--adapted form.

\subsection{Nonholonomic geometric flows and Ricci solitons in canonical
variables}

We consider a family of Riemannian metrics, $g_{\grave{\imath}\grave{j}%
}(\tau )=g_{\grave{\imath}\grave{j}}(\tau ,x^{\grave{k}})$ parameterized by
a temperature like parameter $\tau $ and defined on a 3-d spacelike
hypersurface $\Xi \subset \mathbf{V}$. The Ricci flows of such metrics are
described by standard Hamilton equations
\begin{equation}
\frac{\partial g_{\grave{\imath}\grave{j}}}{\partial \tau }=-2\ R_{\grave{%
\imath}\grave{j}}[\nabla ],  \label{heq1}
\end{equation}%
where the Ricci tensor $R_{\grave{\imath}\grave{j}}$ is determined by
geometric data $(g_{\grave{\imath}\grave{j}},\nabla ).$ Rigorous
mathematical results on such systems of PDEs and related applications for
the proof of the Thurston-Poincar\'{e} conjecture can be found in \cite%
{hamilt1,perelman1,monogrrf1,monogrrf2,monogrrf3}.

The normalizing function in (\ref{canhamiltevol}) can be chosen in such a
form that for self-similar configurations with a fixed evolution parameter $%
\tau _{0}$ such PDEs transform into modified Einstein equations (\ref{deinst}%
) for NES. In h- and v--split components, such equations are written
\begin{equation*}
\widehat{\mathbf{R}}_{ij}=\widehat{\mathbf{\Upsilon }}_{ij};\ \widehat{%
\mathbf{R}}_{ab}=\widehat{\mathbf{\Upsilon }}_{ab};\widehat{\mathbf{R}}_{ia}=%
\widehat{\mathbf{R}}_{ai}=0;\ \widehat{\mathbf{R}}_{ij}=\widehat{\mathbf{R}}%
_{ji};\ \widehat{\mathbf{R}}_{ab}=\widehat{\mathbf{R}}_{ba}.
\end{equation*}%
In geometric flow theory, such equations are called as (nonholonomic) Ricci
solitons if they are of type (\ref{friedsoliton}) but for the canonical
d-connection and another type fixing of the evolution parameter,
\begin{equation}
\widehat{\mathbf{R}}_{\alpha \beta }-\lambda \ \mathbf{g}_{\alpha \beta }=%
\widehat{\mathbf{D}}_{\alpha }\mathbf{v}_{\beta }+\widehat{\mathbf{D}}%
_{\beta }\mathbf{v}_{\alpha }.  \label{canriccisol}
\end{equation}%
Such systems of nonlinear PDEs are determined by some geometric data $(%
\mathbf{g,N,}\widehat{\mathbf{D}})$ and a cosmological constant $\lambda $
and d-vector field $\mathbf{v}_{\alpha }(u).$

The gravitational field equations for nonholonomic Einstein systems, NES,
can be written in similar forms to the nonholonomic Ricci soliton equations
in canonical variables,
\begin{equation}
\widehat{\mathbf{R}}_{\alpha \beta }=\widehat{\mathbf{\Upsilon }}_{\alpha
\beta }.  \label{deinst}
\end{equation}%
Such equations transform into the standard Einstein equation in GR if there
are imposed additional nonholonomic constraints, or found some smooth
limits, for extracting LC--configurations, $\widehat{\mathbf{D}}_{\mid
\widehat{\mathcal{T}}=0}=\nabla ,$ when $\ \widehat{\mathbf{T}}_{\ \alpha
\beta }^{\gamma }=0$ and the sources $\widehat{\mathbf{\Upsilon }}_{\alpha
\beta }$ are constructed as in standard particle physics but with
distortions of linear connections (\ref{distr}). Here we note that a
nonholonomic vacuum Einstein space is characterized by a more rich geometric
structure which can be with generic off-diagonal and locally anisotropic
interactions, encoding nonlinear evolution scenarios etc.

In (\ref{deinst}), there are considered effective and matter fields sources
of type $\widehat{\mathbf{\Upsilon }}_{\mu \nu }=\ ^{e}\widehat{\mathbf{%
\Upsilon }}_{\mu \nu }+\ ^{m}\widehat{\mathbf{\Upsilon }}_{\mu \nu },$ where
$\ ^{e}\widehat{\mathbf{\Upsilon }}_{\mu \nu }$ are effective sources
determined by distortions of the linear connections and effective
Lagrangians for gravitational fields in MGTs \cite%
{kk,odintsov1,gheorghiu16,rajpoot17,bubuianu18,bubuianu19}. Such a source is
not zero even in GR if there are nonzero distortions (\ref{distr}) from $%
\widehat{\mathbf{D}}$ to $\nabla .$ The source for matter field, $\ ^{m}%
\widehat{\mathbf{\Upsilon }}_{\mu \nu },$ can be constructed using a
N--adapted variational calculus for $\ ^{m}\mathcal{L}(\mathbf{g,}\widehat{%
\mathbf{D}},\ \ ^{A}\varphi ),$ when
\begin{equation*}
\ ^{m}\widehat{\mathbf{\Upsilon }}_{\mu \nu }=\varkappa (\ ^{m}\widehat{%
\mathbf{T}}_{\mu \nu }-\frac{1}{2}\mathbf{g}_{\mu \nu }\ ^{m}\widehat{%
\mathbf{T}})\rightarrow \varkappa (\ ^{m}T_{\mu \nu }-\frac{1}{2}\mathbf{g}%
_{\mu \nu }\ ^{m}T)
\end{equation*}%
for [coefficients of $\ \widehat{\mathbf{D}}]$ $\rightarrow $ [coefficients
of $\nabla $] even, in general, $\widehat{\mathbf{D}}\neq \nabla .$ In such
formulas, we consider $\ ^{m}\widehat{\mathbf{T}}=\mathbf{g}^{\mu \nu }\ ^{m}%
\widehat{\mathbf{T}}_{\mu \nu }$ for $\ ^{m}\widehat{\mathbf{T}}_{\alpha
\beta }:=-\frac{2}{\sqrt{|\mathbf{g}_{\mu \nu }|}}\frac{\delta (\sqrt{|%
\mathbf{g}_{\mu \nu }|}\ \ ^{m}\mathcal{L})}{\delta \mathbf{g}^{\alpha \beta
}}.$ In this work, we shall consider only $\ ^{m}\mathcal{L}=\ ^{\phi }%
\mathcal{L}(\mathbf{g,}\widehat{\mathbf{D}},\ \phi )$ determined by a scalar
field $\phi (x,u)$ and/or geometric evolution of scalar fields $\phi (\tau
)=\phi (\tau ,x,u),$ when $\ ^{m}\widehat{\mathbf{T}}_{\alpha \beta }=\
^{\phi }\widehat{\mathbf{T}}_{\alpha \beta }.$

The geometric flow equations (\ref{heq1}) can be generalized for 4-d
Lorentzian metrics describing modified Ricci flows of NES. In N-adapted from
and for the canonical d-connection, such equations are written in the form
\begin{eqnarray}
\frac{\partial \mathbf{g}_{ij}}{\partial \tau } &=&-2\left( \widehat{\mathbf{%
R}}_{ij}-\widehat{\mathbf{\Upsilon }}_{ij}\right) ;\ \frac{\partial \mathbf{g%
}_{ab}}{\partial \tau }=-2\left( \widehat{\mathbf{R}}_{ab}-\widehat{\mathbf{%
\Upsilon }}_{ab}\right) ;  \label{canhamiltevol} \\
\widehat{\mathbf{R}}_{ia} &=&\widehat{\mathbf{R}}_{ai}=0;\ \widehat{\mathbf{R%
}}_{ij}=\widehat{\mathbf{R}}_{ji};\ \widehat{\mathbf{R}}_{ab}=\widehat{%
\mathbf{R}}_{ba};  \notag \\
\partial _{\tau }\ \widehat{f} &=&-\widehat{\square }\widehat{f}+\left\vert
\ \widehat{\mathbf{D}}\ \widehat{f}\right\vert ^{2}-\ \ _{s}\widehat{R}+%
\widehat{\mathbf{\Upsilon }}_{a}^{a},  \label{normalizfunct}
\end{eqnarray}%
where $\ \widehat{\square }(\tau )=\widehat{\mathbf{D}}^{\alpha }(\tau )\
\widehat{\mathbf{D}}_{\alpha }(\tau )$ is used for the geometric flows of
the d'Alambert operator and sources $\widehat{\mathbf{\Upsilon }}_{\alpha
\beta }(\tau )=[\widehat{\mathbf{\Upsilon }}_{ij}(\tau ),\widehat{\mathbf{%
\Upsilon }}_{ab}(\tau )]$ are constructed as for (\ref{deinst}) but for
d-metrics and respective geometric objects with $\tau $-parameter
dependence. A normalization function $\widehat{f}(\tau ,u)$ has to be
introduced for proofs of such systems of nonlinear PDEs from certain
nonholonomically generalized F- and W-functionals (see next section and \cite%
{vacaru09,ruchin13,gheorghiu16,rajpoot17}). Various classes of exact and
parametric solutions describing geometric flow evolution of off-diagonal
stationary and cosmological configurations can be described using the AFDM
\cite{vacaru11,vacaru18tc,bubuianu18,bubuianu19,vacaru19a}.

We emphasize that all systems of PDEs for geometric and physical models on $%
\mathbf{V}$ can be expressed in different equivalent forms using different
geometric data $\ (\mathbf{g,N,}\nabla )\Leftrightarrow (\mathbf{g,N,}%
\widehat{\mathbf{D}})\Leftrightarrow (\mathbf{g,N,D}),$ involving respective
distortion relations. Certain classes of nonholonomic variables (geometric
data) are convenient, for instance, for elaborating canonical methods of
quantization, other type ones can be useful for finding exact solutions. In
this work, we shall give priority to $\widehat{\mathbf{D}}$ because such a
formalism allows us to encode directly into geometric flow and information
theories various classes of generic off-diagonal solutions.

\subsection{G. Perelman functionals and geometric thermodynamic models for
gravity}

The aim of this subsection is to define nonholonomic canonical modifications
of the F- and W-functionals \cite{perelman1} which allow us to prove the
nonholonomic geometric flow equations (\ref{canhamiltevol}). The W-entropy
will be used for constructing an associated statistical thermodynamic model
for geometric flows of NES.

\subsubsection{F- and W-functionals for nonholonomic Einstein systems}

G. Perelman entropic like functionals can be postulated using different
types of nonholonomic variables with conventional 2+2 and 3+1 decomposition
of dimensions or double fibration splitting \cite%
{ruchin13,gheorghiu16,rajpoot17}.

In nonholonomic canonical variables, the relativistic versions of G.
Perelman functionals are postulated
\begin{eqnarray}
\widehat{\mathcal{F}} &=&\int \left( 4\pi \tau \right) ^{-2}e^{-\widehat{%
\underline{f}}}\sqrt{|\mathbf{g}|}d^{4}u(\ _{s}\widehat{R}+|\widehat{\mathbf{%
D}}\widehat{f}|^{2})\mbox{
and }  \label{ffcand} \\
\widehat{\mathcal{W}} &=&\int \widehat{\mu }\sqrt{|\mathbf{g}|}d^{4}u[\tau
(\ _{s}\widehat{R}+|\ \ _{h}\widehat{\mathbf{D}}\widehat{f}|+|\ \ _{v}%
\widehat{\mathbf{D}}\widehat{f}|)^{2}+\widehat{f}-4],  \label{wfcand}
\end{eqnarray}%
where the normalizing function $\widehat{f}(\tau ,u)$ is subjected to the
conditions $\int \widehat{\mu }\sqrt{|\mathbf{g}|}d^{4}u=%
\int_{t_{1}}^{t_{2}}\int_{\Xi _{t}}\widehat{\mu }\sqrt{|\mathbf{g}|}d^{4}u=1$%
, for a classical integration measure$\ \widehat{\mu }=\left( 4\pi \tau
\right) ^{-2}e^{-\widehat{f}}$ and the Ricci scalar $\ _{s}\widehat{R}$ is
taken for the Ricci d-tensor $\widehat{\mathbf{R}}_{\alpha \beta }$ of a
d-connection $\widehat{\mathbf{D}}.$

Applying a N-adapted variational calculus on $\mathbf{g}_{\alpha \beta }$ (%
\ref{ffcand}) or (\ref{wfcand}) (see similar details in \cite%
{ruchin13,gheorghiu16,rajpoot17} and, for Riemannian configurations, in \cite%
{monogrrf1,monogrrf2,monogrrf3}), we can prove the geometric flow evolution
equations (\ref{canhamiltevol}) for NES. Here we note that the functional $%
\widehat{\mathcal{W}}$ (\ref{wfcand}) defines a nonholonomic canonical and
relativistic generalizations of so-called W-entropy introduced in \cite%
{perelman1}. Various types of 4-d - 10-d $\mathcal{W}$--entropies and
associated statistical and quantum thermodynamics values are used for
elaborating models of classical and (commutative and noncommutative/
supersymmetric) quantum geometric flows and geometric information flows, see
\cite{vacaru07,vacaru09b,vacaru09,ruchin13,gheorghiu16,rajpoot17} and our
partner works \cite{vacaru19b,vacaru19c,vacaru19e}.

\subsubsection{Thermodynamic models for (modified) Einstein geometric flows}

\label{ssthmodels}A geometric flow evolution of NES can be characterized by
analogous thermodynamic models. Respective geometric and statistical
thermodynamic values can be defined in nonholonomic canonical variables
(with hats, which is different from tilde, underlined and other type
variables used for GIF and QGIF theories in partner works \cite%
{bubuianu19,vacaru19b,vacaru19c}).\footnote{\label{fndensst} Let us remember
some most important concepts and formulas from statistical thermodynamics. A
partition function $Z=\int \exp (-\beta E)d\omega (E)$ is considered for a
canonical ensemble at temperature $\beta ^{-1}=T$ and a measure $\omega (E) $
as density of states. In standard form, there are computed such canonical
values: average flow energy, $\mathcal{E}=\ \left\langle E\right\rangle
:=-\partial \log Z/\partial \beta ;$\ flow entropy, $\mathcal{S}:=\beta
\left\langle E\right\rangle +\log Z;$\ flow fluctuation, $\eta
:=\left\langle \left( E-\left\langle E\right\rangle \right)
^{2}\right\rangle =\partial ^{2}\log Z/\partial \beta ^{2}.$%
\par
A partition function (equivalently, thermodynamic generating function) $Z$
allows us to define a conventional \textit{state density} (for quantum
models, a \textit{density matrix}) $\ \sigma (\beta ,E)=Z^{-1}e^{-\beta E}.$
The \textit{relative entropy} between two state densities $\rho $ and $%
\sigma $ is defined/computed%
\begin{equation*}
\mathcal{S}(\rho \shortparallel \sigma ):=-\mathcal{S}(\rho )+\int (\beta
\mathcal{E+}\log Z)\rho d\omega (E)=\beta \lbrack \mathcal{E}(\rho )-T%
\mathcal{S}(\rho )]+\log Z.
\end{equation*}%
In this formula, the \textit{average energy} is computed for the density
matrix $\rho ,$ $\mathcal{E}(\rho )=\int \mathcal{E}\rho d\omega (E),$ and
the formula $\log \sigma =-\beta \mathcal{E-}\log Z$ is used. We define the
\textit{free energy} using formula $\mathcal{F}(\rho ):=\mathcal{E}(\rho )-T%
\mathcal{S}(\rho ).$ If $\log Z$ is independent on $\rho ,$ we get $\mathcal{%
S}(\sigma \shortparallel \sigma )=0$ and $\mathcal{S}(\rho \shortparallel
\sigma )=\beta \lbrack \mathcal{F}(\rho )-\mathcal{F}(\sigma )].$
Elaborating geometric flow evolution and analogous thermodynamics systems,
we consider that under evolution it is preserved the thermal equilibrium at
temperature $\beta $ with maps of density states $\rho \rightarrow \rho
^{\prime }$ keeping the same density state $\sigma .$ Such systems are
characterized by inequalities $\mathcal{S}(\rho \shortparallel \sigma )\geq
\mathcal{S}(\rho ^{\prime }\shortparallel \sigma ),$ and $\mathcal{F}(\rho
)\geq \mathcal{F}(\rho ^{\prime }).$}

We associate to (\ref{wfcand}) a respective thermodynamic generating
functions defined in canonical variables
\begin{equation}
\widehat{\mathcal{Z}}[\mathbf{g}(\tau )]=\int (4\pi \tau )^{-2}e^{-\widehat{f%
}}\sqrt{|\mathbf{g}|}d^{4}u(-\widehat{f}+2),\mbox{ for }\mathbf{V.}
\label{genfcanv}
\end{equation}%
Such values are with functional dependence on $\mathbf{g}(\tau )$ (we shall
not write this in explicit forms if it do not result in ambiguities). A
density state defined as in footnote \ref{fndensst} is a functional $%
\widehat{\sigma }[\mathbf{g}(\tau )]=\widehat{\mathcal{Z}}^{-1}e^{-\beta E}.$
We can consider also the geometric evolution densities $\widehat{\rho }[\
_{1}\mathbf{g}]$ and $\widehat{\rho }^{\prime }[\ _{1}\mathbf{g}],$ where
the left label 1 is used in order to distinguish two d-metrics $\mathbf{g}$
and $\ _{1}\mathbf{g}.$

Using (\ref{genfcanv}) and (\ref{wfcand}) and respective 3+1
parameterizations of d-metrics (see formulas in the last footnote and (\ref%
{decomp31}) with a time like coordinate $y^{4}=t$ and temperature like
evolution parameter $\tau ),$ we define and compute analogous thermodynamic
values for geometric evolution flows of NES,%
\begin{eqnarray}
\widehat{\mathcal{E}}\ &=&-\tau ^{2}\int (4\pi \tau )^{-2}e^{-\widehat{f}}%
\sqrt{|q_{1}q_{2}\mathbf{q}_{3}(_{q}N)|}\delta ^{4}u(\ \ _{s}\widehat{R}+|%
\widehat{\mathbf{D}}\widehat{f}|^{2}\mathbf{\ }-\frac{2}{\tau }),
\label{thvcanon} \\
\widehat{\mathcal{S}}\ &=&-\int (4\pi \tau )^{-2}e^{-\widehat{f}}\sqrt{%
|q_{1}q_{2}\mathbf{q}_{3}(_{q}N)|}\delta ^{4}u\left[ \tau \left( \ _{s}%
\widehat{R}+|\widehat{\mathbf{D}}\widehat{f}|^{2}\right) +\widehat{f}-4%
\right] ,  \notag \\
\widehat{\eta }\ &=&-2\tau ^{4}\int (4\pi \tau )^{-2}e^{-\widehat{f}}\sqrt{%
|q_{1}q_{2}\mathbf{q}_{3}(_{q}N)|}\delta ^{4}u[|\ \widehat{\mathbf{R}}%
_{\alpha \beta }+\widehat{\mathbf{D}}_{\alpha }\ \widehat{\mathbf{D}}_{\beta
}\widehat{f}-\frac{1}{2\tau }\mathbf{g}_{\alpha \beta }|^{2}],  \notag
\end{eqnarray}%
where $\delta ^{4}u$ contains N-elongated differentials in order to compute
such integrals in N-adapted forms. Using such values, we can compute in
canonical variables the respective free energy and relative entropy,%
\begin{equation*}
\widehat{\mathcal{F}\ }(\ _{1}\mathbf{g})=\widehat{\mathcal{S}}(\ _{1}%
\mathbf{g})-\beta ^{-1}\widehat{\mathcal{S}}(\ _{1}\mathbf{g})\mbox{ and }%
\widehat{\mathcal{S}}(\ _{1}\mathbf{g}\shortparallel \mathbf{g})=\beta
\lbrack \widehat{\mathcal{F}}(\ _{1}\mathbf{g})-\widehat{\mathcal{F}}(%
\mathbf{g})],\mbox{ where }
\end{equation*}%
\begin{eqnarray}
\widehat{\mathcal{E}}(\ _{1}\mathbf{g}) &=&-\tau ^{2}\int (4\pi \tau
)^{-2}e^{-\widehat{f}}\sqrt{|q_{1}q_{2}\mathbf{q}_{3}(_{q}N)|}\delta ^{4}u[\
_{s}\widehat{R}(\ _{1}\underline{\mathbf{g}})+|\widehat{\mathbf{D}}(\ _{1}%
\mathbf{g})\widehat{f}(\tau ,u)|^{2}\mathbf{\ }-\frac{2}{\tau }],
\label{telativenerg} \\
\widehat{\mathcal{S}}(\ _{1}\mathbf{g}) &=&-\int (4\pi \tau )^{-2}e^{-%
\widehat{f}}\sqrt{|q_{1}q_{2}\mathbf{q}_{3}(_{q}N)|}\delta ^{4}u\left[ \tau
\left( \ _{s}\widehat{R}(\ _{1}\mathbf{g})+|\widehat{\mathbf{D}}(\ _{1}%
\mathbf{g})\widehat{f}(\tau ,u)|^{2}\right) +\widehat{f}(\tau ,u)-4\right] .
\notag
\end{eqnarray}

Finally, we conclude that generating functions (\ref{genfcanv}) and
respective thermodynamical values (\ref{thvcanon}) can be written
equivalently in terms of the canonical d--connections $\widehat{\mathbf{D}}$
and/or for an another type $\mathbf{D}$ if we consider nonholonomic
deformations to certain systems of nonlinear PDEs on $\mathbf{V}.$

\section{Classical and quantum geometric information flows and gravity}

\label{s3} We formulate the theory of (quantum) geometric information flows
(respectively, GIFs and QGIFs) and nonholonomic Einstein systems, NES.

\subsection{The geometric information flow theory of nonholonomic Einstein
systems}

We consider for QGIGs and NES the basic aspects of classical information
theory with fundamental concepts of Shannon\footnote{%
Let $B$ is a random variable taking certain values $b_{\underline{1}},b_{%
\underline{2}},...,b_{\underline{k}},$ for instance, as a long message of
symbols $\underline{N}\gg 1$ containing different $k$ letters. The
respective probabilities to observe such values are denoted $p_{\underline{1}%
},p_{\underline{2}},...,p_{\underline{k}}.$ The Shannon entropy is defined $%
S_{B}:=-\sum\limits_{\underline{j}=1}^{\underline{k}}p_{\underline{j}}\log
p_{\underline{j}}\geq 0$ for $\sum\limits_{\underline{j}=1}^{\underline{k}%
}p_{\underline{j}}=1,$ when $\underline{N}S_{B}$ is the number of bits of
information which can be extracted from a message with $\underline{N}$
symbols. One could be correlations between letters for a more complex random
process. In the "ideal gaze" limit (ignoring correlations), we approximate
the entropy of a long message to be $\underline{N}S$ with the entropy $S$
for a message consisting of only one letter. For a statistical
thermodynamical model and a classical Hamiltonian $H,$ we determin the
probability of a $i$-th symbol $b_{i}$ via formula $p_{\underline{i}%
}=2^{-H(b_{\underline{i}})}.$}, conditional and relative entropies and
applications in modern physics, see \cite%
{preskill,witten18,nielsen,cover,wilde} and references therein. There are
used the W-entropy functional (\ref{wfcand}) and the associated
thermodynamical models elaborated in section \ref{ssthmodels} and partner
works \cite{vacaru19b,vacaru19c}.

For classical GIFs and NES, the canonical thermodynamic values are
determined by data $\left[ \widehat{\mathcal{W}};\widehat{\mathcal{Z}},%
\widehat{\mathcal{E}}, \widehat{\mathcal{S}},\widehat{\eta }\right] ,$ see (%
\ref{wfcand}) and (\ref{thvcanon}). We can introduce probabilities on a
discrete network with random variables, for instance, $\ \widehat{p}_{%
\underline{n}}=2^{-\ \widehat{\mathcal{E}}(b_{\underline{n}})}.$ Continuous
GIF models encoding geometric evolution of NES in canonical covariant
variables are characterized by the thermodynamic entropy $\widehat{\mathcal{S%
}}[\mathbf{g}(\tau )]$ and/or the W-entropy $\widehat{\mathcal{W}}[\mathbf{g}%
(\tau )]$ (\ref{wfcand}) and certain constructions without statistical
thermodynamics values. NES under canonical geometric evolution flows are
denoted in general form as $\widehat{B}=\widehat{B}[\mathbf{g}(\tau )];$
such systems are determined by flows of corresponding canonical d-metrics on
nonholonomic Lorentz spacetimes.

Now, let us consider how to construct models of information thermodynamics
determined by geometric flows of NES.\footnote{%
Let us remember some basic concepts from the classical information theory.
We consider a message with many letters when any letter is a random variable
$X$ taking possible values $x_{\underline{1}},...,x_{\underline{k}}.$ A
receiver get a random variable $Y$ defined by letters $y_{\underline{1}%
},...,y_{\underline{r}}$. The goal is to compute how many bits of
information a receiver will obtain form a message with $N$ letters (with
random variables $X,Y,Z,...$). For one variable, the probability to observe $%
X=x_{\underline{i}}$ is denoted $P_{X}(x_{\underline{i}})$ for $\sum_{%
\underline{i}}$ $P_{X}(x_{\underline{i}})=1.$ A sender and a receiver
communicate via a random process of two variables defined by a joint
distribution $P_{X,Y}(x_{\underline{i}},y_{\underline{j}})$ as the
probabilities, respectively, to send is $X=x_{\underline{i}}$ and to hear is
$Y=y_{\underline{j}}.$ $\ P_{Y}(y_{\underline{j}})=\sum_{\underline{i}%
}P_{X,Y}(x_{\underline{i}},y_{\underline{j}})$ is the probability to receive
$Y=y_{\underline{j}}$ when summation is taken over all choices that could be
send. By definition, the \textit{conditional probability} $P_{X|Y}(x_{%
\underline{i}}|y_{\underline{j}}):=\frac{P_{X,Y}(x_{\underline{i}},y_{%
\underline{j}})}{P_{Y}(y_{\underline{j}})}$ is a value characterizing
receiving $Y=y_{\underline{j}}.$ We estimate the probability that it was
sent $x_{i}.$ For receiver's messages, \newline
$P_{X}(x_{\underline{i}})=\sum_{\underline{j}}$ $P_{X,Y}(x_{\underline{i}%
},y_{\underline{j}})$ or we can consider $P_{X}(x_{\underline{i}})$ as an
independent probability density.
\par
There are defined such important values: $S_{X|Y=y_{j}}:=-\sum_{\underline{i}%
}P_{X|Y}(x_{\underline{i}}|y_{\underline{j}})\log P_{X|Y}(x_{\underline{i}%
}|y_{\underline{j}})$ is the Shanon entropy of the conditional probability; $%
S_{XY}:=-\sum_{\underline{i},\underline{j}}P_{X,Y}(x_{\underline{i}},y_{%
\underline{j}})\log P_{X,Y}(x_{\underline{i}},y_{\underline{j}})$ is the
entropy of the joint distribution;\newline
$S_{Y}:=-\sum_{\underline{i},\underline{j}}P_{X,Y}(x_{\underline{i}},y_{%
\underline{j}})\log P_{Y}(y_{\underline{j}})$ is the total received
information content; $S_{X}:=-\sum_{\underline{i},\underline{j}}P_{X,Y}(x_{%
\underline{i}},y_{\underline{j}})\log P_{X}(x_{\underline{i}})$ is the total
sent information content. Using such formulas, the \textit{conditional
entropy} is by definition $S_{X|Y}:=\sum_{\underline{j}}P_{Y}(y_{\underline{j%
}})S_{X|Y=y_{j}}=S(X|Y)=S_{XY}-S_{Y}\geq 0$. The \textit{mutual information}
between $X$ and $Y$ is defined $\ I(X;Y):=S_{X}-S_{XY}+S_{Y}\geq 0$ which is
a measure of how much we learn about $X$ observing $Y$.} Conventionally, we
work with two such GIFs and NES, $\widehat{A}=\widehat{A}[\mathbf{g}(\tau )]$
and $\widehat{B}=\widehat{B}[\ _{1}\mathbf{g}(\tau )].$ To study conditional
GIFs of gravitational systems we shall use geometric flow models on $\mathbf{%
V}\otimes \mathbf{V}$ when the local coordinates are $(u,\ _{1}u) $ and the
normalizing functions are of type $\ _{AB}\widehat{f}(u,\ _{1}u).$ A
d-metric structure on such tensor products of nonholonomic Lorentz manifolds
can be parameterized in the form$\ _{AB}\mathbf{g}=\{\mathbf{g}=[q_{1},q_{2},%
\mathbf{q}_{3},_{q}N],\ _{1}\mathbf{g}=[\ _{1}q_{1},\ _{1}q_{2},\ _{1}%
\mathbf{q}_{3},_{1q}N]\}.$ Respectively, we can define, for instance, a
canonical d--connection $\ _{AB}\widehat{\mathbf{D}}=\ _{A}\widehat{\mathbf{D%
}}+\ _{B}\widehat{\mathbf{D}}$ and corresponding scalar curvature $\ _{sAB}%
\widehat{R}=\ _{s}\widehat{R}+\ _{s1}\widehat{R}.$

The canonical thermodynamic GIF and NES entropies for respective systems are
$\widehat{\mathcal{S}}[\widehat{A}]$ and $\widehat{\mathcal{S}}[\widehat{B}]$
being respectively defined by $\mathbf{g}(\tau )$ and $\ _{1}\mathbf{g}(\tau
)$ as in (\ref{thvcanon}). They can be considered as analogs of $S_{X}$ and $%
S_{Y}$ used in the last footnote. As an analog of $S_{XY}$ for GIFs, we
introduce the thermodynamic generating function (as a generalization of (\ref%
{genfcanv}))
\begin{equation}
\ _{AB}\widehat{\mathcal{Z}}[\mathbf{g}(\tau ),\ _{1}\mathbf{g}(\tau )]=\
\int \ \ _{1}\int (4\pi \tau )^{-4}e^{-\ _{AB}\widehat{f}}\sqrt{|\ \mathbf{g}%
|}\sqrt{|\ _{1}\mathbf{g}|}d^{4}u\ d^{4}\ _{1}u(-\ _{AB}f+8),\mbox{ for
}\mathbf{V\otimes \mathbf{\mathbf{V}}.}  \label{twogenf}
\end{equation}%
This results in a GIF NES canonical thermodynamic entropy function
\begin{eqnarray}
\ _{AB}\widehat{\mathcal{S}}=\widehat{\mathcal{S}}\ [\widehat{A},\widehat{B}%
] &=&-\ \int \ _{1}\int (4\pi \tau )^{-4}e^{-\ _{AB}\widehat{f}}\sqrt{%
|q_{1}q_{2}\mathbf{q}_{3}(_{q}N)|}\sqrt{|\ _{1}q_{1}\ _{1}q_{2}\ \ _{1}%
\mathbf{q}_{3}(_{1q}N)|}\delta ^{4}u\ \delta ^{4}\ _{1}u  \notag \\
&&\ \left[ \tau \left( \ _{s}\widehat{R}+\ _{s1}\widehat{R}+|\ \widehat{%
\mathbf{D}}\ _{AB}\widehat{f}+\ _{1}\widehat{\mathbf{D}}\ _{AB}\widehat{f}%
|^{2}\right) +\ _{AB}\widehat{f}-8\right] .  \label{twoentr}
\end{eqnarray}%
Using these values, we claim (proofs can be performed in any point of
respective causal curves on Lorentz manifolds) that the formulas for the
conditional entropy and mutual information are respectively generalized for
GIFs of NES,
\begin{equation*}
\ \widehat{\mathcal{S}}\ [\widehat{A}|\widehat{B}]:=\ _{AB}\widehat{\mathcal{%
S}}-\ \widehat{\mathcal{S}}[\widehat{B}]\geq 0\mbox{
and }\ \widehat{\mathcal{J}}\ [\widehat{A};\widehat{B}]:=\ \widehat{\mathcal{%
S}}[\widehat{A}]-\ _{AB}\widehat{\mathcal{S}}+\widehat{\mathcal{S}}[\widehat{%
B}]\geq 0.
\end{equation*}%
Similar claims can be formulated for the W-entropy $\widehat{\mathcal{W}}[%
\mathbf{g}(\tau )]$ (\ref{wfcand}),
\begin{equation*}
\widehat{\mathcal{W}}\ [\widehat{A}|\widehat{B}]:=\ _{AB}\widehat{\mathcal{W}%
}-\ \widehat{\mathcal{W}}[\widehat{B}]\geq 0\mbox{ and }\ \widehat{\mathcal{J%
}}\ [\widehat{A};\widehat{B}]:=\widehat{\mathcal{W}}[\widehat{A}]-\ _{AB}%
\widehat{\mathcal{W}}+\ \widehat{\mathcal{W}}[\widehat{B}]\geq 0.
\end{equation*}%
These formulas can be are computed respectively for the W--entropy instead
of the S-entropy used in the standard probability theory and generalizations
in information theory.

We can define and calculate the relative entropy $S$ and mutual information $%
I$ between two distributions following definitions of the standard
probability and information theory\footnote{%
For convenience, we remember some basic formulas on relative entropy and
mutual information which are necessary for considerations in this paper. In
this paragraph, we do not use "hats" on respective symbols because such
values can be defined in general form not obligatory encoding canonical
nonholonomic variables.\ The relative entropy is introduced for two
probability distributions $P_{X}$ and $Q_{X}.$ Considering $X=x_{\underline{i%
}},$ with \underline{$i$}$=\{1,2,...s\},$ we state $p_{\underline{i}%
}=P_{X}(x_{\underline{i}})$ and $q_{\underline{i}}=Q_{X}(x_{\underline{i}})$
for some long messages with $\underline{N}$ letters. Our goal is to decide
which distribution describes a random process more realistically. Let us
define the relative entropy per observation $S(P_{X}||Q_{X}):=\sum_{%
\underline{i}}p_{\underline{i}}(\log p_{\underline{i}}-\log q_{\underline{i}%
})\geq 1$ under the assumption that $\underline{N}S(P_{X}||Q_{X})\gg 1.$
This value is asymmetric both on $P_{X}$ and $Q_{X}$ and measures the
difference between these two probability distributions (it is considered
that $P_{X}$ is for the correct answer and $Q_{X}$ is taken as an initial
hypothesis).
\par
At the next step, we can consider a pair of random variables $X$ and $Y$ and
respective two probability distributions. The fist one is taken as a
possible correlated joint distribution $P_{X,Y}(x_{\underline{i}},y_{%
\underline{j}})$ and $P_{X}(x_{\underline{i}}):=\sum_{\underline{j}%
}P_{X,Y}(x_{\underline{i}},y_{\underline{j}}),P_{Y}(y_{\underline{j}%
}):=\sum_{\underline{i}}P_{X,Y}(x_{\underline{i}},y_{\underline{j}}).$ We
also use a second probability distribution $Q_{X,Y}(x_{\underline{i}},y_{%
\underline{j}})=P_{X}(x_{\underline{i}})$ $P_{Y}(y_{\underline{j}})$ defined
in a form ignoring correlations between $X$ and $Y.$ In general, $Q_{X,Y}(x_{%
\underline{i}},y_{\underline{j}})$ can be with correlations of type $%
Q_{X}(x_{\underline{i}}):=\sum_{\underline{j}}Q_{X,Y}(x_{\underline{i}},y_{%
\underline{j}}).$ We can introduce three random variables $X,Y,Z$ described
by a joint probability distribution and related values, $P_{X,Y,Z}(x_{%
\underline{i}},y_{\underline{j}},z_{\underline{k}})$ and $P_{X}(x_{%
\underline{i}}):=\sum_{\underline{j},\underline{k}}P_{X,Y,Z}(x_{\underline{i}%
},y_{\underline{j}},z_{\underline{k}}),P_{Y,Z}(y_{\underline{j}},z_{%
\underline{k}}):=\sum_{\underline{i}}P_{X,Y,Z}(x_{\underline{i}},y_{%
\underline{j}},z_{\underline{k}})$. Ignoring the correlations between $X$
and $YZ,$ we define $Q_{X,Y,Z}(x_{\underline{i}},y_{\underline{j}},z_{%
\underline{k}}):=P_{X}(x_{\underline{i}})P_{Y,Z}(y_{\underline{j}},z_{%
\underline{k}}).$ Other type values can be defined for observation of the
subsystem $XY,$ when $P_{X,Y}(x_{\underline{i}},y_{\underline{j}}):=\sum_{%
\underline{k}}P_{X,Y,Z}(x_{\underline{i}},y_{\underline{j}},z_{\underline{k}%
}),Q_{X,Y}(x_{\underline{i}},y_{\underline{j}}):=\sum_{\underline{k}%
}Q_{X,Y,Z}(x_{\underline{i}},y_{\underline{j}},z_{\underline{k}})=P_{X}(x_{%
\underline{i}})P_{Y}(y_{\underline{j}})$.},
\begin{equation*}
S(P_{X}||Q_{X}):=\sum_{i,j}P_{X,Y}(x_{\underline{i}},y_{\underline{j}})[\log
P_{X,Y}(x_{\underline{i}},y_{\underline{j}})-\log (P_{X}(x_{\underline{i}%
})P_{Y}(y_{\underline{j}}))]=S_{X}-S_{XY}+S_{Y}=I(X;Y);
\end{equation*}%
\begin{eqnarray*}
&&S(P_{X,Y}||Q_{X,Y}):=S_{X}-S_{XY}+S_{Y}=I(X;Y); \\
&&S(P_{X,Y,Z}||Q_{X,Y,Z}):=S_{XY}-S_{XYZ}-S_{YZ}=I(X;YZ).
\end{eqnarray*}%
Such values are subjected to important inequalities
\begin{eqnarray*}
&&I(X;Y):=S_{X}+S_{Y}-S_{XY}\geq 0,\mbox{ subadditivity of entropy }; \\
&&S(P_{X,Y}||Q_{X,Y})\geq S(P_{X}||Q_{X}),S(P_{X,Y,Z}||Q_{X,Y,Z})\geq
S(P_{X,Y}||Q_{X,Y}),\mbox{ monotonicity of relative entropy}.
\end{eqnarray*}%
For three random variables, we can introduce also the concepts and condition
of strong subadditivity
\begin{equation*}
S_{X}-S_{XYZ}-S_{YZ}\geq S_{X}-S_{XY}+S_{Y},\mbox{ or }S_{XY}+S_{YZ}\geq
S_{Y}+S_{XYZ},
\end{equation*}%
which is equivalent for the condition of monotonity of mutual information $%
I(X;YZ)\geq I(X;Y).$

Above definitions and formulas for $S$ and $I$ can be generalized
respectively for the relative entropy and mutual information of GIFs and
NES. Proofs can be performed for causal lines and nonholonomic variables
generated by some thermodynamic generating functions$\ _{A}\widehat{\mathcal{%
Z}}:=\widehat{\mathcal{Z}}[\ \mathbf{g}(\tau )]$ and $\ _{B}\widehat{%
\mathcal{Z}}:=\ _{1}\widehat{\mathcal{Z}}[\ _{1}\mathbf{g}(\tau )],$ see (%
\ref{genfcanv}), as analogs of certain values $p_{i}=P_{X}(x_{i})$ and $%
q_{i}=Q_{X}(x_{i}).$ We can consider GIFs of three NES $\widehat{A},\widehat{%
B}$ and $\widehat{C}$ and prove using standard methods in any point of
causal curves and applying explicit integral N-adapted calculations on $%
\mathbf{\mathbf{V}\otimes \mathbf{V}\otimes \mathbf{V}}$ that{\small
\begin{eqnarray*}
\widehat{\mathcal{J}}\ [\widehat{A};\widehat{B}] &:=&\widehat{\mathcal{S}}[%
\widehat{A}]-\ _{AB}\widehat{\mathcal{S}}+\widehat{\mathcal{S}}[\widehat{B}%
]\geq 0,\mbox{ subadditivity of entropy}; \\
\widehat{\mathcal{S}}\ [\ _{AB}\widehat{\mathcal{Z}}||\ _{AB}\widehat{%
\mathcal{Z}}] &\geq &\widehat{\mathcal{S}}\ [\ _{A}\widehat{\mathcal{Z}}||\
_{A}\widehat{\mathcal{Z}}],\widehat{\mathcal{S}}\ [\ _{ABC}\widehat{\mathcal{%
Z}}||\ _{ABC}\widehat{\mathcal{Z}}]\geq \widehat{\mathcal{S}}\ [\ _{AB}%
\widehat{\mathcal{Z}}||\ _{AB}\widehat{\mathcal{Z}}],%
\mbox{ monotonicity of
relative entropy}.
\end{eqnarray*}%
} The conditions of strong subadditivity for GIFs and NES entropies are
stated by formulas
\begin{equation*}
\ _{A}\widehat{\mathcal{S}}-\ _{ABC}\widehat{\mathcal{S}}-\ _{BC}\widehat{%
\mathcal{S}}\geq \ _{A}\widehat{\mathcal{S}}-\ _{AB}\widehat{\mathcal{S}}+\
_{B}\widehat{\mathcal{S}},\mbox{ or }\ _{AB}\widehat{\mathcal{S}}+\ _{BC}%
\widehat{\mathcal{S}}\geq \ _{B}\widehat{\mathcal{S}}+\ _{ABC}\widehat{%
\mathcal{S}}.
\end{equation*}%
In equivalent form, these formulas can be written as the condition of
monotonicity of GIFs and NES mutual information, $\ \widehat{\mathcal{J}}\ [%
\widehat{A};\widehat{B}\widehat{C}]\geq \widehat{\mathcal{J}}\ [\widehat{A};%
\widehat{B}].$

The inequalities claimed above can be proven for any point along causal
curves on $\mathbf{\mathbf{V}.}$ For three systems, there are involved
thermodynamic generating functions generalizing (\ref{genfcanv}) in the
form,
\begin{eqnarray}
\ _{ABC}\widehat{\mathcal{Z}}[\mathbf{g}(\tau ),\ _{1}\mathbf{g}(\tau ),\
_{2}\mathbf{g}(\tau )] &=&\int \ \ _{1}\int \ _{2}\int (4\pi \tau
)^{-6}e^{-\ _{ABC}\widehat{f}}\sqrt{|\ \mathbf{g}|}\sqrt{|\ _{1}\mathbf{g}|}%
\sqrt{|\ _{2}\mathbf{g}|}d^{4}u\ d^{4}\ _{1}u\ d^{4}\ _{2}u  \notag \\
&&(-\ _{ABC}\widehat{f}+6),\mbox{ for }\mathbf{\mathbf{V}\otimes \mathbf{V}%
\otimes \mathbf{V},}  \label{threegenf}
\end{eqnarray}%
with a normalizing function $\ _{ABC}\widehat{f}(\ u,\ _{1}u,\ _{2}u).$ On
such tensor products of nonholonomic Lorentz manifolds, the d-metric
structure is $\ _{ABC}\mathbf{g}=\{\mathbf{g}=[q_{1},q_{2},\mathbf{q}%
_{3},_{q}N],\ _{1}\mathbf{g}=[\ _{1}q_{1},\ _{1}q_{2},\ _{1}\mathbf{q}%
_{3},_{1q}N],\ _{2}\mathbf{g}=[\ _{2}q_{1},\ _{2}q_{2},\ _{2}\mathbf{q}%
_{3},_{2q}N]\}$. We can define a canonical d--connection $\ _{ABC}\widehat{%
\mathbf{D}}=\ \widehat{\mathbf{D}}+\ _{B}\widehat{\mathbf{D}}$ $+\ _{C}%
\widehat{\mathbf{D}}$ and respective scalar curvature $\ _{sABC}\widehat{R}%
=\ _{s}\widehat{R}+\ _{s1}\widehat{R}+\ _{s2}\widehat{R}.$ The resulting
entropy function
\begin{eqnarray}
\ _{ABC}\widehat{\mathcal{S}} &=&\ \widehat{\mathcal{S}}[\widehat{A},%
\widehat{B},\widehat{C}]=-\int \ \ _{1}\int \ _{2}\int (4\pi \tau
)^{-6}e^{-\ _{ABC}\widehat{f}}\sqrt{|\mathbf{g}|}\sqrt{|\ _{1}\mathbf{g}|}%
\sqrt{|\ _{2}\mathbf{g}|}d^{4}u\ d^{4}\ _{1}\underline{u}\ d^{4}\ _{2}u
\notag \\
&&\left[ \tau \left( \ _{s}\widehat{R}+\ _{s1}\widehat{R}+\ _{s2}\widehat{R}%
+|\widehat{\mathbf{D}}\ _{ABC}\widehat{f}+\ _{1}\widehat{\mathbf{D}}\ _{ABC}%
\widehat{f}+\ _{2}\widehat{\mathbf{D}}\ _{ABC}\widehat{f}|^{2}\right) +\
_{ABC}\widehat{f}-12\right] .  \label{threeentr}
\end{eqnarray}

Similar formulas are considered for relativistic mechanical QGIFs in our
partner works \cite{vacaru19b,vacaru19c}.

\subsection{Density matrix for quantum geometric information and
gravitational flows}

The goal of this subsection is to study how GIFs of NES can be generalized
using basic concepts of QM and information theory. We shall elaborate on
QGIFs described in terms of density matrices defined as quantum analogs of
state densities of type $\widehat{\sigma }[\mathbf{g}(\tau )]=\widehat{%
\mathcal{Z}}^{-1}e^{-\beta E}$ with $\widehat{\mathcal{Z}}[\mathbf{g}(\tau
)] $ (\ref{genfcanv}).

Let us consider a thermodynamical model $\widehat{\mathcal{A}}=\left[
\widehat{\mathcal{W}};\widehat{\mathcal{Z}},\widehat{\mathcal{E}},\widehat{%
\mathcal{S}},\widehat{\eta }\right] $ (\ref{thvcanon}). In any point $u\in
\mathbf{V}$ along a causal curve covering an open region with such points,
we associate a typical Hilbert space $\ \widehat{\mathcal{H}}_{\mathcal{A}}.$
A state vector $\psi _{\mathcal{A}}\in $ $\ \widehat{\mathcal{H}}_{\mathcal{A%
}}$ is an infinite dimensional complex vector function. In quantum
information theory, such a value is approximated to a vector in complex
spaces of finite dimension. A vector $\psi _{\mathcal{A}}$ is a solution of
the Schr\"{o}dinger equation with as a well-defined quantum version of a
canonical Hamiltonian $\widehat{\mathcal{H}}_{\mathcal{A}},$ see details in
\cite{preskill,witten18} and, for nonholonomic systems, \cite%
{vacaru19b,vacaru19c}.

For QGIFs of NES, the combined Hilbert space is defined as a tensor product,
$\widehat{\mathcal{H}}_{\mathcal{AB}}=\widehat{\mathcal{H}}_{\mathcal{A}%
}\otimes \widehat{\mathcal{H}}_{\mathcal{B}},$ with an associate Hilbert
space $\ $ $\widehat{\mathcal{H}}_{\mathcal{A}}$ considered for a
complementary system $\widehat{\mathcal{A}}.$ The state vectors for a
combined system are written $\psi _{\mathcal{AB}}=\psi _{\mathcal{A}}\otimes
\psi _{\mathcal{B}}\in \widehat{\mathcal{H}}_{\mathcal{AB}}$ for $\psi _{%
\mathcal{A}}=1_{\mathcal{A}}$ taken as the unity. A pure state $\psi _{%
\mathcal{AB}}\in \widehat{\mathcal{H}}_{\mathcal{AB}}$ may be not only a
tensor product of complex vectors. A quantum system under geometric flow
evolution can be also \textit{entangled} and represented by a matrix of
dimension $\underline{N}\times \underline{M}$ if $\dim \widehat{\mathcal{H}}%
_{\mathcal{A}}=\underline{N}$ and $\dim \widehat{\mathcal{H}}_{\mathcal{B}}=%
\underline{M}$ $\ $(we underline symbols for dimensions in order to avoid
ambiguities with the N-connection symbol $\mathbf{N).}$ A Schmidt
decomposition can be performed for a pure state function,
\begin{equation}
\ \psi _{\mathcal{AB}}=\sum_{\underline{i}}\sqrt{p_{\underline{i}}}\psi _{%
\mathcal{A}}^{\underline{i}}\otimes \psi _{\mathcal{B}}^{\underline{i}},
\label{schmidt}
\end{equation}%
for any index $\underline{i}=1,2,....$(up to a finite value). We can
consider that a state vector $\psi _{\mathcal{A}}^{\underline{i}}$ which is
orthonormal if $<\psi _{\mathcal{A}}^{\underline{i}},\psi _{\mathcal{A}}^{%
\underline{j}}>=<\psi _{\mathcal{B}}^{\underline{i}},\psi _{\mathcal{B}}^{%
\underline{j}}>=\delta ^{\underline{i}\underline{j}},$ where $\delta ^{%
\underline{i}\underline{j}}$ is the Kronecker symbol. Considering $p_{%
\underline{i}}>0$ and $\sum_{\underline{i}}p_{\underline{i}}=1,$ we treat $%
p_{\underline{i}}$ as probabilities. In general, such $\psi _{\mathcal{A}}^{%
\underline{i}}$ and/or $\psi _{\mathcal{B}}^{\underline{i}}$ do not define
bases of $\widehat{\mathcal{H}}_{\mathcal{A}}$ and/or $\widehat{\mathcal{H}}%
_{\mathcal{B}}.$

We define the quantum density matrix for a QGIF of NES $\widehat{\mathcal{A}}
$ as $\ \widehat{\rho }_{\mathcal{A}}:=\sum_{\underline{a}}p_{\underline{a}%
}|\psi _{\mathcal{A}}^{\underline{a}}><\otimes \psi _{\mathcal{A}}^{%
\underline{a}}|$ $\ $as a Hermitian and positive semi-definite operator with
trace $Tr_{\mathcal{H}_{\mathcal{A}}}\widehat{\rho }_{\mathcal{A}}=1.$ The
hat symbol is used in order to emphasize that the constructions are
associated to canonical nonholonomic variables and respective gravitational
systems.

This allows us to compute the \textit{expectation} value of any operator $%
\widehat{\mathcal{O}}_{\mathcal{A}}$ characterizing additionally such a
system,%
\begin{eqnarray}
<\ \widehat{\mathcal{O}}>_{\mathcal{AB}} &=&<\psi _{\mathcal{AB}}|\ \widehat{%
\mathcal{O}}\otimes 1_{\mathcal{B}}|\psi _{\mathcal{AB}}>=\sum_{\underline{i}%
}p_{\underline{i}}<\psi _{\mathcal{A}}^{\underline{i}}|\widehat{\mathcal{O}}%
|\psi _{\mathcal{A}}^{\underline{i}}><\psi _{\mathcal{B}}^{\underline{i}}|1_{%
\mathcal{B}}|\psi _{\mathcal{B}}^{\underline{i}}>=  \notag \\
<\widehat{\mathcal{O}}>_{\mathcal{A}} &=&\sum_{\underline{i}}p_{\underline{i}%
}<\psi _{\mathcal{A}}^{\underline{i}}|\ \widehat{\mathcal{O}}_{\mathcal{A}%
}|\psi _{\mathcal{A}}^{\underline{i}}>=Tr_{\mathcal{H}_{\mathcal{A}}}\
\widehat{\rho }_{\mathcal{A}}\ \widehat{\mathcal{O}}_{\mathcal{A}}.
\label{expectvalues}
\end{eqnarray}%
Here we note that for arbitrary nonholonomic frame transforms and
deformations of d-connection, we can consider a general covariant form when $%
<\mathcal{O}>_{\mathcal{A}}=Tr_{\mathcal{H}_{\mathcal{A}}}\rho _{\mathcal{A}}%
\mathcal{O}_{\mathcal{A}},$ or other type nonholonomic variables with tilde
(for mechanical like variables) are considered, see \cite%
{vacaru19b,vacaru19c}.

To model both quantum information and geometric flow evolution of bipartite
systems we consider GIF and NES of type $\widehat{\mathcal{A}},\ \widehat{%
\mathcal{B}},$ and $\widehat{\mathcal{A}}\ \widehat{\mathcal{B}}$. Such
quantum systems are with both quantum and geometric entanglement defined by
density matrices. In general form, bipartite QGIFs and NES are described by
quantum density matrices of type $\widehat{\rho }_{\mathcal{AB}}.$\footnote{%
In classical theory of probability, a bipartite system $XY$ by a \textit{%
joint probability} distribution $P_{X,Y}(x_{\underline{i}},y_{\underline{j}%
}),$ where $P_{X}(x_{\underline{i}}):=\sum_{\underline{j}}P_{X,Y}(x_{%
\underline{i}},y_{\underline{j}}).$} Considering $\ \widehat{\mathcal{A}}\
\widehat{\mathcal{B}}$ as a bipartite quantum system with Hilbert space $\
\widehat{\mathcal{H}}_{\mathcal{AB}},$ we can define and parameterize a QGIF
NES density matrix:
\begin{equation*}
\ \widehat{\rho }_{\mathcal{AB}}=\sum_{\underline{a},\underline{a}^{\prime },%
\underline{b},\underline{b}^{\prime }}\ \widehat{\rho }_{\underline{a}%
\underline{a}^{\prime }\underline{b}\underline{b}^{\prime }}|\underline{a}>_{%
\mathcal{A}}\otimes |\underline{b}>_{\mathcal{B}}\ _{\mathcal{A}}<\underline{%
a}^{\prime }|\otimes \ _{\mathcal{B}}<\underline{b}^{\prime }|,
\end{equation*}
where $|\underline{a}>_{A},$ $\underline{a}=1,2,...,$\underline{$n$} is an
orthonormal basis of $\mathcal{H}_{\mathcal{A}}$ and $|\underline{b}>_{%
\mathcal{B}},$ $\underline{b}=1,2,...,$\underline{$m$} is an orthonormal
basis of $\widehat{\mathcal{H}}_{\mathcal{B}}.$

A \textit{measurement} of the QGIFs and NES $\ $ $\widehat{\mathcal{H}}$ is
characterized by a \textit{reduced density matrix}
\begin{equation*}
\ \widehat{\rho }_{\mathcal{A}}=Tr_{\mathcal{H}_{\mathcal{B}}}\ \ \widehat{%
\rho }_{\mathcal{AB}}=\sum_{\underline{a},\underline{a}^{\prime },\underline{%
b},\underline{b}}\ \ \widehat{\rho }_{\underline{a}\underline{a}^{\prime }%
\underline{b}\underline{b}}|\underline{a}>_{\mathcal{A}}\ _{\mathcal{A}}<%
\underline{a}^{\prime }|,\mbox{ for }|\underline{b}>_{\mathcal{B}}\ _{%
\mathcal{B}}<\underline{b}|=1.
\end{equation*}%
In a similar form, we can define and compute $\ \widehat{\rho }_{\mathcal{B}%
}=Tr_{\mathcal{H}_{\mathcal{A}}}\ \widehat{\rho }_{\mathcal{AB}}.$ Using
above introduced concepts and formulas, we can elaborate on QGIF models
formulated in canonical variables or in a general covariant form.

Let us analyze the properties of quantum density matrix and von Neumann
entropy for QGIFs and NES. For such systems, the quantum density matrix $\
\widehat{\sigma }_{\mathcal{AB}}$ for a state density $\widehat{\sigma }[%
\mathbf{g}(\tau )]=\widehat{\mathcal{Z}}^{-1}e^{-\beta E}$ can be defined
and computed using formulas (\ref{expectvalues}). We obtain%
\begin{eqnarray}
\widehat{\sigma }_{\mathcal{AB}} &=&<\widehat{\sigma }>_{\mathcal{AB}}=<\psi
_{\mathcal{AB}}|\widehat{\sigma }\otimes 1_{\mathcal{B}}|\psi _{\mathcal{AB}%
}>=\sum_{\underline{i}}p_{\underline{i}}<\psi _{\mathcal{A}}^{\underline{i}}|%
\widehat{\sigma }|\psi _{\mathcal{A}}^{\underline{i}}><\psi _{\mathcal{B}}^{%
\underline{i}}|1_{\mathcal{B}}|\psi _{\mathcal{B}}^{\underline{i}}>=  \notag
\\
\widehat{\sigma }_{\mathcal{A}} &=&<\widehat{\sigma }>_{\mathcal{A}}=\sum_{%
\underline{i}}p_{\underline{i}}<\psi _{\mathcal{A}}^{\underline{i}}|\widehat{%
\sigma }|\psi _{\mathcal{A}}^{\underline{i}}>=Tr_{\mathcal{H}_{\mathcal{A}}}%
\widehat{\rho }_{\mathcal{A}}\widehat{\sigma }.  \label{aux01}
\end{eqnarray}%
The density matrix $\ \ \widehat{\rho }_{\mathcal{A}}$ is taken for
computing the density matrix$\ \widehat{\sigma }_{\mathcal{A}}.$ The values (%
\ref{aux01}) are determined by a state density of the thermodynamical model
for GIFs of a classical NES determined by $\ \widehat{\sigma }.$ We can work
with quantum density matrices $\ \widehat{\sigma }_{\mathcal{AB}},$ $%
\widehat{\sigma }_{\mathcal{A}}=Tr_{\mathcal{H}_{\mathcal{B}}}\ \widehat{%
\sigma }_{\mathcal{AB}}$ and $\widehat{\sigma }_{\mathcal{B}}=Tr_{\mathcal{H}%
_{\mathcal{A}}}\widehat{\sigma }_{\mathcal{AB}}.$ Such formulas can be
written in respective coefficient forms
\begin{equation*}
\ \widehat{\sigma }_{\mathcal{AB}}=\sum_{\underline{a},\underline{a}^{\prime
},\underline{b},\underline{b}^{\prime }}\widehat{\sigma }_{\underline{a}%
\underline{a}^{\prime }\underline{b}\underline{b}^{\prime }}|\underline{a}>_{%
\mathcal{A}}\otimes |\underline{b}>_{\mathcal{B}}\ _{\mathcal{A}}<\underline{%
a}^{\prime }|\otimes \ _{\mathcal{B}}<\underline{b}^{\prime }|\mbox{ and }%
\widehat{\sigma }_{\mathcal{A}}=\sum_{\underline{a},\underline{a}^{\prime },%
\underline{b},\underline{b}}\widehat{\sigma }_{\underline{a}\underline{a}%
^{\prime }\underline{b}\underline{b}}|\underline{a}>_{\mathcal{A}}\ _{%
\mathcal{A}}<\underline{a}^{\prime }|.
\end{equation*}

We conclude that QGIFs of NES can be characterized by quantum analogs of
entropy values used for classical geometric flows. Such values can be
computed using formulas of type (\ref{aux01}) for classical conditional and
mutual entropy, see details for GIFs and in information theory \cite%
{preskill,witten18,vacaru19b}. For instance,
\begin{eqnarray*}
\ _{q}\widehat{\mathcal{W}}_{\mathcal{AB}} &=&Tr_{\mathcal{H}_{\mathcal{AB}%
}}[(\ \widehat{\sigma }_{\mathcal{AB}})(_{\mathcal{AB}}\widehat{\mathcal{W}}%
)]\mbox{ and }\ _{q}\widehat{\mathcal{W}}_{\mathcal{A}}=Tr_{\mathcal{H}_{%
\mathcal{A}}}[(\ \widehat{\sigma }_{\mathcal{A}})(\ _{\mathcal{A}}\widehat{%
\mathcal{W}})],\ _{q}\widehat{\mathcal{W}}_{\mathcal{B}}=Tr_{\mathcal{H}_{%
\mathcal{B}}}[(\ \widehat{\sigma }_{\mathcal{B}})(\ _{\mathcal{B}}\widehat{%
\mathcal{W}})]; \\
\ _{q}\widehat{\mathcal{S}}_{\mathcal{AB}} &=&Tr_{\mathcal{H}_{\mathcal{AB}%
}}[(\ \widehat{\sigma }_{\mathcal{AB}})(\ _{\mathcal{AB}}\widehat{\mathcal{S}%
})]\mbox{ and }\ _{q}\widehat{\mathcal{S}}_{\mathcal{A}}=Tr_{\mathcal{H}_{%
\mathcal{A}}}[(\ \widehat{\sigma }_{\mathcal{A}})(\ _{\mathcal{A}}\widehat{%
\mathcal{S}})],\ _{q}\widehat{\mathcal{S}}_{\mathcal{B}}=Tr_{\mathcal{H}_{%
\mathcal{B}}}[(\ \widehat{\sigma }_{\mathcal{B}})(\ _{\mathcal{B}}\widehat{%
\mathcal{S}})].
\end{eqnarray*}%
Such values describe additional entropic properties of quantum NES with rich
geometric structure under QGIFs.

Let us consider quantum generalizations of the concept of W- and
thermodynamic entropy of GIFs of NES. We describe such systems in standard
QM form for the von Neumann entropy determined by $\ \widehat{\sigma }_{%
\mathcal{A}}$ (\ref{aux01}) \ as a probability distribution,%
\begin{equation}
\ _{q}\widehat{\mathcal{S}}(\ \widehat{\sigma }_{\mathcal{A}}):=Tr\ \
\widehat{\sigma }_{\mathcal{A}}\log \ \widehat{\sigma }_{\mathcal{A}}.
\label{neumgfentr}
\end{equation}%
We can consider generalizations of this concept of quantum entropy for $\
\widehat{\mathcal{A}}\ \widehat{\mathcal{B}}$ and $\ \widehat{\mathcal{C}}\ $
systems, respectively,
\begin{equation*}
\ _{q}\widehat{\mathcal{S}}(\ \widehat{\sigma }_{\mathcal{AB}}):=Tr\ \
\widehat{\sigma }_{\mathcal{AB}}\log \ \widehat{\sigma }_{\mathcal{AB}}%
\mbox{ and }\ \ _{q}\widehat{\mathcal{S}}(\ \widehat{\sigma }_{\mathcal{A}%
}):=Tr\ \ \widehat{\sigma }_{\mathcal{A}}\log \ \ \widehat{\sigma }_{%
\mathcal{A}},\ \ _{q}\widehat{\mathcal{S}}(\ \widehat{\sigma }_{\mathcal{B}%
})\ :=Tr\ \ \widehat{\sigma }_{\mathcal{B}}\log \ \widehat{\sigma }_{%
\mathcal{B}}.
\end{equation*}

The von Neumann entropy for QGIFs of NES, $\ _{q}\widehat{\mathcal{S}}(\
\widehat{\sigma }_{\mathcal{A}}),$ has a purifying property not existing for
classical analogs. For instance, considering bipartite systems $\psi _{%
\mathcal{AB}}=\sum_{\underline{i}}\sqrt{p_{\underline{i}}}\psi _{\mathcal{A}%
}^{\underline{i}}\otimes \psi _{\mathcal{B}}^{\underline{i}}$ and $\widehat{%
\sigma }_{\mathcal{A}}:=\sum_{\underline{i}}p_{\underline{i}}|\psi _{%
\mathcal{A}}^{\underline{i}}>\otimes <\psi _{\mathcal{A}}^{\underline{i}}|,$
we compute
\begin{equation}
\widehat{\sigma }_{\mathcal{A}}:= \sum_{\underline{a},\underline{a}^{\prime
},\underline{b},\underline{b}}\ \sum_{\underline{k}}^{{}}\ \widehat{\sigma }%
_{\underline{a}\underline{a}^{\prime }\underline{b}\underline{b}}p_{%
\underline{k}}\ _{\mathcal{A}}<\underline{a}^{\prime }||\psi _{\mathcal{A}}^{%
\underline{k}}><\otimes \psi _{\mathcal{A}}^{\underline{k}}||\underline{a}>_{%
\mathcal{A}},\ \ \widehat{\sigma }_{\mathcal{B}}:= \sum_{\underline{a},%
\underline{a}^{\prime },\underline{b},\underline{b}}\ \sum_{\underline{k}%
}^{{}}\widehat{\sigma }_{\underline{a}\underline{a}^{\prime }\underline{b}%
\underline{b}}p_{\underline{k}}\ _{\mathcal{B}}<\underline{a}^{\prime
}||\psi _{\mathcal{B}}^{\underline{k}}><\otimes \psi _{\mathcal{B}}^{%
\underline{k}}||\underline{b}>_{\mathcal{B}}.  \label{aux03}
\end{equation}

Using $\ _{q}\widehat{\mathcal{S}}(\ \widehat{\sigma }_{\mathcal{A}})$ (\ref%
{neumgfentr}) and respective formulas (\ref{aux01}) and (\ref{aux03}) for
classical conditional and mutual entropy considered for GIFs and NES and in
information theory, there are defined and computed respectively%
\begin{eqnarray*}
\ _{q}\widehat{\mathcal{W}}_{\mathcal{AB}} &=&Tr_{\mathcal{H}_{\mathcal{AB}%
}}[(\ \widehat{\sigma }_{\mathcal{AB}})(\ _{\mathcal{AB}}\widehat{\mathcal{W}%
})]\mbox{ and }\ _{q}\widehat{\mathcal{W}}_{\mathcal{A}}=Tr_{\mathcal{H}_{%
\mathcal{A}}}[(\ \widehat{\sigma }_{\mathcal{A}})(\ _{\mathcal{A}}\widehat{%
\mathcal{W}})],\ _{q}\widehat{\mathcal{W}}_{\mathcal{B}}=Tr_{\mathcal{H}_{%
\mathcal{B}}}[(\ \widehat{\sigma }_{\mathcal{B}})(\ _{\mathcal{B}}\widehat{%
\mathcal{W}})]; \\
\ _{q}\widehat{\mathcal{S}}_{\mathcal{AB}} &=&Tr_{\mathcal{H}_{\mathcal{AB}%
}}[(\ \widehat{\sigma }_{\mathcal{AB}})(\ _{\mathcal{AB}}\widehat{\mathcal{S}%
})]\mbox{ and }\ _{q}\widehat{\mathcal{S}}_{\mathcal{A}}=Tr_{\mathcal{H}_{%
\mathcal{A}}}[(\ \widehat{\sigma }_{\mathcal{A}})(\ _{\mathcal{A}}\widehat{%
\mathcal{S}})],\ _{q}\widehat{\mathcal{S}}_{\mathcal{B}}=Tr_{\mathcal{H}_{%
\mathcal{B}}}[(\ \widehat{\sigma }_{\mathcal{B}})(\ _{\mathcal{B}}\widehat{%
\mathcal{S}})].
\end{eqnarray*}%
Such values describe complimentary entropic properties of quantum NES
systems with rich geometric structure under quantum GIF evolution.

\subsection{Geometric flows of nonholonomic Einstein systems with
entanglement}

We study entanglement of QGIFs and NES using the notion of bipartite
entanglement introduced for pure states and density matrices in description
of finite-dimensional QM systems \cite{preskill,witten18,aolita14,nishioka18}%
. For thermodynamic and QM analogs of gravitational GIFs, we consider a
series of relevant entropic values related to G. Perelman's W-entropy. A set
of inequalities involving GIFs, NES, and entanglement entropies playing a
crucial role in definition and description of such systems will be
formulated.

\subsubsection{Bipartite entanglement for quantum geometric information
flows and gravity}

For a NES and various type MGTs, we can consider various canonical,
(relativistic) mechanic, continuous or lattice models of QFT, thermofield
theory, QGIF models etc. A QM model can be characterized by a pure ground
state $|\widehat{\Psi }>$ for a total Hilbert space $\ _{t}\widehat{\mathcal{%
H}}.$ In this section, "hat" variables are used for all types of classical
and quantum thermodynamic systems described in canonical nonholonomic
variables. The density matrix
\begin{equation}
\ _{t}\widehat{\rho }=|\widehat{\Psi }><\widehat{\Psi }|  \label{pureground}
\end{equation}%
can be normalized following the conditions $<\widehat{\Psi }|\widehat{\Psi }%
>=1$ and total trace $\ _{t}tr(\ _{t}\widehat{\rho })=1.$ We suppose that
such a total quantum system is divided into a two subsystems $\widehat{%
\mathcal{A}}$ and $\widehat{\mathcal{B}}$ associated to some analogous GIF
and NES thermodynamic models. For instance, $\widehat{\mathcal{A}}(\mathbf{g}%
)=\left[ \widehat{\mathcal{W}}(\mathbf{g});\widehat{\mathcal{Z}}(\mathbf{g}),%
\widehat{\mathcal{E}}(\mathbf{g}),\widehat{\mathcal{S}}(\mathbf{g}),\widehat{%
\eta }(\mathbf{g})\right] $ (\ref{thvcanon}) is determined by canonical
functionals on a d-metric $\mathbf{g.}$ Similarly, a second subsystem is
generated by a d-metric $\ _{1}\mathbf{g}$ and respective $\widehat{\mathcal{%
B}}(\ _{1}\mathbf{g})=\left[ \widehat{\mathcal{W}}(\ _{1}\mathbf{g});%
\widehat{\mathcal{Z}}(\ _{1}\mathbf{g}),\widehat{\mathcal{E}}(\ _{1}\mathbf{g%
}),\widehat{\mathcal{S}}(\ _{1}\mathbf{g}),\widehat{\eta }(\ _{1}\mathbf{g})%
\right] .$ We can elaborate on models with the same $\ \mathbf{g}$ but with $%
\widehat{\mathcal{A}}$ and $\widehat{\mathcal{B}}$ associated to different
causal regions of a nonholonomic Lorentz spacetime $\mathbf{V}.$ Two
subsystems $\widehat{\mathcal{A}}$ and $\widehat{\mathcal{B}}=\overline{%
\widehat{\mathcal{A}}}$ are complimentary to each other if there is a common
boundary $\partial \widehat{\mathcal{A}}=\partial \widehat{\mathcal{B}}$ of
codimension 2 and when the non-singular flow evolution $\widehat{\mathcal{A}}
$ transforms into a necessary analytic class of flows on $\overline{\widehat{%
\mathcal{A}}}.$ For such bipartite NES and QGIFs, we define $\ _{t}\widehat{%
\mathcal{H}}\mathcal{=}\ \widehat{\mathcal{H}}_{\mathcal{AB}}=$ $\widehat{%
\mathcal{H}}_{\mathcal{A}}\otimes \widehat{\mathcal{H}}_{\mathcal{B}}.$

We introduce the measure of entanglement of a QGIF and NES as the von
Neumann entropy $\ _{q}\widehat{\mathcal{S}}$ (\ref{neumgfentr}) but defined
for the above considered total reduced density matrix $\ \widehat{\rho }_{%
\mathcal{A}}=Tr_{\mathcal{H}_{\mathcal{B}}}(\ \ _{t}\widehat{\rho }).$ For
such canonical nonholonomic systems, it is possible to define and compute
the \textit{entanglement entropy} of $\underline{\mathcal{A}}$ as
\begin{equation}
\ _{qt}\widehat{\mathcal{S}}(\ \widehat{\rho }_{\mathcal{A}}):=Tr(\ \widehat{%
\rho }_{\mathcal{A}}\ \log \widehat{\rho }_{\mathcal{A}}),
\label{entangentr}
\end{equation}%
when $\ \ \widehat{\rho }_{\mathcal{A}}$ is associated to a state density of
type $\ \widehat{\rho }(\beta ,\ \widehat{\mathcal{E}}\ ,\mathbf{g})=$ $\
\widehat{\rho }[\mathbf{g}(\tau )]=\widehat{\mathcal{Z}}^{-1}e^{-\beta E}$
with $\widehat{\mathcal{Z}}[\mathbf{g}(\tau )]$ (\ref{genfcanv}). We note
that the total entropy $\ _{qt}\widehat{\mathcal{S}}=0$ for a pure grand
state (\ref{pureground}) associated to $\mathbf{V}.$ \

\subsubsection{Separable and entangled gravitational and quantum geometric
information flows}

We can extend for NES in canonical nonholonomic variables the concepts were
introduced for QGIFs in our partner works \cite{vacaru19b,vacaru19c} (those
papers are based on analogous thermodynamic models standard constructions in
quantum information theory \cite{preskill,witten18,aolita14,nishioka18}).
Let us consider $\{|\underline{a}>_{\mathcal{A}};\underline{a}%
=1,2,...k_{a}\}\in \widehat{\mathcal{H}}_{\mathcal{A}}$ and $\{|\underline{b}%
>_{\mathcal{B}};\underline{b}=1,2,...k_{b}\}\in \widehat{\mathcal{H}}_{%
\mathcal{B}}$ as orthonormal bases when a pure total ground state is
parameterized in the form%
\begin{equation}
|\widehat{\Psi }>=\sum_{\underline{a}\underline{b}}\widehat{C}_{\underline{a}%
\underline{b}}|\underline{a}>_{\mathcal{A}}\otimes |\underline{b}>_{\mathcal{%
B}},  \label{groundstate}
\end{equation}%
where $\widehat{C}_{\underline{a}\underline{b}}$ is a complex matrix of
dimension $\dim \widehat{\mathcal{H}}_{\mathcal{A}}\times \dim \widehat{%
\mathcal{H}}_{\mathcal{B}}.$ If such coefficients factorize, $\widehat{C}_{%
\underline{a}\underline{b}}=\widehat{C}_{\underline{a}}\widehat{C}_{%
\underline{b}},$ there are defined separable ground states (equivalently,
pure product states), when
\begin{equation*}
|\widehat{\Psi }>=|\widehat{\Psi }_{\mathcal{A}}>\otimes |\widehat{\Psi }_{%
\mathcal{B}}>,\mbox{ for }|\widehat{\Psi }_{\mathcal{A}}>=\sum_{\underline{a}%
}\widehat{C}_{\underline{a}}|\underline{a}>_{\mathcal{A}}\mbox{ and }|%
\widehat{\Psi }_{\mathcal{B}}>=\sum_{\underline{b}}\widehat{C}_{\underline{b}%
}|\underline{b}>_{\mathcal{B}}.
\end{equation*}%
The entanglement entropy vanishes, $\ _{q}\widehat{\mathcal{S}}(\ \widehat{%
\rho }_{\mathcal{A}})=0,$ if and only if the pure ground state is separable
(we omit the label t as total considering that such quantum systems may
involve bi- or multi-partition and respective total spaces). For NES and
QGIFs, such definitions are motivated because all sub-systems are described
by an effective thermodynamics energy, $\ _{\mathcal{A}}\widehat{\mathcal{E}}
$ and $\ _{\mathcal{B}}\widehat{\mathcal{E}}$ as in (\ref{thvcanon}).
Similar values can be defined and computed for a W-entropy $\widehat{%
\mathcal{W}}[\mathbf{g}(\tau )]$ (\ref{wfcand}).

We say that a ground state $|\widehat{\Psi }>$ (\ref{groundstate}) is
\textit{entangled (inseparable )} if $\widehat{C}_{\underline{a}\underline{b}%
}\neq \widehat{C}_{\underline{a}}\widehat{C}_{\underline{b}}$. For such a
state, the entanglement entropy is positive, $\ _{q}\widehat{\mathcal{S}}(\
\widehat{\rho }_{\mathcal{A}}):>0.$ Contrary, certain QGIF and NES are
considered un-physical. Using quantum Schmidt decompositions (\ref{schmidt}%
), we can prove for any point along a causal curve on $\mathbf{V}$ that
\begin{equation}
\ \ _{q}\widehat{\mathcal{S}}=\mathcal{-}\sum_{\underline{a}}^{\min (%
\underline{a},\underline{b})}p_{\underline{a}}\log p_{\underline{a}}%
\mbox{
and }\ _{q}\underline{\mathcal{S}}_{|\max }=\log \min (\underline{a},%
\underline{b})\mbox{ for }\sum_{\underline{a}}p_{\underline{a}}=1\mbox{ and }%
p_{\underline{a}}=1/\min (\underline{a},\underline{b}),\forall a.
\label{aux04}
\end{equation}%
It should be noted that such quantum entropy is associated to a
thermodynamic model for geometric/ information flows and not directly for a
curved spacetime and possible geometric evolution. Nevertheless, d-metrics
can be used for elaborating on causal physical states defined along timelike
curves and respective 3+1 splitting.

An entangled state of NES and QGIFs is a superposition of several quantum
states associated to respective GIFs of gravitational systems. This means
that an observer having access only to a quantum subsystem $\widehat{%
\mathcal{A}}$ will find him/ herself in a mixed state when the total ground
state $|\widehat{\Psi }>$ is entangled following such conditions:\ $|%
\widehat{\Psi }>:\mbox{ separable }\longleftrightarrow \widehat{\rho }_{%
\mathcal{A}}: \mbox{ pure state},\mbox{ or }|\widehat{\Psi }>:%
\mbox{entangled }\longleftrightarrow \widehat{\rho }_{\mathcal{A}}:%
\mbox{
mixed state}$.

We conclude that the von Neumann entanglement entropy for QGIF and NES $\
_{qt}\widehat{\mathcal{S}}(\ \widehat{\rho }_{\mathcal{A}})\mathcal{\ }$(\ref%
{entangentr}) encodes four types of information data: 1) how the geometric
evolution in canonical nonholonomic variables is quantum flow correlated; 2)
how much a given QGIF canonical nonholonomic state differs from a separable
associated QM state; 3) how NES, or other type MGT models, are subjected to
quantum flow evolution; and 4) in which forms such GIFs and NES are modelled
in canonical nonholonomic variables. A maximum value of quantum correlations
is reached when a given QGIF NES state is a superposition of all possible
quantum states with an equal weight. Such constructions are derived for
associated thermodynamic models. There are also additional GIF and NES
properties which are characterized by the W-entropy, $\widehat{\mathcal{W}}$
(\ref{wfcand}), and thermodynamic entropy, $\ \widehat{\mathcal{S}}$ (\ref%
{thvcanon}),$\ $ which can be computed in certain quasi-classical QM limits
for a 3+1 splitting and along a time like curve, see similar constructions
in \cite{palmer12}.

\subsubsection{Thermofield double states for gravity and geometric
information flows}

In our works \cite{bubuianu19,vacaru19b,vacaru19c}, we consider an evolution
parameter $\beta =T^{-1}$ is treated as temperature similarly to G.
Perelman's approach \cite{perelman1}. This allows us to elaborate on GIF and
NES theories as relativistic classical and/or quantum thermofield models. A
nontrivial example with entanglement and a thermofield double state is
defined by a ground state (\ref{groundstate}) parameterized in the form
\begin{equation}
|\widehat{\Psi }>=\widehat{Z}^{-1/2}\sum\limits_{\underline{k}}e^{-\beta E_{%
\underline{k}}/2}|\underline{k}>_{\mathcal{A}}\otimes |\underline{k}>_{%
\mathcal{B}},  \label{dtfst}
\end{equation}%
for a partition function $\widehat{Z}=\sum\limits_{\underline{k}}e^{-\beta
E_{\underline{k}}/2}$. Such values are associated to the thermodynamic
generating function $\widehat{\mathcal{Z}}\ [\mathbf{g}(\tau )]$ (\ref%
{genfcanv}) and state density matrix $\ \widehat{\rho }(\beta ,\widehat{%
\mathcal{E}}\ ,\mathbf{g})=$ $\ \widehat{\rho }[\mathbf{g}(\tau )]=\widehat{%
\mathcal{Z}}^{-1}e^{-\beta E}.$ Values of energy $\ \widehat{\mathcal{E}}_{%
\mathcal{A}}=\{E_{\underline{k}}\}\ $\ is considered quantized with a
discrete spectrum for a QGIF and NES $\widehat{\mathcal{A}}=[\widehat{%
\mathcal{W}};\widehat{\mathcal{Z}},\widehat{\mathcal{E}},\widehat{\mathcal{S}%
},\widehat{\eta }]$ (\ref{thvcanon}). The thermodynamic values are computed
via integrals and measures determine by $\mathbf{g}(\tau )$ and canonical
nonholonomic variables. We compute the density matrix for such a thermofield
subsystem determining a Gibbs state, \
\begin{equation*}
\ \widehat{\rho }_{\mathcal{A}}=\widehat{Z}^{-1}\sum\limits_{\underline{k}%
}e^{-\beta E_{\underline{k}}/2}|\underline{k}>_{\mathcal{A}}\otimes \ _{%
\mathcal{A}}<\underline{k}|=\widehat{Z}^{-1}e^{-\beta \widehat{\mathcal{E}}_{%
\mathcal{A}}}.
\end{equation*}%
In above formulas, we consider $\ \widehat{\mathcal{E}}$ as a (modular)
Hamiltonian $\widehat{\mathcal{E}}_{\mathcal{A}}$ such that $\widehat{%
\mathcal{E}}_{\mathcal{A}}|\underline{k}>_{\mathcal{A}}=E_{\underline{k}}|%
\underline{k}>_{\mathcal{A}}.$

Thermofield double states are certain entanglement purifications of thermal
states with Boltzmann weight $p_{k}=\widehat{Z}^{-1}\sum\limits_{\underline{k%
}}e^{-\beta E_{\underline{k}}}.$ Transferring state vectors $\{|\underline{k}%
>_{\mathcal{B}}\}$ from $\widehat{\mathcal{H}}_{\mathcal{A}}$ to $\widehat{%
\mathcal{H}}_{\mathcal{B}},$ we can purify $\widehat{\mathcal{A}}$ in the
extended Hilbert space $\widehat{\mathcal{H}}_{\mathcal{A}}$ $\otimes $ $%
\widehat{\mathcal{H}}_{\mathcal{B}}.$ So, every expectation of local
operators in $\widehat{\mathcal{A}}$ can be represented using the
thermofield double state $|\widehat{\Psi }>$ (\ref{dtfst}) of the total
system $\widehat{\mathcal{A}}\mathcal{\cup }\widehat{\mathcal{B}}.$ \ The
entanglement entropy can be treated as a measure of the thermal entropy of
the subsystem $\widehat{\mathcal{A}},$
\begin{equation*}
\widehat{\mathcal{S}}(\widehat{\rho }_{\mathcal{A}})=-tr_{\mathcal{A}}[%
\widehat{\rho }_{\mathcal{A}}(-\beta \widehat{\mathcal{E}}_{\mathcal{A}%
}-\log \widehat{Z})]=\beta (<\widehat{\mathcal{E}}_{\mathcal{A}}>-\ \widehat{%
\mathcal{F}}_{\mathcal{A}}).
\end{equation*}%
In this formula, $\widehat{\mathcal{F}}_{\mathcal{A}}=-\log \widehat{Z}$ is
the thermal free energy.

For thermofield values, we omit the label "q" considered, for instance, for $%
\ _{qt}\widehat{\mathcal{S}}(\ \widehat{\rho }_{\mathcal{A}})\mathcal{\ }$(%
\ref{entangentr}). Here it should be noted that thermofield GIF and NES
configurations are also characterized by W-entropy $\widehat{\mathcal{W}}$ (%
\ref{wfcand}), see examples how to compute such values for the gravitational
configurations generated by the AFDM in \cite{ruchin13}.

\subsection{Inequalities for entropies of NES QGIFs}

We study certain important inequalities and properties of the entanglement
entropy (\ref{entangentr}) using the density matrix $\widehat{\rho }_{%
\mathcal{A}}=Tr_{\mathcal{H}_{\mathcal{B}}}(\ _{t}\widehat{\rho }).$ Proofs
with entanglement entropy are similar to those presented in \cite{nielsen10}%
. Concerning geometric analysis technique \cite%
{perelman1,monogrrf1,monogrrf2,monogrrf3}, we refer readers to
generalizations for nonholonomic manifolds and applications in modern in
modern gravity and particle physics theories in \cite%
{ruchin13,gheorghiu16,bubuianu19}.

\subsubsection{(Strong) subadditivity of entangled NES systems}

There are three important properties of QGIFs and NES related to strong
subadditivity property of entanglement and Perelman's entropies.

\textit{Entanglement entropy for complementary QGIF and gravitational
subsystems: \ } If $\ \widehat{\mathcal{B}}=\overline{\widehat{\mathcal{A}}}%
, $ we have such a condition for entropies $\ _{q}\widehat{\mathcal{S}}_{%
\mathcal{A}}=\ _{q}\widehat{\mathcal{S}}_{\overline{\mathcal{A}}},$ which
can be proven using formulas (\ref{aux04}) for a pure ground state wave
function. Similar equalities for the W-entropy $\widehat{\mathcal{W}}$ (\ref%
{wfcand}) and/or thermodynamic entropy $\ \widehat{\mathcal{S}}$ (\ref%
{thvcanon}) can be proven if we use the same d-metric $\underline{\mathbf{g}}
$ and respective normalization on $\widehat{\mathcal{A}}$ and $\overline{%
\widehat{\mathcal{A}}}.$ For quantum models of GIF thermodynamic systems, $\
_{q}\widehat{\mathcal{S}}_{\mathcal{A}}\neq \ _{q}\widehat{\mathcal{S}}_{%
\mathcal{B}}$ if $\widehat{\mathcal{A}}\mathcal{\cup }\widehat{\mathcal{B}}$
is a mixed state. In result, we have general inequalities,
\begin{equation*}
\ _{q}\widehat{\mathcal{S}}_{\mathcal{A}}\neq \ _{q}\widehat{\mathcal{S}}_{%
\mathcal{B}}\mbox{ and }\ \ _{q}\widehat{\mathcal{W}}_{\mathcal{A}}\neq \
_{q}\widehat{\mathcal{W}}_{\mathcal{B}},
\end{equation*}%
which can be also proven in any quasi-classical limit, for instance, in the
WKB approximation as in \cite{palmer12}. For some special subclasses of
nonholonomic deformations and certain classes of normalizing functions such
conditions may transform in equalities.

\textit{Subadditivity} conditions are satisfied for \ disjoint subsystems $%
\widehat{\mathcal{A}}$ and $\widehat{\mathcal{B}},$
\begin{equation}
\ _{q}\widehat{\mathcal{S}}_{\mathcal{A\cup B}}\leq \ \ _{q}\widehat{%
\mathcal{S}}_{\mathcal{A}}+\ \ _{q}\widehat{\mathcal{S}}_{\mathcal{B}}%
\mbox{
and }|\ \ _{q}\widehat{\mathcal{S}}_{\mathcal{A}}-\ \ _{q}\widehat{\mathcal{S%
}}_{\mathcal{B}}|\leq \ \ _{q}\widehat{\mathcal{S}}_{\mathcal{A\cup B}}.
\label{subaditcond}
\end{equation}
Similar conditions hold for the W-entropy $\ \widehat{\mathcal{W}}$ (\ref%
{wfcand}) and respective quantum versions,%
\begin{equation*}
\ _{q}\widehat{\mathcal{W}}_{\mathcal{A\cup B}}\leq \ _{q}\widehat{\mathcal{W%
}}_{\mathcal{A}}+\ _{q}\widehat{\mathcal{W}}_{\mathcal{B}}\mbox{
and }|\ \ _{q}\widehat{\mathcal{W}}_{\mathcal{A}}-\ _{q}\widehat{\mathcal{W}}%
_{\mathcal{B}}|\leq \ \ _{q}\widehat{\mathcal{W}}_{\mathcal{A\cup B}}.
\end{equation*}%
Such NES flow evolution and QM scenarios are elaborated for mixed geometric,
gravitational and quantum probabilistic information flows.

\textit{Strong subadditivity} is considered for three disjointed QGIF
gravitational subsystems $\widehat{\mathcal{A}},\widehat{\mathcal{B}}$ and $%
\widehat{\mathcal{C}}$ and conditions of convexity of a function built from
respective density matrices and unitarity of systems \cite%
{lieb73,narnhofer85,witten18,nishioka18}. One follow such inequalities:%
\begin{equation*}
\ _{q}\widehat{\mathcal{S}}_{\mathcal{A\cup B\cup C}}+\ _{q}\widehat{%
\mathcal{S}}_{\mathcal{B}}\leq \ \ _{q}\widehat{\mathcal{S}}_{\mathcal{A\cup
B}}+\ _{q}\widehat{\mathcal{S}}_{B\mathcal{\cup C}}\mbox{ and }\ \ _{q}%
\widehat{\mathcal{S}}_{\mathcal{A}}+\ _{q}\widehat{\mathcal{S}}_{\mathcal{C}%
}\leq \ \ _{q}\widehat{\mathcal{S}}_{\mathcal{A\cup B}}+\ _{q}\widehat{%
\mathcal{S}}_{B\mathcal{\cup C}}.
\end{equation*}%
The conditions of subadditivity (\ref{subaditcond}) consist certain
particular cases defined by strong subadditivity. Similar formulas can be
proven for the W-entropy and small quantum perturbations,%
\begin{equation*}
\ _{q}\widehat{\mathcal{W}}_{\mathcal{A\cup B\cup C}}+\ _{q}\widehat{%
\mathcal{W}}_{\mathcal{B}}\leq \ \ _{q}\widehat{\mathcal{W}}_{\mathcal{A\cup
B}}+\ _{q}\widehat{\mathcal{W}}_{B\mathcal{\cup C}}\mbox{ and }\ \ _{q}%
\widehat{\mathcal{W}}_{\mathcal{A}}+\ _{q}\widehat{\mathcal{W}}_{\mathcal{C}%
}\leq \ _{q}\widehat{\mathcal{W}}_{\mathcal{A\cup B}}+\ _{q}\widehat{%
\mathcal{W}}_{B\mathcal{\cup C}}.
\end{equation*}

\subsubsection{ Relative entropy and mutual information of NES and QGIFs}

The concept of\textit{\ relative entropy} is defined in canonical
nonholonomic variables as in geometric information theories,
\begin{equation}
\ \widehat{\mathcal{S}}(\widehat{\rho }_{\mathcal{A}}\shortparallel \widehat{%
\sigma }_{\mathcal{A}})=Tr_{\mathcal{H}_{\mathcal{B}}}[\widehat{\rho }_{%
\mathcal{A}}(\log \widehat{\rho }_{\mathcal{A}}-\log \widehat{\sigma }_{%
\mathcal{A}})],  \label{relativentr}
\end{equation}%
where $\ \widehat{\mathcal{S}}(\widehat{\rho }_{\mathcal{A}}\shortparallel
\widehat{\rho }_{\mathcal{A}})=0.$ This value allow us to define a measure
of "distance" between two QGIFs and NES with a norm $||\widehat{\rho }_{%
\mathcal{A}}||=tr(\sqrt{(\widehat{\rho }_{\mathcal{A}})(\widehat{\rho }_{%
\mathcal{A}}^{\dag })}),$ see details in reviews \cite%
{preskill,witten18,nishioka18}.

Two QGIFs and NES are characterized by some important formulas and
conditions for relative entropy:

\begin{itemize}
\item tensor products of density matrices,
\begin{equation*}
\ \ \widehat{\mathcal{S}}(\ _{1}\widehat{\rho }_{\mathcal{A}}\otimes \ _{2}%
\widehat{\rho }_{\mathcal{A}}\shortparallel \ _{1}\widehat{\sigma }_{%
\mathcal{A}}\otimes \ _{2}\widehat{\sigma }_{\mathcal{A}})=\ \ \widehat{%
\mathcal{S}}(\ _{1}\widehat{\rho }_{\mathcal{A}}\shortparallel \ _{1}%
\widehat{\sigma }_{\mathcal{A}})+\ \ \widehat{\mathcal{S}}(\ _{2}\widehat{%
\rho }_{\mathcal{A}}\shortparallel \ _{2}\widehat{\sigma }_{\mathcal{A}});
\end{equation*}

\item positivity, $\ \ \widehat{\mathcal{S}}(\ \widehat{\rho }_{\mathcal{A}%
}\shortparallel \widehat{\sigma }_{\mathcal{A}})\geq \frac{1}{2}||\ \widehat{%
\rho }_{\mathcal{A}}-\widehat{\sigma }_{\mathcal{A}}||^{2},$ i.e.$\ \
\widehat{\mathcal{S}}(\ \widehat{\rho }_{\mathcal{A}}\shortparallel \widehat{%
\sigma }_{\mathcal{A}})\geq 0;$

\item monotonicity, $\ \widehat{\mathcal{S}}(\ \widehat{\rho }_{\mathcal{A}%
}\shortparallel \widehat{\sigma }_{\mathcal{A}})\geq \ \widehat{\mathcal{S}}%
(tr_{s}\ \widehat{\rho }_{\mathcal{A}}|tr_{s}\ \widehat{\sigma }_{\mathcal{A}%
}),$ where $tr_{s}$ denotes the trace for a subsystem of $\widehat{\mathcal{A%
}}.$
\end{itemize}

The positivity formula and (Schwarz) inequality $||X||\geq tr(XY)/||X||$
result in $\ 2\ \widehat{\mathcal{S}}(\ \widehat{\rho }_{\mathcal{A}%
}\shortparallel \widehat{\sigma }_{\mathcal{A}})\geq (\langle \mathcal{O}%
\rangle _{\rho }-\langle \mathcal{O}\rangle _{\sigma })^{2}/||\mathcal{O}%
||^{2},$ for any expectation value $\langle \mathcal{O}\rangle _{\rho }$ of
an operator $\mathcal{O}$ computed (\ref{expectvalues}) with the density
matrix $\widehat{\rho }_{\mathcal{A}}.$ The relative entropy $\widehat{%
\mathcal{S}}(\widehat{\rho }_{\mathcal{A}}\shortparallel \widehat{\sigma }_{%
\mathcal{A}})$ (\ref{relativentr}) is related to the entanglement entropy $\
_{q}\ \widehat{\mathcal{S}}(\widehat{\rho }_{\mathcal{A}})$ (\ref{entangentr}%
) using formula $\widehat{\mathcal{S}}(\widehat{\rho }_{\mathcal{A}%
}\shortparallel 1_{\mathcal{A}}/k_{\mathcal{A}})=\log k_{\mathcal{A}}-\
_{q}\ \widehat{\mathcal{S}}(\widehat{\rho }_{\mathcal{A}}),$ where $1_{%
\mathcal{A}}$ is the $k_{\mathcal{A}}\times k_{\mathcal{A}}$ unit matrix for
a $k_{\mathcal{A}}$-dimensional Hilbert space associated to the region $%
\widehat{\mathcal{A}}.$

Let us denote by $\widehat{\rho }_{\mathcal{A\cup B\cup C}}$ the density
matrix of $\widehat{\mathcal{A}}\mathcal{\cup }\widehat{\mathcal{\mathcal{B}}%
}\mathcal{\cup }\widehat{\mathcal{C}}$ when $\ \widehat{\rho }_{\mathcal{%
A\cup B}}$ is written for its restriction on $\widehat{\mathcal{A}}\mathcal{%
\cup }\widehat{\mathcal{B}}$ and $\widehat{\rho }_{\mathcal{B}}$ is stated
for its restriction on $\widehat{\mathcal{B}}.$ Using $\ tr_{\mathcal{A\cup
B\cup C}}[\ \widehat{\rho }_{\mathcal{A\cup B\cup C}}(\mathcal{O}_{\mathcal{%
A\cup B}}\otimes 1_{\mathcal{C}}/k_{\mathcal{C}})]=tr_{\mathcal{A\cup B}}(\
\widehat{\rho }_{\mathcal{A\cup B}}\mathcal{O}_{\mathcal{A\cup B}}),$ we
prove such identities
\begin{eqnarray*}
\widehat{\mathcal{S}}(\widehat{\rho }_{\mathcal{A\cup B\cup C}%
}\shortparallel 1_{\mathcal{A\cup B\cup C}}/k_{\mathcal{A\cup B\cup C}}) &=&%
\widehat{\mathcal{S}}(\widehat{\rho }_{\mathcal{A\cup B}}\shortparallel 1_{%
\mathcal{A\cup B}}/k_{\mathcal{A\cup B}})+ \widehat{\mathcal{S}}(\widehat{%
\rho }_{\mathcal{A\cup B\cup C}}\shortparallel \widehat{\rho }_{\mathcal{%
A\cup B}}\otimes 1_{\mathcal{C}}/k_{\mathcal{C}}), \\
\widehat{\mathcal{S}}(\widehat{\rho }_{\mathcal{B\cup C}}\shortparallel 1_{%
\mathcal{B\cup C}}/k_{\mathcal{B\cup C}}) &=&\ \widehat{\mathcal{S}}(%
\widehat{\rho }_{\mathcal{B}}\shortparallel 1_{\mathcal{B}}/k_{\mathcal{B}%
})+\ \ \widehat{\mathcal{S}}(\widehat{\rho }_{\mathcal{B\cup C}%
}\shortparallel \widehat{\rho }_{\mathcal{B}}\otimes 1_{\mathcal{C}}/k_{%
\mathcal{C}});
\end{eqnarray*}%
\begin{eqnarray*}
&&\mbox{ and inequalities }\ \widehat{\mathcal{S}}(\widehat{\rho }_{\mathcal{%
A\cup B\cup C}}\shortparallel \widehat{\rho }_{\mathcal{A\cup B}}\otimes 1_{%
\mathcal{C}}/k_{\mathcal{C}})\geq \ \ \widehat{\mathcal{S}}(\widehat{\rho }_{%
\mathcal{B\cup C}}\shortparallel \widehat{\rho }_{\mathcal{B}}\otimes 1_{%
\mathcal{C}}/k_{\mathcal{C}}), \\
&& \widehat{\mathcal{S}}(\widehat{\rho }_{\mathcal{A\cup B\cup C}}
\shortparallel 1_{\mathcal{A\cup B\cup C}}/k_{\mathcal{A\cup B\cup C}})+\ \
\widehat{\mathcal{S}}(\widehat{\rho }_{\mathcal{B}}\shortparallel 1_{%
\mathcal{B}}/k_{\mathcal{B}})\geq \ \ \widehat{\mathcal{S}}(\widehat{\rho }_{%
\mathcal{A\cup B}}\shortparallel 1_{\mathcal{A\cup B}}/k_{\mathcal{A\cup B}%
})+ \widehat{\mathcal{S}}(\widehat{\rho }_{\mathcal{B\cup C}}\shortparallel
1_{\mathcal{B\cup C}}/k_{\mathcal{B\cup C}}).
\end{eqnarray*}

The correlation between two QGIFs and NES $\widehat{\mathcal{A}}$ and $%
\widehat{\mathcal{B}}$ (it can be involved also a third system $\widehat{%
\mathcal{C}}$) is characterized by \textit{mutual information}
\begin{equation*}
\widehat{\mathcal{J}}(\widehat{\mathcal{A}},\widehat{\mathcal{B}}):=\
\widehat{\mathcal{S}}_{\mathcal{A}}+\ \widehat{\mathcal{S}}_{\mathcal{B}}-\
\widehat{\mathcal{S}}_{\mathcal{A\cup B}}\geq 0\mbox{ and }\ \widehat{%
\mathcal{J}}(\widehat{\mathcal{A}},\widehat{\mathcal{B}}\mathcal{\cup }%
\underline{\mathcal{C}})\leq \widehat{\mathcal{J}}(\widehat{\mathcal{A}},%
\widehat{\mathcal{B}}).
\end{equation*}%
Using formula $\widehat{\mathcal{J}}(\widehat{\mathcal{A}},\widehat{\mathcal{%
B}})=\ \widehat{\mathcal{S}}(\widehat{\rho }_{\mathcal{A\cup B}%
}\shortparallel \widehat{\rho }_{\mathcal{A}}\otimes \widehat{\rho }_{%
\mathcal{B}}),$ we can prove important inequalities for the entanglement of
QGIFs and NES,%
\begin{equation*}
\ _{q}\widehat{\mathcal{J}}(\widehat{\mathcal{A}},\widehat{\mathcal{B}}):=\
_{q}\ \widehat{\mathcal{S}}_{\mathcal{A}}+\ _{q}\ \widehat{\mathcal{S}}_{%
\mathcal{B}}-\ _{q}\ \widehat{\mathcal{S}}_{\mathcal{A\cup B}}\geq 0%
\mbox{
and }\ _{q}\widehat{\mathcal{J}}(\widehat{\mathcal{A}},\widehat{\mathcal{B}}%
\mathcal{\cup }\underline{\mathcal{C}})\leq \ _{q}\widehat{\mathcal{J}}(%
\widehat{\mathcal{A}},\widehat{\mathcal{B}}),\mbox{ for }\ _{q}\widehat{%
\mathcal{J}}(\widehat{\mathcal{A}},\widehat{\mathcal{B}})=\ _{q}\ \widehat{%
\mathcal{S}}(\widehat{\rho }_{\mathcal{A\cup B}}\shortparallel \widehat{\rho
}_{\mathcal{A}}\otimes \widehat{\rho }_{\mathcal{B}}).
\end{equation*}

The mutual information between two QGIFs and NES is a measure how much the
density matrix $\widehat{\rho }_{\mathcal{A\cup B}}$ differs from a
separable state $\ \widehat{\rho }_{\mathcal{A}}\otimes \widehat{\rho }_{%
\mathcal{B}}.$ Quantum correlations entangle even spacetime disconnected
regions of the phase spacetime under geometric flow evolution. For GIF and
NES flows in respective regions, $\ 2\widehat{\mathcal{J}}(\widehat{\mathcal{%
A}},\widehat{\mathcal{B}})\geq (\langle \mathcal{O}_{\mathcal{A}}\mathcal{O}%
_{\mathcal{B}}\rangle -\langle \mathcal{O}_{\mathcal{A}}\rangle \langle
\mathcal{O}_{\mathcal{B}}\rangle )^{2}/||\mathcal{O}_{\mathcal{A}}||^{2}||%
\mathcal{O}_{\mathcal{B}}||^{2},$ for bounded operators $\mathcal{O}_{%
\mathcal{A}}$ and $\mathcal{O}_{\mathcal{B}}$.

\subsubsection{The R\'{e}nyi entropy for NES QGIFs}

The concept of R\'{e}nyi entropy \cite{renyi61} is used for computing the
entanglement entropy of QFTs by developing the replica method (see section
IV of \cite{nishioka18} and further generalizations in \cite{bao19}).
Similar constructions are possible for QGIF and NES because the
thermodynamic generating function $\widehat{\mathcal{Z}}[\mathbf{g}(\tau )]$
(\ref{genfcanv}) with $\widehat{\sigma }[\mathbf{g}(\tau )]=\widehat{%
\mathcal{Z}}^{-1}e^{-\beta E}$ as statistical density $\widehat{\rho }(\beta
,\widehat{\mathcal{E}}\ ,\mathbf{g})$ used for defining $\widehat{\sigma }_{%
\mathcal{A}}$ (\ref{aux01}) as a probability distribution. We extend the
replica method to G. Perelman's thermodynamical model and related classical
and quantum information theories for NES. Considering an integer $r$
(replica parameter), the R\'{e}nyi entropy is
\begin{equation}
\ _{r}\ \widehat{\mathcal{S}}(\widehat{\mathcal{A}}):=\frac{1}{1-r}\log [tr_{%
\mathcal{A}}(\widehat{\rho }_{\mathcal{A}})^{r}]  \label{renentr}
\end{equation}%
for a QGIF\ and NES determined by a density matrix $\widehat{\rho }_{%
\mathcal{A}}.$ A replica computational formalism is elaborated for an
analytic continuation of $r$ to a real number with a defined limit $\ _{q}%
\widehat{\mathcal{S}}(\widehat{\rho }_{\mathcal{A}})=\lim_{r\rightarrow 1}\
\ _{r}\ \widehat{\mathcal{S}}(\widehat{\mathcal{A}})$ and normalization $tr_{%
\mathcal{A}}(\widehat{\rho }_{\mathcal{A}})$ for $r\rightarrow 1,$ when (\ref%
{renentr}) reduces to the entanglement entropy (\ref{entangentr}).

Considering similar formulas proven in \cite{zycz03}, there are introduced
such important inequalities for the derivative on the replica parameter, $%
\partial _{r}.$ We have%
\begin{equation}
\partial _{r}(\ _{r}\ \widehat{\mathcal{S}}\mathcal{)}\leq 0,\ \partial
_{r}\left( \frac{r-1}{r}\ _{r}\ \widehat{\mathcal{S}}\right) \geq 0,\
\partial _{r}[(r-1)\ _{r}\ \widehat{\mathcal{S}}\mathcal{]}\geq 0,\ \partial
_{rr}^{2}[(r-1)](\ _{r}\ \widehat{\mathcal{S}}\mathcal{)}\leq 0.
\label{aux07}
\end{equation}%
A usual thermodynamical interpretation of such formulas is possible for GIF
and NES with a conventional modular Hamiltonian $\widehat{H}_{\mathcal{A}}=\
\widehat{\mathcal{E}}$ and effective statistical density $\widehat{\rho }_{%
\mathcal{A}}:=e^{-2\pi \widehat{H}_{\mathcal{A}}}.$ The value $\beta
_{r}=2\pi r$ is considered as the inverse temperature and the effective
"thermal" statistical generation (partition) function is defined $\ _{r}%
\widehat{\mathcal{Z}}(\beta _{r}):=tr_{\mathcal{A}}(\widehat{\rho }_{%
\mathcal{A}})^{r}=tr_{\mathcal{A}}(e^{-\beta _{r}\widehat{H}_{\mathcal{A}}})$
similarly to $\widehat{\mathcal{Z}}\ [\mathbf{g}(\tau )]$ (\ref{genfcanv}).
We compute using canonical relations such statistical mechanics values%
\begin{eqnarray*}
\mbox{modular energy} :&&\ _{r}\widehat{\mathcal{E}}(\beta _{r}):=-\partial
_{\beta _{r}}\log [\ _{r}\widehat{\mathcal{Z}}(\beta _{r})]\geq 0; \\
\mbox{modular entropy} :&&\ _{r}\mathcal{\breve{S}}(\beta _{r}):=\left(
1-\beta _{r}\partial _{\beta _{r}}\right) \log [\ _{r}\widehat{\mathcal{Z}}%
(\beta _{r})]\geq 0; \\
\mbox{modular capacity} :&&\ _{r}\widehat{\mathcal{C}}(\beta _{r}):=\beta
_{r}^{2}\partial _{\beta _{r}}^{2}\log [\ _{r}\widehat{\mathcal{Z}}(\beta
_{r})]\geq 0.
\end{eqnarray*}%
These inequalities are equivalent to the conditions stated in the second
line in (\ref{aux07}) and characterize the stability of GIFs and NES
considered as a thermal system with replica parameter regarded as the
inverse temperature for a respective modular Hamiltonian. Such replica
criteria of stability define a new direction for the theory of geometric
flows, QGIFs, and applications in modern physics and cosmology, see \cite%
{vacaru09,rajpoot17,ruchin13,gheorghiu16,vacaru19,vacaru19a,vacaru19b,bubuianu18}
and references therein.

The constructions with the modular entropy can be transformed into models
derived for GIFs and associated thermodynamic models and with the R\'{e}nyi
entropy and inversely. Such transforms in canonical nonholonomic can be
performed using formulas $\ _{r}\mathcal{\breve{S}}:=r^{2}\partial
_{r}\left( \frac{r-1}{r}\ _{r}\widehat{\mathcal{S}}\right)$ and, inversely, $%
\ _{r}\widehat{\mathcal{S}}\mathcal{=}\frac{r}{r-1}\int_{1}^{r}dr^{\prime }%
\frac{\ _{r^{\prime }}\mathcal{\breve{S}}}{(r^{\prime })^{2}}$. The
implications of the inequalities for the R\'{e}nyi entropy were analyzed for
the quantum information and gravitational systems with holographic
description, see reviews \cite{bub21,preskill,nishioka18}. In this
work, the approach is generalized for G.\ Perelman entropies, QGIFs and NES.

The concept of relative entropy $\widehat{\mathcal{S}}(\ \ \widehat{\rho }_{%
\mathcal{A}}\shortparallel \widehat{\sigma }_{\mathcal{A}})$ (\ref%
{relativentr}) introduced for nonholonomic geometric information flows can
be extended to that of relative R\'{e}nyi entropy (for a review, see section
II.E.3b in \cite{nishioka18}). For a system QGIFs with two density matrices $%
\widehat{\rho }_{\mathcal{A}}$ and $\widehat{\sigma }_{\mathcal{A}},$ we
compute
\begin{eqnarray}
\ _{r}\widehat{\mathcal{S}}(\ \widehat{\rho }_{\mathcal{A}}\shortparallel
\widehat{\sigma }_{\mathcal{A}}) &=&\frac{1}{r-1}\log \left[ tr\left( (\
\widehat{\sigma }_{\mathcal{A}})^{(1-r)/2r}\ \widehat{\rho }_{\mathcal{A}}(\
\widehat{\sigma }_{\mathcal{A}})^{(1-r)/2r}\right) ^{r}\right] ,\mbox{ for }%
r\in (0,1)\cup (1,\infty );  \label{relatrenyi} \\
\mbox{ or }\ _{1}\widehat{\mathcal{S}}(\ \widehat{\rho }_{\mathcal{A}%
}\shortparallel \ \widehat{\sigma }_{\mathcal{A}}) &=&\widehat{\mathcal{S}}%
(\ \widehat{\rho }_{\mathcal{A}}\shortparallel \ \widehat{\sigma }_{\mathcal{%
A}})\mbox{ and }\ _{\infty }\widehat{\mathcal{S}}(\ \widehat{\rho }_{%
\mathcal{A}}\shortparallel \ \widehat{\sigma }_{\mathcal{A}})=\log ||(\
\widehat{\sigma }_{\mathcal{A}})^{-1/2}\ \widehat{\rho }_{\mathcal{A}}(%
\widehat{\sigma }_{\mathcal{A}})^{-1/2}||_{\infty }.  \notag
\end{eqnarray}%
In any point of causal curves, one prove monotonic properties, $\ _{r}%
\widehat{\mathcal{S}}(\widehat{\rho }_{\mathcal{A}}\shortparallel \widehat{%
\sigma }_{\mathcal{A}})\geq \ _{r}\widehat{\mathcal{S}}(tr_{s}\ \widehat{%
\rho }_{\mathcal{A}}|tr_{s}\ \widehat{\sigma }_{\mathcal{A}})$ and $\partial
_{r}[\ _{r}\widehat{\mathcal{S}}(\widehat{\rho }_{\mathcal{A}}\shortparallel
\widehat{\sigma }_{\mathcal{A}})]\geq 0,$ and to reduce the relative R\'{e}%
nyi entropy to the R\'{e}nyi entropy using \ formula $\ _{r}\widehat{%
\mathcal{S}}(\widehat{\rho }_{\mathcal{A}}\shortparallel 1_{\mathcal{A}}/k_{%
\mathcal{A}})=\log k_{\mathcal{A}}-\ \ _{r}\underline{\mathcal{S}}(\widehat{%
\mathcal{A}}).$

The values (\ref{relatrenyi}) do not allow a naive extension of the concept
of mutual information and interpretation as an entanglement measure of
quantum information for QGIF systems and possible applications in gravity
theory. There are possible negative values of relative R\'{e}nyi entropy for
$r\neq 1$ even for standard quantum information models \cite{adesso12}. This
problem can be solved for various classes of solutions and MGTs if it is
introduced the concept of the $r$-R\'{e}nyi mutual information \cite{beigi13}%
,
\begin{equation*}
\ _{r}\widehat{\mathcal{J}}(\widehat{\mathcal{A}},\widehat{\mathcal{B}})\
:=\min_{\ ^{\shortmid }\sigma _{\mathcal{B}}}\ _{r}\widehat{\mathcal{S}}(\ \
\widehat{\rho }_{\mathcal{A\cup B}}\shortparallel \ \widehat{\rho }_{%
\mathcal{A}}\otimes \ \widehat{\sigma }_{\mathcal{B}})\geq 0,
\end{equation*}%
for a minimum taken over all $\ \widehat{\sigma }_{\mathcal{B}}.$ Such
conditions can be satisfied for corresponding nonholonomic distributions
with causality and respective subclasses of generating functions. We obtain
the standard definition of mutual information for $r=1.$ In result, we can
elaborate a self--consistent geometric--information thermodynamic theory for
QGIFs and NES which consists a more general approach than the constructions
with area horizon, holographic and conformal field entropic models.

\section{Decoupling and integrability of geometric flow equations for NES}

\label{s4}The goal of this section is to prove that the system of nonlinear
PDEs (\ref{canhamiltevol}) describing nonholonomic geometric evolution of
NES can be decoupled and integrated in very general forms when the
generating functions and (effective and/or matter field) sources are
determined by geometric data for nonlinear waves, solitonic hierarchies and
black hole, BH, configurations. For corresponding parameterizations, the
coefficients of such generic off-diagonal metrics and canonical
d-connections (in particular, for LC--configurations) may depend on all space like coordinates and run on a temperature like parameter $\tau $. The geometric / physical objects for such effective statistical thermodynamical systems and models of GIFs and
QGIFs corresponding to such solutions are determined by generating and
integration functions and (effective) matter sources encoding solitonic
configurations, see details and examples in \cite%
{vacaru01,vacaru10,anco06,vacaru15}. In this work, we restrict our
considerations only to stationary configurations or families of such generic
off-diagonal solutions which in certain canonical systems of reference do not depend on time like coordinate but may  run on a geometric flow parameter $\tau $ used
also for parametrization of curve flows and related solitonic hierarchies.
Various types of locally anisotropic cosmological solutions in MGTs and GR
and related inflation and dark matter and dark energy models were studied in
\cite{vacaru18tc,bubuianu18,vacaru19e}. More general GIFs and QGIFs constructions and applications to cosmological thermodynamic models will be elaborated in our further works.

\subsection{Why and how GIFs can be encoded into solitonic hierarchies?}

Geometric flow and dynamical field equations in MGTs and GR are described by systems of nonlinear PDEs. In general, the solutions of such systems do not possess any geometric/ physical symmetries but we have to prescribe certain symmetries in order to select and study some classes of solutions in explicit form. For nonlinear systems, it is  not satisfy a principle of superposition and the solutions can not be expressed as Fourier series on certain finite regions of $\mathbf{V},T\mathbf{V}$ and $T^{\ast}\mathbf{V}$. Nevertheless, any pseudo-Riemannian metric and  mentioned type physically important solutions can be encoded into certain classes of solitonic hierarchies with associated bi-Hamilton structures if there are satisfied some very general assumptions on the smooth classes of metrics and connections under consideration, see details in \cite{vacaru10,anco06,vacaru15} and references therein. On some different solitonic models and alternative approaches see \cite{kdp70,belinski01}. The constructions can be generalized for MGTs (Einstein-Dirac structures, Finsler like commutative and noncommutative or fractional derivative models,nonsymmetric metrics, string gravity, black hole and black ring deformations etc.)  \cite{vacaru01a,vacaru02,vacaru09c,baleanu11b,rajpoot15,vacaru06,vacaru07b,vacaru08}.  Recent results on soliton, quasiperiodic and pattern structures and ellipsoid--solitonic deformations in string and modified massive gravity and quantum anomalies can be found in \cite{bubuianu16,gheorghiu14,vacaru13}.

We model geometric evolution of a a 'prime' metric, $\mathbf{\mathring{g}}$, into a family 'target' d-metrics $\mathbf{g}(\tau )$ (\ref{dm1}), with transforms $\mathbf{\mathring{g}\rightarrow g}(\tau ),$ using so-called $\eta $-polarization functions when
\begin{eqnarray}
g(\tau ) &=&\eta _{\alpha }(\tau ,x^{k},y^{b})\mathring{g}_{\alpha }\mathbf{e%
}^{\alpha }[\eta ]\otimes \mathbf{e}^{\alpha }[\eta ] =\eta _{i}(\tau ,x^{k})%
\mathring{g}_{i}dx^{i}\otimes dx^{i}+\eta _{a}(\tau ,x^{k},y^{b})\mathring{h}%
_{a}e^{a}[\eta ]\otimes e^{a}[\eta ],  \notag \\
\mathbf{e}^{\alpha }[\eta ] &=&(dx^{i},\mathbf{e}^{a}=dy^{a}+\eta _{i}^{a}%
\mathring{N}_{i}^{a}dx^{i}),  \label{dme}
\end{eqnarray}%
where the target N-connection coefficients are parameterized $N_{i}^{a}(\tau
,u)=\eta _{i}^{a}(\tau ,x^{k},y^{b})\mathring{N}_{i}^{a}(\tau ,x^{k},y^{b})$.\footnote{we do not consider summation on repeating indices if they are not written as a contraction of "up-low" ones} The values $\eta _{i}(\tau )=\eta _{i}(\tau ,x^{k}),\eta _{a}(\tau )=\eta _{a}(\tau ,x^{k},y^{b})$ and $\eta _{i}^{a}(\tau )=\eta _{i}^{a}(\tau ,x^{k},y^{b})$ are called respectively geometric flow, or gravitational, polarization functions, or $\eta $%
-polarizations. Any $\mathbf{g}(\tau )$ is subjected to the condition that it defines a solution of the N-adapted Hamilton equations in canonical variables (\ref{canhamiltevol}), or for relativistic nonholonomic Ricci soliton equations (\ref{deinst}) with $\tau =\tau _{0}$ as modified Einstein equations for NES.

A prime metric $\mathbf{\mathring{g}}=\mathring{g}_{\alpha \beta
}(x^{i},y^{a})du^{\alpha }\otimes du^{\beta }$ can be parameterized in a
general coordinate form with off-diagonal N-coefficients (\ref{ofdans}) and
represented equivalently in N-adapted form
\begin{eqnarray}
\mathbf{\mathring{g}} &=&\mathring{g}_{\alpha }(u)\mathbf{\mathring{e}}%
^{\alpha }\otimes \mathbf{\mathring{e}}^{\beta }=\mathring{g}%
_{i}(x)dx^{i}\otimes dx^{i}+\mathring{g}_{a}(x,y)\mathbf{\mathring{e}}%
^{a}\otimes \mathbf{\mathring{e}}^{a},  \label{primedm} \\
\mbox{for}&& \mathbf{\mathring{e}}^{\alpha }=(dx^{i},\mathbf{e}^{a}=dy^{a}+%
\mathring{N}_{i}^{a}(u)dx^{i}),\mbox{ and }\mathbf{\mathring{e}}_{\alpha }=(%
\mathbf{\mathring{e}}_{i}=\partial /\partial y^{a}-\mathring{N}%
_{i}^{b}(u)\partial /\partial y^{b},\ {e}_{a}=\partial /\partial y^{a}).
\notag
\end{eqnarray}%
We consider that such a d-metric $\mathbf{\mathring{g}(\tau }_{0}\mathbf{)=}%
\mathring{g}_{\alpha }(u)$ can be, or not, a solution of some gravitational
field equations in a MGT or GR but under geometric evolution it transform
into a target metric (\ref{dme}) subjected to the condition to define an
exact or parametric solution of certain geometric flow evolution equations
of NES, or some nonholonomic Ricci soliton / (modified) Einstein equations.

\subsubsection{Generating Solitonic Hierarchies}

To geometric evolution of a d-metric $\mathbf{g}(\tau )$ we can associate a non--stretching curve $\gamma (\tau ,\mathbf{l})$ on a (modified) Einstein manifold $\mathbf{V,}$ where $\tau $ is a real parameter (it can be identified with the geometric flow parameter. The value $\mathbf{l}$ is the arclength of the curve on $\mathbf{V}$ which is defined by such evolution
d--vector $\mathbf{Y}=\gamma _{\tau }$ and tangent d--vector $\mathbf{X}%
=\gamma _{\mathbf{l}}$ that $\mathbf{g(X,X)=}1$. Such a curve $%
\gamma (\tau ,\mathbf{l})$ swept out a two--dimensional surface in $%
T_{\gamma (\tau ,\mathbf{l})}\mathbf{V}\subset T\mathbf{V,}$ see details in
\cite{vacaru10,anco06,vacaru15}. We consider a coframe $\mathbf{e}\in
T_{\gamma }^{\ast }\mathbf{V}_{\mathbf{N}}\otimes (h\mathfrak{p\oplus }v%
\mathfrak{p}),$ which is a N--adapted $\left( SO(n)\mathfrak{\oplus }%
SO(m)\right) $--parallel basis along $\gamma$. We use a label $\mathbf{N}$ in order to emphasize that the geometric constructions are performed for nontrivial N-connection structures. In this work, we consider that $n=m=4$ and model the evolution of 4-d Lorentzian d-metrics. The symbols $n$ and $m$ will be kept in order to distinguish N-adapted decompositions into $h$- and $v$-, or $cv$-components.

We can associate a canonical d-connection $\widehat{\mathbf{D}}$ with a linear connection 1--form is $\widehat{\mathbf{\Gamma }}\in
T_{\gamma }^{\ast }\mathbf{V}_{\mathbf{N}}\otimes (\mathfrak{so}(n)\mathfrak{%
\oplus so}(m)).$ Similar 1-forms can be associated to other types of
d-connections or to a LC-connection. We parameterize frame bases by 1-forms $\mathbf{e}_{%
\mathbf{X}}=\mathbf{e}_{h\mathbf{X}}+\mathbf{e}_{v\mathbf{X}},$ where (for $%
(1,\overrightarrow{0})\in \mathbb{R}^{n},\overrightarrow{0}\in \mathbb{R}%
^{n-1}$ and $(1,\overleftarrow{0})\in \mathbb{R}^{m},\overleftarrow{0}\in
\mathbb{R}^{m-1}),$ for
\begin{equation*}
\mathbf{e}_{h\mathbf{X}}=\gamma _{h\mathbf{X}}\rfloor h\mathbf{e=}\left[
\begin{array}{cc}
0 & (1,\overrightarrow{0}) \\
-(1,\overrightarrow{0})^{T} & h\mathbf{0}%
\end{array}%
\right] ,\mathbf{e}_{v\mathbf{X}}=\gamma _{v\mathbf{X}}\rfloor v\mathbf{e=}%
\left[
\begin{array}{cc}
0 & (1,\overleftarrow{0}) \\
-(1,\overleftarrow{0})^{T} & v\mathbf{0}%
\end{array}%
\right] .
\end{equation*}
For a $n+m$ splitting, $\widehat{\mathbf{\Gamma }}=\left[ \widehat{\mathbf{%
\Gamma }}_{h\mathbf{X}},\widehat{\mathbf{\Gamma }}_{v\mathbf{X}}\right] $,
with 
$\widehat{\mathbf{\Gamma }}_{h\mathbf{X}}=\gamma _{h\mathbf{X}}\rfloor
\widehat{\mathbf{L}}\mathbf{=}\left[
\begin{array}{cc}
0 & (0,\overrightarrow{0}) \\
-(0,\overrightarrow{0})^{T} & \widehat{\mathbf{L}}%
\end{array}%
\right] \in \mathfrak{so}(n+1)$, 
where $\widehat{\mathbf{L}}=\left[
\begin{array}{cc}
0 & \overrightarrow{v} \\
-\overrightarrow{v}^{T} & h\mathbf{0}%
\end{array}%
\right] \in \mathfrak{so}(n),~\overrightarrow{v}\in \mathbb{R}^{n-1},~h%
\mathbf{0\in }\mathfrak{so}(n-1)$; and 
$\widehat{\mathbf{\Gamma }}_{v\mathbf{X}}=\gamma _{v\mathbf{X}}\rfloor
\widehat{\mathbf{C}}=\left[
\begin{array}{cc}
0 & (0,\overleftarrow{0}) \\
-(0,\overleftarrow{0})^{T} & \widehat{\mathbf{C}}%
\end{array}%
\right] \in \mathfrak{so}(m+1)$, 
where $\widehat{\mathbf{C}}=\left[
\begin{array}{cc}
0 & \overleftarrow{v} \\
-\overleftarrow{v}^{T} & v\mathbf{0}%
\end{array}%
\right] \in \mathfrak{so}(m),~\overleftarrow{v}\in \mathbb{R}^{m-1},~v%
\mathbf{0\in }\mathfrak{so}(m-1).$

Using the canonical d--connection $\widehat{\mathbf{D}},$ we can define some d-matrices being  decomposed with respect to the flow direction:\ in the h--direction, 
$\mathbf{e}_{h\mathbf{Y}}=\gamma _{\tau }\rfloor h\mathbf{e}=\left[
\begin{array}{cc}
0 & \left( h\mathbf{e}_{\parallel },h\overrightarrow{\mathbf{e}}_{\perp
}\right) \\
-\left( h\mathbf{e}_{\parallel },h\overrightarrow{\mathbf{e}}_{\perp
}\right) ^{T} & h\mathbf{0}%
\end{array}%
\right] $, 
when \newline
$\mathbf{e}_{h\mathbf{Y}}\in h\mathfrak{p,}\left( h\mathbf{e}_{\parallel },h%
\overrightarrow{\mathbf{e}}_{\perp }\right) \in \mathbb{R}^{n} $ and $h%
\overrightarrow{\mathbf{e}}_{\perp }\in \mathbb{R}^{n-1},$ and
\begin{equation*}
\widehat{\mathbf{\Gamma }}_{h\mathbf{Y}}\mathbf{=}\gamma _{h\mathbf{Y}%
}\rfloor \widehat{\mathbf{L}}\mathbf{=}\left[
\begin{array}{cc}
0 & (0,\overrightarrow{0}) \\
-(0,\overrightarrow{0})^{T} & h\mathbf{\varpi }_{\tau }%
\end{array}%
\right] \in \mathfrak{so}(n+1),
\end{equation*}%
where $\ h\mathbf{\varpi }_{\tau }\mathbf{=}\left[
\begin{array}{cc}
0 & \overrightarrow{\varpi } \\
-\overrightarrow{\varpi }^{T} & h\widehat{\mathbf{\Theta }}%
\end{array}%
\right] \in \mathfrak{so}(n),~\overrightarrow{\varpi }\in \mathbb{R}^{n-1},~h%
\widehat{\mathbf{\Theta }}\mathbf{\in }\mathfrak{so}(n-1).$ Similar
parameterizations can be performed in the v--direction, 
$\mathbf{e}_{v\mathbf{Y}}=\gamma _{\tau }\rfloor v\mathbf{e=}\left[
\begin{array}{cc}
0 & \left( v\mathbf{e}_{\parallel },v\overleftarrow{\mathbf{e}}_{\perp
}\right) \\
-\left( v\mathbf{e}_{\parallel },v\overleftarrow{\mathbf{e}}_{\perp }\right)
^{T} & v\mathbf{0}%
\end{array}%
\right] $, 
when $\mathbf{e}_{v\mathbf{Y}}\in v\mathfrak{p,}\left( v\mathbf{e}%
_{\parallel },v\overleftarrow{\mathbf{e}}_{\perp }\right) \in \mathbb{R}^{m}$
and $v\overleftarrow{\mathbf{e}}_{\perp }\in \mathbb{R}^{m-1},$ and
\begin{equation*}
\widehat{{\mathbf{\Gamma }}}_{v\mathbf{Y}}\mathbf{=}\gamma _{v\mathbf{Y}%
}\rfloor \widehat{\mathbf{C}}\mathbf{=}\left[
\begin{array}{cc}
0 & (0,\overleftarrow{0}) \\
-(0,\overleftarrow{0})^{T} & v\widehat{\mathbf{\varpi }}_{\tau }%
\end{array}%
\right] \in \mathfrak{so}(m+1),
\end{equation*}%
where $v\mathbf{\varpi }_{\tau }\mathbf{=}\left[
\begin{array}{cc}
0 & \overleftarrow{\varpi } \\
-\overleftarrow{\varpi }^{T} & v\widehat{\mathbf{\Theta }}%
\end{array}%
\right] \in \mathfrak{so}(m),~\overleftarrow{\varpi }\in \mathbb{R}^{m-1},~v%
\widehat{\mathbf{\Theta }}\mathbf{\in }\mathfrak{so}(m-1).$

Summarizing the results proven in  \cite{vacaru10,anco06,vacaru15} for parameterizations related to geometric flows of 4-d Lorentzian metrics, we formulate such

\textbf{Main Results: } For any solution of N-adapted Hamilton equations in
canonical variables (\ref{canhamiltevol}), or for relativistic nonholonomic
Ricci soliton equations (\ref{deinst}), there is a canonical hierarchy of
N--adapted flows of curves ${\gamma }(\tau ,\mathbf{l})=h{\gamma }(\tau ,%
\mathbf{l})+v{\gamma }(\tau ,\mathbf{l})$ described by geometric
nonholonomic map equations:
\begin{itemize}
\item The $0$ flows are convective (travelling wave) maps ${\gamma }_{\tau }=%
{\gamma }_{\mathbf{l}}$ distinguished as $\left( h{\gamma }\right) _{\tau
}=\left( h{\gamma }\right) _{h\mathbf{X}}$ and $\left( v{\gamma }\right)
_{\tau }=\left( v{\gamma }\right) _{v\mathbf{X}}$. The classification of
such maps depend on the type of d-connection structure.

\item There are +1 flows defined as non--stretching mKdV maps (see details
and examples in \cite{vacaru10,anco06,vacaru15})
\begin{equation*}
-\left( h{\gamma }\right) _{\tau }=\widehat{\mathbf{D}}_{h\mathbf{X}%
}^{2}\left( h{\gamma }\right) _{h\mathbf{X}}+\frac{3}{2}|\widehat{\mathbf{D}}%
_{h\mathbf{X}}\left( h{\gamma }\right) _{h\mathbf{X}}|_{h\mathbf{g}%
}^{2}~\left( h{\gamma }\right) _{h\mathbf{X}},-\left( v{\gamma }\right)
_{\tau }=\widehat{\mathbf{D}}_{v\mathbf{X}}^{2}\left( v{\gamma }\right) _{v%
\mathbf{X}}+\frac{3}{2}|\widehat{\mathbf{D}}_{v\mathbf{X}}\left( v{\gamma }%
\right) _{v\mathbf{X}}|_{v\mathbf{g}}^{2}~\left( v{\gamma }\right) _{v%
\mathbf{X}},
\end{equation*}%
and the +2,... flows as higher order analogs.

\item Finally, the -1 flows are defined by the kernels of the canonical
recursion h--operator,
\begin{equation*}
h\widehat{{\mathfrak{R}}}=\widehat{\mathbf{D}}_{h\mathbf{X}}\left( \widehat{%
\mathbf{D}}_{h\mathbf{X}}+\widehat{\mathbf{D}}_{h\mathbf{X}}^{-1}\left(
\overrightarrow{v}\cdot \right) \overrightarrow{v}\right) +\overrightarrow{v}%
\rfloor \widehat{\mathbf{D}}_{h\mathbf{X}}^{-1}\left( \overrightarrow{v}%
\wedge \widehat{\mathbf{D}}_{h\mathbf{X}}\right) ,
\end{equation*}%
and of the canonical recursion v--operator, 
$v\widehat{{\mathfrak{R}}}=\widehat{\mathbf{D}}_{v\mathbf{X}}\left( \widehat{%
\mathbf{D}}_{v\mathbf{X}}+\widehat{\mathbf{D}}_{v\mathbf{X}}^{-1}\left(
\overleftarrow{v}\cdot \right) \overleftarrow{v}\right) +\overleftarrow{v}%
\rfloor \widehat{\mathbf{D}}_{v\mathbf{X}}^{-1}\left( \overleftarrow{v}%
\wedge \widehat{\mathbf{D}}_{v\mathbf{X}}\right)$, 
inducing non--stretching maps $\widehat{\mathbf{D}}_{h\mathbf{Y}}\left( h{%
\gamma }\right) _{h\mathbf{X}}=0$ and $\widehat{\mathbf{D}}_{v\mathbf{Y}%
}\left( v{\gamma }\right) _{v\mathbf{X}}=0$.
\end{itemize}

The canonical recursion d-operator $\widehat{{\mathfrak{R}}}=(h\widehat{{%
\mathfrak{R}}},v\widehat{{\mathfrak{R}}})$ is related to respective
bi-Hamiltonian structures in our case determined by geometric flows and
respective GIF models.

\subsubsection{Examples of solitonic space like stationary distributions and
nonlinear waves}

Using the Main Results from previous section, we conclude that the geometric
flow evolution of any d-metric on a Lorentz manifold and related models on
(co) tangent Lorentz bundles can be encoded into solitonic hierarchies. In
this work, we shall study stationary exact and parametric solutions $\mathbf{%
g}(\tau )=\mathbf{g}(\tau ,x^{i},y^{3})=[h\mathbf{g}(\tau ,x^{i}),v\mathbf{g}%
(\tau ,x^{i},y^{3})],$ with Killing symmetry on $\partial _{4}=$ $\partial
_{t}$ when in adapted coordinates the coefficients of d-metrics do not
depend on the time like coordinate $y^{4}=t.$

\paragraph{Stationary solitonic distributions: \newline
}

We shall use distributions $\ \iota =\iota (r,\vartheta ,\varphi )$ as
solutions of a respective class of solitonic 3-d equations
\begin{eqnarray}
\partial _{rr}^{2}\iota +\epsilon \partial _{\varphi }(\partial _{\vartheta
}\iota +6\iota \partial _{\varphi }\iota +\partial _{\varphi \varphi \varphi
}^{3}\iota )=0,\ \partial _{rr}^{2}\iota +\epsilon \partial _{\vartheta
}(\partial _{\varphi }\iota +6\iota \partial _{\vartheta }\iota +\partial
_{\vartheta \vartheta \vartheta }^{3}\iota ) &=&0,  \label{solitdistr} \\
\partial _{\vartheta \vartheta }^{2}\iota +\epsilon \partial _{\varphi
}(\partial _{r}\iota +6\iota \partial _{\varphi }\iota +\partial _{\varphi
\varphi \varphi }^{3}\iota )=0,\partial _{\vartheta \vartheta }^{2}\iota
+\epsilon \partial _{r}(\partial _{\varphi }\iota +6\iota \partial _{r}\iota
+\partial _{rrr}^{3}\iota ) &=&0,  \notag \\
\partial _{\varphi \varphi }^{2}\iota +\epsilon \partial _{r}(\partial
_{\vartheta }\iota +6\iota \partial _{r}\iota +\partial _{rrr}^{3}\iota
)=0,\ \partial _{\varphi \varphi }^{2}\iota +\epsilon \partial _{\vartheta
}(\partial _{r}\iota +6\iota \partial _{\vartheta }\iota +\partial
_{\vartheta \vartheta \vartheta }^{3}\iota ) &=&0,  \notag
\end{eqnarray}%
for $\epsilon =\pm 1$. The label "v" states that such a function is defined
as a "solitonic distribution" when in N-adapted frames a function $\iota (u)$
is a solution of an equation (\ref{solitdistr}) and does not depend on the
time coordinate. These equations and their solutions can be redefined via
frame/coordinate transforms for stationary generating functions
parameterized in non-spherical coordinates and labeled in the  form $\iota =\iota
(x^{i},y^{3})$.

\paragraph{Generating nonlinear solitonic waves: \newline
}

Stationary geometric flow evolution on a Lorentz manifold can be
characterized by 3-d solitonic waves with explicit dependence flow parameter
$\tau $ defined by functions $\iota (\tau ,u)$ as solutions of such
nonlinear PDEs:
\begin{equation}
\ \ \iota =\left\{
\begin{array}{ccc}
\iota (\tau ,\vartheta ,\varphi ) & \mbox{ as a solution of } & \partial
_{\tau \tau }^{2}\ \ \iota +\epsilon \frac{\partial }{\partial \varphi }%
[\partial _{\vartheta }\ \ \iota +6\ \ \iota \frac{\partial }{\partial
\varphi }\ \ \iota +\frac{\partial ^{3}}{(\partial \varphi )^{3}}\ \ \iota
]=0; \\
\iota (\vartheta ,\tau ,\varphi ) & \mbox{ as a solution of } & \partial
_{\vartheta \vartheta }^{2}\ \iota +\epsilon \frac{\partial }{\partial
\varphi }[\partial _{t}\ \iota +6\ \iota \frac{\partial }{\partial \varphi }%
\ \iota +\frac{\partial ^{3}}{(\partial \varphi )^{3}}\ \iota ]=0; \\
\iota (\tau ,r,\varphi ) & \mbox{ as a solution of } & \partial _{\tau \tau
}^{2}\ \iota +\epsilon \frac{\partial }{\partial \varphi }[\partial _{r}\
\iota +6\ \iota \frac{\partial }{\partial \varphi }\ \iota +\frac{\partial
^{3}}{(\partial \varphi )^{3}}\ \iota ]=0; \\
\iota (r,\tau ,\varphi ) & \mbox{ as a solution of } & \partial
_{rr}^{2}\iota +\epsilon \frac{\partial }{\partial \varphi }[\partial _{\tau
}\iota +6\iota \frac{\partial }{\partial \varphi }\iota +\frac{\partial ^{3}%
}{(\partial \varphi )^{3}}\iota ]=0; \\
\iota (\tau ,\varphi ,\vartheta ) & \mbox{ as a solution of } & \partial
_{\tau \tau }^{2}\iota +\epsilon \frac{\partial }{\partial \vartheta }%
[\partial _{\varphi }\iota +6\iota \frac{\partial }{\partial \vartheta }%
\iota +\frac{\partial ^{3}}{(\partial \vartheta )^{3}}\iota ]=0; \\
\ \iota (\varphi ,\tau ,\vartheta ) & \mbox{ as a solution of } & \partial
_{\varphi \varphi }^{2}\ \iota +\epsilon \frac{\partial }{\partial \vartheta
}[\partial _{\tau }\ \iota +6\ \iota \frac{\partial }{\partial \vartheta }\
\iota +\frac{\partial ^{3}}{(\partial \vartheta )^{3}}\ q\iota ]=0.%
\end{array}%
\right.  \label{swaves}
\end{equation}%
Applying general frame/coordinate transforms on respective solutions (\ref%
{swaves}), we construct solitonic waves parameterized by functions labled in the form $\
\iota =\iota (\tau ,x^{i}),$ $=\iota (\tau ,x^{1},y^{3}),$ or $=\iota (\tau
,x^{2},y^{3}).$

In a similar form, we can consider other types of solitonic stationary
configurations determined, for instance, by sine-Gordon and various types of
nonlinear wave configurations characterized by geometric curve flows as the
equations outlined in Main Results. Any such solitonic hierarchy
configuration, nonlinear wave and solitonic distribution of type $\iota
(\tau ,u)$ (\ref{swaves}) or $\iota =\iota (x^{i},y^{3})$ (\ref{solitdistr})
\ can be can be used as generating functions for certain classes of
nonholonomic deformations of stationary, or cosmological metrics, and as
generating sources. for geometric flow and MGT generic off-diagonal
solutions, see details in \cite%
{vacaru10,anco06,vacaru15,vacaru01a,vacaru02,vacaru09c,baleanu11b,rajpoot15,vacaru06,vacaru07b,vacaru08,bubuianu16,gheorghiu14,vacaru13}%
. In this work, we denote such d-metrics of type $\mathbf{g}(\tau )$ (\ref%
{dm1}) or (\ref{dme}) as d-tensor functionals of type
\begin{equation}
\mathbf{g}(\tau )=\mathbf{g[}\iota (\tau ,u\mathbf{)]=g[}\iota \mathbf{]=}%
(g_{i}[\iota ],g_{a}[\iota ])  \label{solitondm}
\end{equation}%
with polarization functions $\eta _{i}(\tau )=\eta _{i}(\tau ,x^{k})=\eta
_{i}[\iota ],\eta _{a}(\tau )=\eta _{a}(\tau ,x^{k},y^{b})=\eta _{a}[\iota ]$
and $\eta _{i}^{a}(\tau )=\eta _{i}^{a}(\tau ,x^{k},y^{b})=\eta
_{i}^{a}[\iota ].$ In general, a functional dependence $\mathbf{[}\iota
\mathbf{]}$ may be on multiple types of solitonic hierarchies (for instance,
on some different solutions of equations of type (\ref{swaves}), (\ref%
{solitdistr})) which can be written conventionally in the form $[\iota ]=[\
_{1}\iota ,\ _{2}\iota ,...]$ where the left label is use for numbering the
type of solitonic hierarchies. We shall construct in explicit form such
stationary solutions for geometric flows of NES in next section.

\subsubsection{Table 1 with ansatz for stationary geometric flows and
solitonic hierarchies}

In this work, we use brief notations of partial derivatives $\partial
_{\alpha }q=\partial q/\partial u^{\alpha }$ when a function $%
q(x^{k},y^{a}), $
\begin{eqnarray*}
\partial _{1}q &=&q^{\bullet }=\partial q/\partial x^{1},\partial
_{2}q=q^{\prime }=\partial q/\partial x^{2},\partial _{3}q=\partial
q/\partial y^{3}=\partial q/\partial \varphi =q^{\diamond },\partial
_{4}q=\partial q/\partial t=\partial _{t}q=q^{\ast }, \\
\partial _{33}^{2} &=&\partial ^{2}q/\partial \varphi ^{2}=\partial
_{\varphi \varphi }^{2}q=q^{\diamond \diamond },\partial _{44}^{2}=\partial
^{2}q/\partial t^{2}=\partial _{tt}^{2}q=q^{\ast \ast }.
\end{eqnarray*}%
Partial derivatives on a flow parameter will be written in the form $%
\partial _{\tau }=\partial /\partial \tau .$

\paragraph{Ansatz for geometric and curve flows of d-metrics with stationary
solitonic hierarchies: \newline
}

Using frame transforms, the $\tau $-evolution of any d--metric $\mathbf{g}%
(\tau )$ of type (\ref{dm1}), (\ref{dme}) and (\ref{solitondm}) \ can be
parameterized for respective spherical symmetric coordinates $u^{\alpha }=({%
r,\theta },y^{3}=\varphi ,t)$ or some general local coordinates $%
(x^{k},y^{4}=t)$ and a common geometric flow evolution and/or curve flows
parameter $\tau ,$
\begin{eqnarray}
g_{i}(\tau ) &=&e^{\psi {(\tau ,r,\theta )}}=e^{\psi \lbrack \ _{1}\iota
]},\,\,g_{a}(\tau )=\omega ({\tau ,r,\theta },y^{b})h_{a}({\tau ,r,\theta }%
,\varphi )=\omega \lbrack \ _{3}\iota ]h_{a}[\ _{2}\iota ],  \notag \\
\ N_{i}^{3}(\tau ) &=&w_{i}({\tau ,r,\theta },\varphi )=w_{i}[\ _{2}\iota
],\,\,\,\,N_{i}^{4}(\tau )=n_{i}({\tau ,r,\theta },\varphi )=n_{i}[\
_{4}\iota ,\ _{2}\iota ],  \label{statf}
\end{eqnarray}%
taking $\omega =1$ for a large class of stationary configurations. The AFDM
results in more simple and explicit (still very general classes) of
solutions if we work with nonholonomic configurations possessing at least
one Killing symmetry, for instance, on $\partial _{4}=\partial _{t}$ for
stationary solutions.

We can construct for physically important systems on nonlinear PDEs (\ref{canhamiltevol}) or (\ref{deinst}) various classes of exact and parametric off-diagonal solutions generically depending on $\tau $ and all spacetime coordinates $(x^{k},y^{a})$ but that would result in hundreds of pages with a cumbersome formulas for respective geometric techniques, see details and examples in \cite{vacaru10,anco06,vacaru15,vacaru01a,vacaru02,vacaru09c,baleanu11b,rajpoot15,
vacaru06,vacaru07b,vacaru08,bubuianu16,gheorghiu14,vacaru13} and references therein.

\paragraph{Ansatz for flow evolution of (effective) sources with stationary
solitonic hierarchies: \newline
}

Using nonholonomic frame transforms and tetradic (vierbein) fields, we can
introduce effective sources for geometric flows of NES (\ref{canhamiltevol})
or (\ref{deinst}) which in N--adapted form are parameterized in the form
\begin{equation*}
\ ^{eff}\widehat{\mathbf{\Upsilon }}_{\mu \nu }(\tau )=\mathbf{e}_{\ \mu
}^{\mu ^{\prime }}(\tau )\mathbf{e}_{\nu }^{\ \nu ^{\prime }}(\tau )[~\
\widehat{\mathbf{\Upsilon }}_{\mu ^{\prime }\nu ^{\prime }}(\tau )+\frac{1}{2%
}~\partial _{\tau }\mathbf{g}_{\mu ^{\prime }\nu ^{\prime }}(\tau )]=[~\ _{h}%
\widehat{\mathbf{\Upsilon }}(\tau ,{x}^{k})\delta _{j}^{i},\widehat{\mathbf{%
\Upsilon }}(\tau ,x^{k},y^{c})\delta _{b}^{a}].
\end{equation*}%
Such families of vielbein transforms $\mathbf{e}_{\ \mu ^{\prime }}^{\mu
}(\tau )=\mathbf{e}_{\ \mu ^{\prime }}^{\mu }(\tau ,u^{\gamma })$ and their
dual $\mathbf{e}_{\nu }^{\ \nu ^{\prime }}(\tau ,u^{\gamma })$, when $%
\mathbf{e}_{\ }^{\mu }=\mathbf{e}_{\ \mu ^{\prime }}^{\mu }du^{\mu ^{\prime
}}$ can be chosen in the form (\ref{nadif}) and/or any frame transforms of a
N-connection structure (\ref{ncon}). The values $\ _{h}\widehat{\mathbf{%
\Upsilon }}(\tau ,{x})$ and $\widehat{\mathbf{\Upsilon }}(\tau ,x,y)$\ can
be taken as functionals of certain solutions of solitonic equations and then
considered as generating data for (effective) matter sources and certain
forms compatible with solitonic hierarchies for d-metrics (\ref{statf}).
Prescribing geometric and physically motivated data, we impose certain
nonholonomic frame constraints on geometric evolution and self-similar
configurations of stationary NES. In such cases, we write
\begin{equation}
\widehat{\Im }[\iota ]=\ \ ^{eff}\widehat{\mathbf{\Upsilon }}_{\ \nu }^{\mu
}(\tau )=[~\ _{h}\widehat{\Im }[\ _{1}\iota ]=~\ _{h}\widehat{\mathbf{%
\Upsilon }}(\tau ,{r,\theta })\delta _{j}^{i},~\ _{v}\widehat{\Im }[\
_{2}\iota ]=\widehat{\mathbf{\Upsilon }}(\tau ,{r,\theta },\varphi )\delta
_{b}^{a}].  \label{dsourcparam}
\end{equation}%
There are used "hat" symbols in order to emphasize that such values are
considered systems on nonlinear PDEs involving a canonical d-connection.

In canonical nonholonomic variables with functional dependence of d-metrics
and effective sources on some prescribed classes of solitonic hierarchie,
the system of nonholonomic entropic R. Hamilton equations (\ref%
{canhamiltevol}) can be written in the form (\ref{deinst}) but with
geometric objects depending additionally on a temperature like parameter $%
\tau $ and for effective source (\ref{dsourcparam}),%
\begin{equation}
\widehat{\mathbf{R}}_{\alpha \beta }[\iota ]=\widehat{\Im }_{\alpha \beta
}[\iota ].  \label{solitonhierarcheq}
\end{equation}%
We note that such geometric evolution equations are for an undetermined
normalization function $f(\tau )=f(\tau ,u^{\gamma })$ which can be defined
explicitly for respective classes of exact or parametric solutions for
prescribed solitonic hierarchies. For self-similar point $\tau =\tau _{0}$
configurations with $\partial _{\tau }\mathbf{g}_{\mu \nu }(\tau _{0})=0,$
this system of nonlinear PDEs transforms into the canonical nonholonomic
Ricci soliton equations (\ref{canriccisol}).

Let us summarize in Table 1 below the data on nonholonomic 2+2 variables and
corresponding ansatz which allow us to transform relativistic geometric flow
equations and/or nonholonomic Ricci solitons (and equivalent gravitational
field equations in MGTs and GR) into respective systems of nonlinear
ordinary differential equations, ODEs, and partial differential equations,
PDEs, determined by respective solitonic hierarchies. All formulas will be
proven in next sections, see details in \cite%
{vacaru10,rajpoot15,vacaru06,vacaru07b,vacaru08,bubuianu16,gheorghiu14,vacaru13}%
. Our goal is to show that such systems of nonlinear PDEs can be decoupled
in general forms for generating functions and effective sources determined
by solitonic hierarchies.


{\scriptsize
\begin{eqnarray*}
&&%
\begin{tabular}{l}
\hline\hline
\begin{tabular}{lll}
& {\ \textsf{Table 1:\ Geometric flows and solitonic modified Einstein eqs
as systems of nonlinear PDEs}} &  \\
& and the Anholonomic Frame Deformation Method, \textbf{AFDM}, &  \\
& \textit{for constructing generic off-diagonal exact, parametric, and
stationary solutions} &
\end{tabular}%
\end{tabular}
\\
&&{%
\begin{tabular}{lll}
\hline
diagonal ansatz: PDEs $\rightarrow $ \textbf{ODE}s &  & AFDM: \textbf{PDE}s
\textbf{with decoupling; \ generating functions} \\
radial coordinates $u^{\alpha }=(r,\theta ,\varphi ,t)$ & $u=(x,y):$ &
\mbox{  2+2
splitting, } $u^{\alpha }=(x^{1},x^{2},y^{3},y^{4}=t);$%
\mbox{  flow
parameter  }$\tau $ \\
LC-connection $\mathring{\nabla}$ & [connections] & $%
\begin{array}{c}
\mathbf{N}:T\mathbf{V}=hT\mathbf{V}\oplus vT\mathbf{V,}\mbox{ locally }%
\mathbf{N}=\{N_{i}^{a}(x,y)\} \\
\mbox{ canonical connection distortion }\widehat{\mathbf{D}}=\nabla +%
\widehat{\mathbf{Z}}%
\end{array}%
$ \\
$%
\begin{array}{c}
\mbox{ diagonal ansatz  }g_{\alpha \beta }(u) \\
=\left(
\begin{array}{cccc}
\mathring{g}_{1} &  &  &  \\
& \mathring{g}_{2} &  &  \\
&  & \mathring{g}_{3} &  \\
&  &  & \mathring{g}_{4}%
\end{array}%
\right)%
\end{array}%
$ & $\mathbf{\mathring{g}}\Leftrightarrow \mathbf{g}(\tau )$ & $%
\begin{array}{c}
g_{\alpha \beta }(\tau )=%
\begin{array}{c}
g_{\alpha \beta }(\tau ,x^{i},y^{a})\mbox{ general frames / coordinates} \\
\left[
\begin{array}{cc}
g_{ij}+N_{i}^{a}N_{j}^{b}h_{ab} & N_{i}^{b}h_{cb} \\
N_{j}^{a}h_{ab} & h_{ac}%
\end{array}%
\right] ,\mbox{ 2 x 2 blocks }%
\end{array}
\\
\mathbf{g}_{\alpha \beta }(\tau )=[g_{ij}(\tau ),h_{ab}(\tau )], \\
\mathbf{g}(\tau )=\mathbf{g}_{i}(\tau ,x^{k})dx^{i}\otimes dx^{i}+\mathbf{g}%
_{a}(\tau ,x^{k},y^{b})\mathbf{e}^{a}\otimes \mathbf{e}^{b}%
\end{array}%
$ \\
$\mathring{g}_{\alpha \beta }=\mathring{g}_{\alpha }(r)\mbox{ for BHs}$ &
[coord.frames] & $g_{\alpha \beta }(\tau )=g_{\alpha \beta }(\tau ,r,\theta
,y^{3}=\varphi )\mbox{ stationary
configurations}$ \\
$%
\begin{array}{c}
\mbox{coord.tranfsorms }e_{\alpha }=e_{\ \alpha }^{\alpha ^{\prime
}}\partial _{\alpha ^{\prime }}, \\
e^{\beta }=e_{\beta ^{\prime }}^{\ \beta }du^{\beta ^{\prime }},\mathring{g}%
_{\alpha \beta }=\mathring{g}_{\alpha ^{\prime }\beta ^{\prime }}e_{\ \alpha
}^{\alpha ^{\prime }}e_{\ \beta }^{\beta ^{\prime }} \\
\begin{array}{c}
\mathbf{\mathring{g}}_{\alpha }(x^{k},y^{a})\rightarrow \mathring{g}_{\alpha
}(r),\mbox{ or }\mathring{g}_{\alpha }(t), \\
\mathring{N}_{i}^{a}(x^{k},y^{a})\rightarrow 0.%
\end{array}%
\end{array}%
$ & [N-adapt. fr.] & $\left\{
\begin{array}{cc}
\mathbf{g}_{i}(\tau ,r,\theta )=\mathbf{g}_{i}[\ _{1}\iota ],\mathbf{g}%
_{a}(\tau ,r,\theta ,\varphi )=\mathbf{g}_{a}[\ _{2}\iota ], &
\mbox{
d-metrics } \\
N_{i}^{3}(\tau )=w_{i}[\ _{2}\iota ],N_{i}^{4}=n_{i}[\ _{4}\iota ,\
_{2}\iota ], & \mbox{N-connections}%
\end{array}%
\right. $ \\
$\mathring{\nabla},$ $Ric=\{\mathring{R}_{\ \beta \gamma }\}$ & Ricci tensors
& $\widehat{\mathbf{D}},\ \widehat{\mathcal{R}}ic=\{\widehat{\mathbf{R}}_{\
\beta \gamma }\}$ \\
$~^{m}\mathcal{L[\mathbf{\phi }]\rightarrow }\ ^{m}\mathbf{T}_{\alpha \beta }%
\mathcal{[\mathbf{\phi }]}$ & sources & $%
\begin{array}{cc}
\widehat{\Im }[\iota ]=\widehat{\mathbf{\Upsilon }}_{\ \nu }^{\mu }(\tau )=%
\mathbf{e}_{\ \mu ^{\prime }}^{\mu }\mathbf{e}_{\nu }^{\ \nu ^{\prime }}%
\widehat{\mathbf{\Upsilon }}_{\ \nu ^{\prime }}^{\mu ^{\prime }} &  \\
=[~\ _{h}\widehat{\Im }[\ _{1}\iota ]\delta _{j}^{i},\widehat{\Im }[\
_{2}\iota ]\delta _{b}^{a}], & \mbox{stationary conf.}%
\end{array}%
$ \\
trivial equations for $\mathring{\nabla}$-torsion & LC-conditions & $%
\widehat{\mathbf{D}}_{\mid \widehat{\mathcal{T}}\rightarrow 0}=\mathbf{%
\nabla }\mbox{
extracting new classes of solutions in GR}$ \\ \hline\hline
\end{tabular}%
}
\end{eqnarray*}%
}

In this paper, we consider a physically important cases when $\mathbf{%
\mathring{g}}$ (\ref{primedm}) defines a BH solution (for instance, a vacuum
Kerr, or Schwarzschild, Kerr-(anti) de Sitter metric). For diagonalizable
via coordinate transforms prime metrics, we can always find a coordinate
system when $\mathring{N}_{i}^{b}=0.$ To study non-singular noholonomic
deformations is convenient to construct exact solutions with nontrivial
functions $\eta _{\alpha }=(\eta _{i},\eta _{a}),\eta _{i}^{a},$ and nonzero
coefficients $\mathring{N}_{i}^{b}(u)$. For a d-metric (\ref{dme}), we can
analyze the conditions of existence and geometric/ physical properties of
some target and/or prime solutions, for instance, when $\eta _{\alpha
}\rightarrow 1$ and $N_{i}^{a}\rightarrow \mathring{N}_{i}^{a}$. The values $%
\eta _{\alpha }=1$ and/or $\mathring{N}_{i}^{a}=0$ can be imposed as some
special nonholonomic constraints.

\subsection{Decoupling of GIF flow equations into stationary solitonic hierarchies}

In this subsection, we prove that the system of nonlinear PDEs (\ref%
{solitonhierarcheq}) describing geometric flow evolution of stationary NES
encoding solitonic hierarchies can be decoupled in general form. Such
geometric and information flow systems possess an important type of
nonlinear symmetries relating generating functions and effective generating
source which will be applied for computing effective thermodynamic values.

\subsubsection{Canonical Ricci d-tensors for geometric flows of NES encoding
solitonic hierarchies}

\label{ssdecst}We can chose certain systems of reference/ coordinates when
coefficients of the d-metrics (\ref{statf}) and derived geometric objects do
not depend on $y^{4}=t$ with respect to a class of N-adapted frames. Using
parameterizations for a d-metric with $\omega =1$ and a source $\widehat{\Im
}[\iota ]=[~\ _{h}\widehat{\Im }[\ _{1}\iota ],~\ _{v}\widehat{\Im }[\
_{2}\iota ]]$ (\ref{dsourcparam}) , we obtain such nontrivial N--adapted
coefficients of the Ricci d-tensor, which allow to write the geometric flow
modified Einstein equations (\ref{solitonhierarcheq}) in the form
\begin{eqnarray}
\widehat{\mathbf{R}}_{1}^{1}[\ \iota ] &=&\widehat{\mathbf{R}}_{2}^{2}[\
\iota ]=-\ _{h}\widehat{\Im }[\ _{1}\iota ]\mbox{ i.e.}\frac{g_{1}^{\bullet
}g_{2}^{\bullet }}{2g_{1}}+\frac{\left( g_{2}^{\bullet }\right) ^{2}}{2g_{2}}%
-g_{2}^{\bullet \bullet }+\frac{g_{1}^{\prime }g_{2}^{\prime }}{2g_{2}}+%
\frac{(g_{1}^{\prime })^{2}}{2g_{1}}-g_{1}^{\prime \prime }=-2g_{1}g_{2}\
_{h}\widehat{\Im };  \label{eq1a} \\
\widehat{\mathbf{R}}_{3}^{3}[\ \iota ] &=&\widehat{\mathbf{R}}_{4}^{4}[\
\iota ]=-~\ _{v}\widehat{\Im }[\ _{2}\iota ]\mbox{
i.e. }\frac{\left( h_{4}^{\diamond }\right) ^{2}}{2h_{4}}+\frac{%
h_{3}^{\diamond }\ h_{4}^{\diamond }}{2h_{3}}-h_{4}^{\diamond \diamond
}=-2h_{3}h_{4}~\ _{v}\widehat{\Im };  \label{eq2a} \\
\widehat{\mathbf{R}}_{3k}(\tau ) &=&-w_{k}\left[ \left( \frac{%
h_{4}^{\diamond }}{2h_{4}}\right) ^{2}+\frac{h_{3}^{\diamond }\ }{2h_{3}}%
\frac{h_{4}^{\diamond }}{2h_{4}}-\frac{h_{4}^{\diamond \diamond }}{2h_{4}}%
\right] +\frac{h_{4}^{\diamond }}{2h_{4}}(\frac{\partial _{k}h_{3}}{2h_{3}}+%
\frac{\partial _{k}h_{4}}{2h_{4}})-\frac{\partial _{k}h_{4}^{\diamond }}{%
2h_{4}}=0,  \label{eq3a} \\
\widehat{\mathbf{R}}_{4k}(\tau ) &=&\frac{h_{4}}{2h_{4}}n_{k}^{\diamond
\diamond }+(\frac{3}{2}h_{4}^{\diamond }-\frac{h_{4}}{h_{3}}h_{3}^{\diamond
})\frac{n_{k}^{\diamond }}{2h_{3}}=0.  \label{eq4a}
\end{eqnarray}%
This system of nonlinear PDEs has a very important decoupling property:
Using (\ref{eq1a}), we can find $g_{1}$ (or, inversely, $g_{2}$) for any
prescribed functional of solitonic hierarchies encoded into a h-source $\ \
_{h}\widehat{\Im }[\ _{1}\iota ]$ and any given coefficient $g_{2}(\tau
,r,\theta )=g_{2}[\ \iota ]$ (or, inversely, $g_{1}(\tau ,r,\theta )=g_{1}[\
\iota ]$) when the solitonic hierarchies for the coefficients of a h-metric
can be different from the geometric/ solitonic data for effective sources.
Then we can integrate on $y^{3}$ in (\ref{eq2a}) and define $h_{3}(\tau ,{%
r,\theta },\varphi )$ as a solution of first order PDE for any prescribed
v-source $~\ _{v}\widehat{\Im }[\ _{2}\iota ]$ and given coefficient $%
h_{4}(\tau ,{r,\theta },\varphi )=h_{4}[\ _{2}\iota ].$ Inversely, we can
define $h_{4}(\tau ,{r,\theta },\varphi )$ if $h_{3}(\tau ,{r,\theta }%
,\varphi )=h_{3}[\ _{2}\iota ]$ is given but in such cases we have to solve
a second order PDE. The coefficients of v-metrics involve, in general,
different types of solitonic hierarchies comparing to those prescribed for
the effective v-source. If the values of $h_{3}$ and $h_{4}$ are defined,
the equations (\ref{eq3a}) transform into a system of algebraic linear
equations for $w_{k}(\tau ,{r,\theta },\varphi )=w_{k}[\ \iota ].$ We have
to integrate two times on $y^{3}$ in (\ref{eq4a}) in order to compute $%
n_{k}(\tau ,{r,\theta },\varphi )=n_{k}[\ \iota ]$ for any defined $h_{3}$
and $h_{4}$. The solitonic hierarchies encoded in the coefficients of a
N-connection are different (in general) from those encoded in the
coefficients of d-metric and nontrivial effective sources.

Using the decoupling property of nonlinear systems of type (\ref{eq1a})--(%
\ref{eq4a}), we can integrate such PDEs step by step by prescribing
respectively the effective sources, the h-coefficients, $g_{i},$ and
v--coefficients, $h_{a},$ for geometric flowd of d-metrics $[\mathbf{g}%
_{i}(\tau )=g_{i}(\tau ),\mathbf{g}_{a}(\tau )=g_{a}(\tau )]$ and for the
N-connection coefficients, $N_{i}^{a}(\tau )=[w_{i}(\tau ),n_{i}(\tau )],$
see formulas in (\ref{dme}) and/or (\ref{statf}). The geometric evolution
of such solutions involves a prescribed nonholonomic constraint on $%
~\partial _{\tau }\mathbf{g}_{\mu ^{\prime }\nu ^{\prime }}(\tau )$ included
in $\widehat{\Im }[\iota ].$

\subsubsection{Nonlinear symmetries for solitonic generating functions and sources}

Introducing the coefficients $\alpha _{i}=(\partial _{\varphi }h_{4})\
(\partial _{i}\varpi ),\ \beta =(\partial _{\varphi }h_{4})\ (\partial
_{\varphi }\varpi ),\ \gamma =\partial _{\varphi }\left( \ln
|h_{4}|^{3/2}/|h_{3}|\right) ,$ where
\begin{equation}
\varpi {=\ln |\partial _{3}h_{4}/\sqrt{|h_{3}h_{4}|}|}  \label{aux02}
\end{equation}%
for nonsingular values for $\partial _{3}h_{a}\neq 0$ and $\partial
_{t}\varpi \neq 0,$\footnote{%
we can construct nontrivial solutions if such conditions are not satisfied;
we omit in this work considerations for more special geometric evolution
models; it is possible to introduce such frame/coorinate transforms when
necessary type conditions are satisfied} we obtain
\begin{equation}
\psi ^{\bullet \bullet }+\psi ^{\prime \prime }=2~\ \ \ _{h}\widehat{\Im }[\
_{1}\iota ],\quad \varpi ^{\diamond }\ h_{4}^{\diamond }=2h_{3}h_{4}\ \ _{v}%
\widehat{\Im }[\ _{2}\iota ],\ \beta w_{i}-\alpha _{i}=0,\quad
n_{k}^{\diamond \diamond }+\gamma n_{k}^{\diamond }=0.  \label{estatsimpl}
\end{equation}%
Such a system can be integrated in explicit form (see details in \cite%
{vacaru10,rajpoot15,vacaru06,vacaru07b,vacaru08,bubuianu16,gheorghiu14,vacaru13}
and a series of examples will be provided in next sections) if there are
prescribed a generating function $\Psi (\tau )=\Psi (\tau ,x^{i},y^{3})=\Psi
\lbrack \ \iota ]:=e^{\varpi }$ and generating sources $\ _{h}\widehat{\Im }$
and $\ \ _{v}\widehat{\Im }.$\footnote{\label{lcconda} The LC--conditions (\ref%
{lccond}) for stationary configurations transform into a system of 1st order
PDEs,
\begin{equation*}
\partial _{\varphi }w_{i}=(\partial _{i}-w_{i}\partial _{\varphi })\ln \sqrt{%
|h_{3}|},(\partial _{i}-w_{i}\partial _{\varphi })\ln \sqrt{|h_{4}|}%
=0,\partial _{k}w_{i}=\partial _{i}w_{k},\partial _{\varphi
}n_{i}=0,\partial _{i}n_{k}=\partial _{k}n_{i},
\end{equation*}%
imposing additional constraints on off-diagonal coefficients of metrics of
type (\ref{statf}).}

We have a system of two equations for $\varpi $ in (\ref{aux02}) and (\ref%
{estatsimpl}) involving four functions ($h_{3},h_{4},\ _{v}\widehat{\Im },$
and $\Psi ).$ By straightforward computation we can check that there an
important nonlinear symmetry which allows to redefine the generating
function and effective source (in particular, to introduce a family of
effective cosmological constants $\Lambda (\tau )\neq 0,\Lambda (\tau
_{0})=const,$ not depending on spacetime coordinates $u^{\alpha }).$ We can
consider nonlinear transforms $(\Psi (\tau ),\ _{v}\widehat{\Im }(\tau
))\iff (\Phi (\tau ),\Lambda (\tau ))$ defined by formulas
\begin{equation}
\Lambda (\ \Psi ^{2}[\ _{1}\iota ])^{\diamond }=|\ _{v}\widehat{\Im }[\ \
_{2}\iota ]|(\Phi ^{2}[\ \iota ])^{\diamond },\mbox{
or  }\Lambda \ \Psi ^{2}[\ _{1}\iota ]=\Phi ^{2}[\ \iota ]|\ _{v}\widehat{%
\Im }[\ \ _{2}\iota ]|-\int dy^{3}\ \Phi ^{2}[\ \iota ]|\ _{v}\widehat{\Im }%
[\ _{2}\iota ]|^{\diamond },  \label{nsym1a}
\end{equation}%
which allow us to introduce families of new generating functions $\Phi (\tau
,x^{i},y^{3})=\Phi \lbrack \ \iota ]$ and families of (effective)
cosmological constants$.$ The values $\Lambda (\tau )$ can be chosen from
certain physical considerations when the geometric/physical data for $\Phi $
encode nonlinear symmetries and solitonic hierarchies for $_{v}\widehat{\Im }%
,$ and $\Psi .$ Solutions with $\Lambda =0$ have to be studied by applying
special methods, see details and examples in \cite%
{gheorghiu14,bubuianu18,vacaru18tc}. Using nonlinear symmetries, we can
describe nonlinear systems of PDEs by two equivalent sets of generating data
$(\Psi ,\Upsilon )$ or $(\Phi ,\Lambda )$ but in all cases the symmetries of
solitonic hierarchies are encoded into functionals with respective partial
derivations on $\partial _{3}$ and/or integration on $dy^{3}.$ To generate
certain classes of solutions, we can work with effective cosmological
constants but for other ones we have to consider generating sources. Such
alternatives are convenient for constructing more general classes of exact
solutions, prescribe necessary types of solitonic symmetries, and to
elaborate on realistic physical models. Modules in formulas (\ref{nsym1a})
should be taken in certain forms resulting in physically motivated nonlinear
symmetries, relativistic causal models which are compatible with
observational data.

\subsection{Integrability of geometric flow equations with solitonic hierarchies}

We integrate in explicit form and study properties of some classes of generic off-diagonal stationary solutions of (\ref{solitonhierarcheq}) determined by generated functions and sources with solitonic hierarchies.

\subsubsection{Stationary solutions for off-diagonal metrics and
N--coefficients}

\label{ssintst}By straightforward computations we can prove that integrating
"step by step" the system (\ref{eq1a})--(\ref{eq4a}) \ represented in the
form (\ref{estatsimpl}) (see similar details and rigorous proofs in \cite%
{gheorghiu14, bubuianu18}) one generates exact stationary solutions of
geometric flow and/or modified Einstein equations if the d--metric and
respective N--connection coefficients are computed {\small
\begin{eqnarray}
\ g_{i}(\tau )&=&e^{\ \psi (\tau ,x^{k})}%
\mbox{ as a solution of 2-d
Poisson eqs. }\psi ^{\bullet \bullet }+\psi ^{\prime \prime }=2~~\ _{h}%
\widehat{\Im }[\ _{1}\iota ];  \notag \\
g_{3}[\ _{3}\iota ] &=&h_{3}(\tau ,{r,\theta },\varphi )=-\frac{(\Psi
^{\diamond }[\iota ])^{2}}{4(\ _{v}\widehat{\Im }[\ _{2}\iota ])^{2}h_{4}[\
_{4}\iota ]}=-\frac{(\partial _{3}\Psi )^{2}}{4(\ _{v}\widehat{\Im }%
)^{2}\left( h_{4}^{[0]}(\tau ,x^{k})-\int dy^{3}(\Psi ^{2})^{\diamond }/4\
_{v}\widehat{\Im }\right) }  \label{offdstat} \\
&=&-\frac{(\Phi ^{2})(\Phi ^{2})^{\diamond }}{h_{4}|\Lambda (\tau )\int
dy^{3}\ _{v}\widehat{\Im }[\Phi ^{2}]^{\diamond }|}=-\frac{[\partial
_{3}(\Phi ^{2})]^{2}}{4[h_{4}^{[0]}(\tau ,x^{k})-\Phi ^{2}/4\Lambda (\tau
)]|\int dy^{3}\ \ _{v}\widehat{\Im }\partial _{3}[\Phi ^{2}]|};  \notag \\
g_{4}[\ _{4}\iota ] &=&h_{4}(\tau ,{r,\theta },\varphi )=h_{4}^{[0]}(\tau
,x^{k})-\int dy^{3}\frac{(\Psi ^{2})^{\diamond }}{4\ _{v}\widehat{\Im }}%
=h_{4}^{[0]}(\tau ,x^{k})-\Phi ^{2}/4\Lambda (\tau );  \notag \\
\ N_{i}^{3}[\iota ] &=&w_{i}(\tau ,{r,\theta },\varphi )=\frac{\partial
_{i}\ \Psi }{\partial _{3}\Psi }\ =\frac{\partial _{i}\ \Psi ^{2}}{\partial
_{3}\Psi ^{2}}\ =\frac{\partial _{i}[\int dy^{3}\ \ _{v}\widehat{\Im }(\Phi
^{2})^{\diamond }]}{\ _{v}\widehat{\Im }(\Phi ^{2})^{\diamond }};  \notag
\end{eqnarray}%
\begin{eqnarray*}
N_{k}^{4}[\ _{5}\iota ] &=&n_{k}(\tau ,{r,\theta },\varphi )=\
_{1}n_{k}(\tau ,x^{i})+\ _{2}n_{k}(\tau ,x^{i})\int dy^{3}\frac{(\Psi
^{\diamond })^{2}}{\ _{v}\widehat{\Im }^{2}|h_{4}^{[0]}(\tau ,x^{i})-\int
dy^{3}(\Psi ^{2})^{\diamond }/4\ _{v}\widehat{\Im }|^{5/2}} \\
&=&\ _{1}n_{k}(\tau ,x^{i})+\ _{2}n_{k}(\tau ,x^{i})\int dy^{3}\frac{(\Phi
^{\diamond })^{2}}{4|\Lambda (\tau )\int dy^{3}\ _{v}\widehat{\Im }[\Phi
^{2}]^{\diamond }||h_{4}|^{5/2}}.
\end{eqnarray*}%
} In these formulas, there are stated different sets of solitonic
hierarchies which is motivated by the facts that there are integration
functions $h_{3}^{[0]}(\tau ,x^{k}),$ $\ _{1}n_{k}(\tau ,x^{i}),$ and $\
_{2}n_{k}(\tau ,x^{i})$ encoding (non) commutative parameters and
integration constants but also nonlinear evolution scenarios on $\tau .$
Such values, together with symmetries of solitonic hierarchies generating
geometric evolution data $(\Psi ,\Upsilon ),$ or $(\Phi ,\Lambda ),$ related
by nonlinear differential / integral transforms (\ref{nsym1a}) can be stated
in explicit form following certain topology/ symmetry / asymptotic
conditions. The coefficients (\ref{offdstat}) define generic off-diagonal
stationary solitonic solutions with associated bi Hamilton structures if the
corresponding anholonomy coefficients are not trivial. Such geometric flow
solutions are with nontrivial nonholonomically induced d-torsion solitonic
hierarchies determined by evolution of N-adapted coefficients of d-metric
structures. We can impose additional nonholonomic constraints (\ref{lccond})
in order to extract LC-configurations under geometric flow evolution.

\subsubsection{Quadratic line elements for off-diagonal stationary solitonic
hierarchies}

We can consider as a generating function any coefficient $h_{4}[\ _{4}\iota
]=h_{4}^{[0]}-\Phi ^{2}/4\Lambda ,h_{4}^{\diamond }(\tau )\neq 0$ and write
formulas $\Phi ^{2}(\tau )=4\Lambda \left( h_{4}[\ _{4}\iota
]-h_{4}^{[0]}\right) ,(\Phi ^{2})^{\diamond }=4\Lambda (h_{4})^{\diamond }$
and $(\Phi ^{\diamond })^{2}=\Lambda (h_{4})^{\diamond }(\frac{h_{4}}{%
h_{4}^{[0]}}-1).$ Using (\ref{nsym1a}), find $\ (\Psi ^{2})^{\diamond }=4\
\left\vert \ _{v}\widehat{\Im }[\ _{2}\iota ]\right\vert (h_{4})^{\diamond }$
and $\Psi ^{2}=4\left\vert \ _{v}\widehat{\Im }\right\vert h_{4}-4\int
dy^{3}\left\vert \ _{v}\widehat{\Im }\right\vert ^{\diamond }h_{4},$ which
allows to construct functionals $\Psi \lbrack \ _{v}\widehat{\Im }%
,h_{4},h_{4}^{[0]}]$ and $\Phi \lbrack \Lambda ,h_{4},h_{4}^{[0]}].$ Then,
we can introduce such values into respective formulas for $h_{a},N_{i}^{b}$
and $\ _{v}\widehat{\Im }$ in (\ref{offdstat}) and and express possible
generating functions and the \ d--metric (\ref{dm1}) with stationary data (%
\ref{statf}) in terms of $h_{4},$ integration functions and effective
sources for geometric evolutions. {\small
\begin{eqnarray*}
g_{3}[\ _{3}\iota ] &=&h_{3}(\tau ,{r,\theta },\varphi )=-\frac{(\Phi
^{2})(\Phi ^{2})^{\diamond }}{h_{4}|\Lambda (\tau )\int dy^{3}\ _{v}\widehat{%
\Im }[\Phi ^{2}]^{\diamond }|}=\frac{4|(h_{4})^{\diamond }|}{|\int dy^{3}\
_{v}\widehat{\Im }(h_{4})^{\diamond }|}; \\
g_{4}[\ _{4}\iota ] &=&h_{4}(\tau ,{r,\theta },\varphi )=h_{4}^{[0]}(\tau
,x^{k})-\int dy^{3}\frac{(\Psi ^{2})^{\diamond }}{4\ _{v}\widehat{\Im }}%
=h_{4}^{[0]}(\tau ,x^{k})-\Phi ^{2}/4\Lambda (\tau ); \\
\ N_{i}^{3}[\iota ] &=&w_{i}(\tau ,{r,\theta },\varphi )=\frac{\partial
_{i}\ \Psi }{\partial _{3}\Psi }\ =\frac{\partial _{i}\ \Psi ^{2}}{\partial
_{3}\Psi ^{2}}\ =\frac{\partial _{i}[\int dy^{3}\ \ _{v}\widehat{\Im }(\Phi
^{2})^{\diamond }]}{\ _{v}\widehat{\Im }(\Phi ^{2})^{\diamond }}=\frac{%
\partial _{i}\left( \left\vert \ _{v}\widehat{\Im }\right\vert h_{4}-\int
dy^{3}\left\vert \ _{v}\widehat{\Im }\right\vert ^{\diamond }h_{4}\right) }{%
\ \left\vert \ _{v}\widehat{\Im }[\ _{2}\iota ]\right\vert h_{4}^{\diamond }}%
;\ \  \\
N_{k}^{4}[\ _{5}\iota ] &=&n_{k}(\tau ,{r,\theta },\varphi )=\
_{1}n_{k}(\tau ,x^{i})+\ _{2}n_{k}(\tau ,x^{i})\int dy^{3}\frac{(\Phi
^{\diamond })^{2}}{4|\Lambda (\tau )\int dy^{3}\ _{v}\widehat{\Im }[\Phi
^{2}]^{\diamond }||h_{4}|^{5/2}}= \\
&=&\ _{1}n_{k}(\tau ,x^{i})+\ _{2}\widetilde{n}_{k}(\tau ,x^{i})\int dy^{3}%
\frac{(h_{4})^{\diamond }(1-h_{4}/h_{4}^{[0]})}{|\Lambda \int dy^{3}\ _{v}%
\widehat{\Im }(h_{4})^{\diamond }||h_{4}|^{5/2}}.
\end{eqnarray*}%
}

In result, we can express the respective quadratic elements in three
equivalent forms: {\small
\begin{eqnarray}
ds^{2} &=&e^{\ \psi (\tau ,x^{k})}[(dx^{1})^{2}+(dx^{2})^{2}]+
\label{gensolstat} \\
&&\left\{
\begin{array}{cc}
\begin{array}{c}
-\frac{4|(h_{4})^{\diamond }|}{|\int dy^{3}\ _{v}\widehat{\Im }%
(h_{4})^{\diamond }|}[dy^{3}+\frac{\partial _{i}\left( \left\vert \ _{v}%
\widehat{\Im }\right\vert h_{4}-\int dy^{3}\left\vert \ _{v}\widehat{\Im }%
\right\vert ^{\diamond }h_{4}\right) }{\ \left\vert \ _{v}\widehat{\Im }[\
_{2}\iota ]\right\vert h_{4}^{\diamond }}dx^{i}]- \\
h_{4}[dt+(\ \ _{1}n_{k}(\tau ,x^{i})+\ _{2}\widetilde{n}_{k}(\tau
,x^{i})\int dy^{3}\frac{(h_{4})^{\diamond }(1-h_{4}/h_{4}^{[0]})}{|\Lambda
\int dy^{3}\ _{v}\widehat{\Im }(h_{4})^{\diamond }||h_{4}|^{5/2}})dx^{k}],
\\
\mbox{ or }%
\end{array}
&
\begin{array}{c}
\mbox{gener.  funct.}h_{4}, \\
\mbox{ source }_{v}\widehat{\Im },\mbox{ or }\Lambda ;%
\end{array}
\\
\begin{array}{c}
\frac{\partial _{\varphi }(\Psi ^{2})}{4(\ _{v}\widehat{\Im }%
)^{2}(h_{4}^{[0]}-\int dy^{3}\frac{(\Psi ^{2})^{\diamond }}{4\ _{v}\widehat{%
\Im }})}[dy^{3}+\frac{\partial _{i}\ \Psi }{\ \partial _{3}\Psi }%
dx^{i}]-(h_{4}^{[0]}-\int dy^{3}\frac{(\Psi ^{2})^{\diamond }}{4\ \ _{v}%
\widehat{\Im }}) \\
\lbrack dt+(_{1}n_{k}+\ _{2}n_{k}\int dy^{3}\frac{(\Psi ^{\diamond })^{2}}{%
4(\ _{v}\widehat{\Im })^{2}|h_{4}^{[0]}-\int dy^{3}\frac{(\Psi
^{2})^{\diamond }}{4\ \ _{v}\widehat{\Im }}|^{5/2}})dx^{k}], \\
\mbox{ or }%
\end{array}
&
\begin{array}{c}
\mbox{gener.  funct.}\Psi , \\
\mbox{source }\ _{v}\widehat{\Im };%
\end{array}
\\
\begin{array}{c}
-\frac{[(\Phi ^{2})^{\diamond }]^{2}}{4|\Lambda \int dy^{3}\ \ _{v}\widehat{%
\Im }[(\Phi )^{2}]^{\diamond }|\ (h_{4}^{[0]}-\frac{\Phi ^{2}}{4\Lambda })}%
[dy^{3}+\frac{\partial _{i}[\int dy^{3}\ \ _{v}\widehat{\Im }(\Phi
^{2})^{\diamond }]}{\ \ \ _{v}\widehat{\Im }\ (\Phi ^{2})^{\diamond }}%
dx^{i}]-(h_{4}^{[0]}-\frac{\Phi ^{2}}{4\Lambda }) \\
\lbrack dt+(_{1}n_{k}+\ _{2}n_{k}\int dy^{3}\frac{[(\Phi ^{2})^{\diamond
}]^{2}}{|\ 4\Lambda \int dy^{3}\ \ _{v}\widehat{\Im }[(\Phi )^{2}]^{\diamond
}|}|h_{4}^{[0]}-\frac{\Phi ^{2}}{4\Lambda }|^{-5/2})dx^{k}],%
\end{array}
&
\begin{array}{c}
\mbox{gener.  funct.}\Phi \\
\mbox{effective }\Lambda \mbox{ for }\ _{v}\widehat{\Im }.%
\end{array}%
\end{array}%
\right.  \notag
\end{eqnarray}%
} Formulas (\ref{offdstat}) \ and (\ref{gensolstat}) encode solitonic
hierarchies determined by generating functions but a generating source $\ _{v}\widehat{\Im }$ and effective cosmological constant $\Lambda $ do not involve (in general) any solitonic behaviour. Nonlinear symmetries (\ref{nsym1a}) mix
different solitonic structures of generating functions and any functional for source.\footnote{%
We can consider nonholonomic solitonic deformations of a primary d-metric $%
\mathbf{\mathring{g}}$ into a target stationary one $\ \mathbf{g(\tau )=}%
[g_{\alpha }(\tau )=\eta _{\alpha }(\tau )\mathring{g}_{\alpha },\ \eta
_{i}^{a}(\tau )\mathring{N}_{i}^{a}]$ with Killing symmetry on $\partial
_{t} $ and respective bi-Hamilton structures. Formulas for the coefficients
of d-metrics and N-connections presented above can be re-written
equivalently in terms of $\eta $--polarization functions, $\eta _{\alpha }$
and $\eta _{i}^{a},$ determined by generation and integration functions and
respective sources and encoding primary data $[\mathring{g}_{\alpha },%
\mathring{N}_{i}^{a}].$ For instance, we can consider a d-metric $\mathbf{%
\mathring{g}}$ for a BH solution in GR and study hierarchies of solitonic
deformations by geometric flows or for certain nonholonomic Ricci solition
configurations which result in a stationary target d-metric $\mathbf{g}(\tau
).$ Off-diagonal nonholonomic deformations of the metric and (non) linear
connection structures and sources may preserve the singular structure of a
primary metric with certain possible deformations of the horizons, for
certain classes of solutions encoding nonsingular solitonic hierarchies and
nonsingular distributions of effective sources. For more general classes of
solutions with singular solitonic configurations, deformations of horizons,
nonlinear symmetries etc., there are possible scenarios eliminating the
singular structure, generating new symmetries, or changing the topology of
target solutions.}

\subsubsection{Off-diagonal Levi-Civita stationary solitonic hierarchies}

The equations (\ref{lccond}) for zero torsion conditions, see also footnote %
\ref{lcconda}, can be solved for a special class of generating functions and
sources. For instance, we can take a $\Psi (\tau )=\check{\Psi}(\tau
,x^{i},\varphi )$ for which $(\partial _{i}\check{\Psi})^{\diamond
}=\partial _{i}(\check{\Psi}^{\diamond })$ and fix $\ _{v}\widehat{\Im }%
(\tau ,x^{i},y^{3})=\ _{v}\widehat{\Im }[\check{\Psi}]=\ _{v}\check{\Im}%
(\tau ),$ or $\ _{v}\widehat{\Im }=const,$ modifying the nonlinear
symmetries (\ref{nsym1a}) to $\Lambda \ \check{\Psi}^{2}=\check{\Phi}^{2}|\
_{v}\widehat{\Im }|-\int dy^{3}\ \Phi ^{2}|\ _{v}\widehat{\Im }|^{\diamond },%
\check{\Phi}^{2}=-4\Lambda \check{h}_{4}(\tau ,x^{i},y^{3}),\check{\Psi}%
^{2}=\int dy^{3}\ \ _{v}\widehat{\Im }\check{h}_{4}^{\diamond }.$ The
coefficient $h_{4}(\tau )=\check{h}_{4}(\tau ,x^{i},y^{3})$ can be
considered also as generating function when $h_{3}$ and N-connection
coefficients are computed using certain nonlinear symmetries and
nonholonomic constraints. For zero torsion solitonic hierarchies, we find
some functions $\check{A}(\tau )=\check{A}(\tau ,x^{i},y^{3})$ and $n(\tau
)=n(\tau ,x^{i})$ when the coefficients of N-connection are
\begin{equation*}
w_{i}(\tau )=\partial _{i}\check{A}(\tau )=\frac{\partial _{i}(\int dy^{3}\
_{v}\check{\Im}\ \check{h}_{4}^{\diamond }{}])}{\ \ _{v}\check{\Im}\ \check{h%
}_{4}^{\diamond }{}}=\frac{\partial _{i}\check{\Psi}}{\check{\Psi}^{\diamond
}}=\frac{\partial _{i}[\int dy^{3}\ \ \ _{v}\check{\Im}(\check{\Phi}%
^{2})^{\diamond }]}{\ \ _{v}\check{\Im}(\check{\Phi}^{2})^{\diamond }}%
\mbox{ and
}n_{k}(\tau )=\check{n}_{k}(\tau )=\partial _{k}n(\tau ,x^{i}).
\end{equation*}%
In result, we can construct new classes of off-diagonal zero
torsion stationary solutions encoding solitonic hierarchies and defined as
subclasses of solutions (\ref{gensolstat}), {\small
\begin{equation}
ds^{2}=e^{\ \psi (\tau ,x^{k})}[(dx^{1})^{2}+(dx^{2})^{2}]-\left\{
\begin{array}{cc}
\begin{array}{c}
\frac{(\check{h}_{4}^{\diamond }{})^{2}}{|\int dy^{3}\ _{v}\check{\Im}\check{%
h}_{4}^{\diamond }|\ \check{h}_{4}}[dy^{3}+(\partial _{i}\check{A})dx^{i}]+%
\check{h}_{4}\left[ dt+(\partial _{k}n)dx^{k}\right] , \\
\mbox{ or }%
\end{array}
&
\begin{array}{c}
\mbox{gener.  funct.}\check{h}_{4}, \\
\mbox{ source }\ \ _{v}\check{\Im},\mbox{ or }\Lambda ;%
\end{array}
\\
\begin{array}{c}
\frac{\partial _{\varphi }(\check{\Psi}^{2})}{4\ (\ _{v}\check{\Im}%
)^{2}(h_{4}^{[0]}-\int dy^{3}\frac{(\check{\Psi}^{2})^{\diamond }}{4\ \ _{v}%
\check{\Im}})}[dy^{3}+(\partial _{i}\check{A})dx^{i}]+ \\
(h_{4}^{[0]}-\int dy^{3}\frac{(\check{\Psi}^{2})^{\diamond }}{4\ \ _{v}%
\check{\Im}})\left[ dt+(\partial _{k}n)dx^{k}\right] , \\
\mbox{ or }%
\end{array}
&
\begin{array}{c}
\mbox{gener.  funct.}\check{\Psi}, \\
\mbox{source }\ \ _{v}\check{\Im};%
\end{array}
\\
\begin{array}{c}
\frac{\lbrack (\check{\Phi}^{2})^{\diamond }]^{2}}{4|\Lambda \int dy^{3}\ \
_{v}\check{\Im}(\check{\Phi}^{2})^{\diamond }|\ (h_{4}^{[0]}-\frac{\check{%
\Phi}^{2}}{4\Lambda })}[dy^{3}+(\partial _{i}\check{A})dx^{i}] \\
+(h_{4}^{[0]}-\frac{\check{\Phi}^{2}}{4\Lambda })\left[ dt+(\partial
_{k}n)dx^{k}\right] ,%
\end{array}
&
\begin{array}{c}
\mbox{gener.  funct.}\ \check{\Phi} \\
\mbox{effective }\Lambda \mbox{ for }\ \ _{v}\check{\Im}.%
\end{array}%
\end{array}%
\right.  \label{lcsolstat}
\end{equation}%
} For any value of flow parameter $\tau ,$ such stationary metrics are
generic off-diagonal and define new classes of solutions which are different,
for instance, from the Kerr metric (defined by rotation coordinates, or
other equivalent ones). We may check if the anholonomy coefficients $%
C_{\alpha \beta }^{\gamma }=\{C_{ia}^{b}=\partial _{a}N_{i}^{b},C_{ji}^{a}=%
\mathbf{e}_{j}N_{i}^{a}-\mathbf{e}_{i}N_{j}^{a}\}$ are not zero for
solitonic values of $N_{i}^{3}=\partial _{i}\check{A}$ and $%
N_{k}^{4}=\partial _{k}n$ and understand if certain metrics are or not
generic off-diagonal. We can \ fix and analyze certain nonholonomic
solitonic configurations determined, for instance, by data $(\ _{v}\check{\Im%
},\check{\Psi},h_{4}^{[0]},\check{n}_{k}),$ with $w_{i}=\partial _{i}\check{A%
}\rightarrow 0$ and $\partial _{k}n\rightarrow 0.$

\section{Stationary geometric flows of BH and solitonic hierarchies}

\label{s5} The goal of this section is to provide applications of the anholonomic frame deformation method (AFDM, outlined in previous section) for constructing in explicit form exact and parametric stationary generic off-diagonal solutions describing solitonic geometric flow deformations of prime BH metrics.

\subsection{Table 2: AFDM for constructing solitonic stationary flows of NES}

We consider $h_{4}(\tau )=h_{4}(\tau ,x^{i},y^{3})$ $=h_{4}(\tau ,{r,\theta }%
,\varphi )$ (\ref{gensolstat}) as a generating function (it can be also
determined by a family of solitonic hierarchies, $h_{4}(\tau )=h_{4}[\
_{4}\iota ]$) and construct a deformation procedure for constructing a class
of off--diagonal stationary solutions with Killing symmetry on $\partial
_{t} $ determined by solitonic hierarchies $\widehat{\Im }[\iota ]=[~\ _{h}%
\widehat{\Im }[\ _{1}\iota ],~\ _{v}\widehat{\Im }[\ _{2}\iota ]]$ (\ref%
{dsourcparam}) and a parametric running cosmological constant $\Lambda (\tau
),$ {\small
\begin{eqnarray*}
ds^{2} &=&e^{\ \psi (\tau ,x^{k})}[(dx^{1})^{2}+(dx^{2})^{2}]-\frac{%
[h_{4}^{\diamond }{}(\tau )]^{2}}{|\int dy^{3}\ \ _{v}\widehat{\Im }(\tau
)(h_{4}{}^{\diamond }(\tau ))|\ h_{4}}[dy^{3}+\frac{\partial _{i}(\int
d\varphi \ \ _{v}\widehat{\Im }(\tau )h_{4}^{\diamond }{}])}{\ _{v}\widehat{%
\Im }(\tau )\ h_{4}^{\diamond }{}(\tau )}dx^{i}] \\
&&+h_{4}(\tau )[dt+(\ _{1}n_{k}+4\ _{2}n_{k}\int d\varphi \frac{%
(h_{4}^{\diamond }{}(\tau ))^{2}}{|\int dy^{3}\ \ _{v}\widehat{\Im }(\tau
)(h_{4}^{\diamond }(\tau ))|\ (h_{4}(\tau ))^{5/2}})dx^{k}].
\end{eqnarray*}%
}

Such solutions involve different types of solitonic hierarchies and, in
general, are with nontrivial nonholonomically induced torsion which can be
nonholonomically constrained to LC-configurations (\ref{lcsolstat}).


{\scriptsize
\begin{eqnarray*}
&&%
\begin{tabular}{l}
\hline\hline
\begin{tabular}{lll}
& {\large \textsf{Table 2:\ Off-diagonal stationary flows with solitonic
hierarchies}} &  \\
& Exact solutions of $\widehat{\mathbf{R}}_{\mu \nu }(\tau )=\widehat{\Im }%
_{\mu \nu }(\tau )$ (\ref{solitonhierarcheq}) transformed into a system of
nonlinear PDEs (\ref{eq1a})-(\ref{eq4a}) &
\end{tabular}
\\
\end{tabular}
\\
&&%
\begin{tabular}{lll}
\hline\hline
$%
\begin{array}{c}
\mbox{d-metric ansatz with} \\
\mbox{Killing symmetry }\partial _{4}=\partial _{t}%
\end{array}%
$ &  & $%
\begin{array}{c}
ds^{2}=g_{i}(\tau )(dx^{i})^{2}+g_{a}(\tau )(dy^{a}+N_{i}^{a}(\tau
)dx^{i})^{2},\mbox{ for } \\
g_{i}=e^{\psi {(\tau ,r,\theta )}},\,\,\,\,g_{a}=h_{a}(\tau ,{r,\theta }%
,\varphi ),\ N_{i}^{3}=w_{i}(\tau ,{r,\theta },\varphi
),\,\,\,\,N_{i}^{4}=n_{i}(\tau ,{r,\theta },\varphi ),%
\end{array}%
$ \\
Effective matter sources &  & $\widehat{\Im }_{\ \nu }^{\mu }(\tau )=[~\ _{h}%
\widehat{\Im }(\tau ,{r,\theta })\delta _{j}^{i},~\ _{v}\widehat{\Im }(\tau ,%
{r,\theta },\varphi )\delta _{b}^{a}];x^{1}=r,x^{2}=\theta ,y^{3}=\varphi
,y^{4}=t$ \\ \hline
Nonlinear PDEs (\ref{estatsimpl}) &  & $%
\begin{array}{c}
\psi ^{\bullet \bullet }+\psi ^{\prime \prime }=2~\ \ _{h}\widehat{\Im }[\
_{1}\iota ]; \\
\varpi ^{\diamond }\ h_{4}^{\diamond }=2h_{3}h_{4}~\ _{v}\widehat{\Im }[\
_{2}\iota ]; \\
\beta w_{i}-\alpha _{i}=0; \\
n_{k}^{\diamond \diamond }+\gamma n_{k}^{\diamond }=0;%
\end{array}%
$ for $%
\begin{array}{c}
\varpi {=\ln |\partial _{3}h_{4}/\sqrt{|h_{3}h_{4}|}|,} \\
\alpha _{i}=(\partial _{\varphi }h_{4})\ (\partial _{i}\varpi ),\ \beta
=(\partial _{\varphi }h_{4})\ (\partial _{\varphi }\varpi ), \\
\ \gamma =\partial _{\varphi }\left( \ln |h_{4}|^{3/2}/|h_{3}|\right) , \\
\partial _{1}q=q^{\bullet },\partial _{2}q=q^{\prime },\partial
_{3}q=\partial q/\partial \varphi =q^{\diamond }%
\end{array}%
$ \\ \hline
$%
\begin{array}{c}
\mbox{ Generating functions:}\ h_{4}[\ _{4}\iota ], \\
\Psi (\tau ,{r,\theta },\varphi )=e^{\varpi },\Phi \lbrack \iota ]; \\
\mbox{integration functions:}\ h_{4}^{[0]}(\tau ,x^{k}),\  \\
_{1}n_{k}(\tau ,x^{i}),\ _{2}n_{k}(\tau ,x^{i})%
\end{array}%
$ &  & $%
\begin{array}{c}
\ (\Psi ^{2})^{\diamond }=-\int dy^{3}\ ~\ _{v}\widehat{\Im }h_{4}^{\
\diamond },\Phi ^{2}=-4\Lambda (\tau )h_{4}, \\
\mbox{ see nonlinear symmetries }(\ref{nsym1a}); \\
h_{4}(\tau )=h_{4}^{[0]}-\Phi ^{2}/4\Lambda (\tau ),h_{4}^{\diamond }\neq
0,\Lambda (\tau )\neq 0=const%
\end{array}%
$ \\ \hline
Off-diag. solutions, $%
\begin{array}{c}
\mbox{d--metric} \\
\mbox{N-connec.}%
\end{array}%
$ &  & $%
\begin{array}{c}
\ g_{i}(\tau )=e^{\ \psi (\tau ,x^{k})}%
\mbox{ as a solution of 2-d Poisson
eqs. }\psi ^{\bullet \bullet }+\psi ^{\prime \prime }=2~\ _{h}\widehat{\Im }%
(\tau ); \\
h_{3}(\tau )=-(\Psi ^{\diamond })^{2}/4(\ _{v}\widehat{\Im })^{2}h_{4},%
\mbox{ see }(\ref{offdstat}); \\
h_{4}(\tau )=h_{4}^{[0]}-\int dy^{3}(\Psi ^{2})^{\diamond }/4\ _{v}\widehat{%
\Im }=h_{4}^{[0]}-\Phi ^{2}/4\Lambda (\tau ); \\
w_{i}(\tau )=\partial _{i}\ \Psi /\ \partial _{\varphi }\Psi =\partial _{i}\
\Psi ^{2}/\ |\partial _{\varphi }\Psi ^{2}|; \\
n_{k}(\tau )=\ _{1}n_{k}+\ _{2}n_{k}\int dy^{3}(\Psi ^{\diamond })^{2}/(\
_{v}\widehat{\Im })^{2}|h_{4}^{[0]}-\int dy^{3}(\Psi ^{2})^{\diamond }/4\
_{v}\widehat{\Im }|^{5/2}.%
\end{array}%
$ \\ \hline
LC-configurations (\ref{lcconda}) &  & $%
\begin{array}{c}
\partial _{\varphi }w_{i}=(\partial _{i}-w_{i}\partial _{\varphi })\ln \sqrt{%
|h_{3}(\tau )|},(\partial _{i}-w_{i}\partial _{\varphi })\ln \sqrt{%
|h_{4}(\tau )|}=0, \\
\partial _{k}w_{i}(\tau )=\partial _{i}w_{k}(\tau ),\partial _{\varphi
}n_{i}(\tau )=0,\partial _{i}n_{k}(\tau )=\partial _{k}n_{i}(\tau );\Psi =%
\check{\Psi}[\iota ],(\partial _{i}\check{\Psi})^{\diamond }=\partial _{i}(%
\check{\Psi}^{\diamond }) \\
\mbox{ and }\ _{v}\widehat{\Im }(\tau ,x^{i},\varphi )=\ _{v}\widehat{\Im }[%
\check{\Psi}]=\ _{v}\check{\Im},\mbox{ or }\ _{v}\widehat{\Im }=const.%
\end{array}%
$ \\ \hline
N-connections, zero torsion &  & $w_{i}(\tau )=\partial _{i}\check{A}(\tau
)=\left\{
\begin{array}{c}
\partial _{i}(\int d\varphi \ \check{\Im}\ \check{h}_{4}{}^{\diamond }])/%
\check{\Im}\ \check{h}_{4}{}^{\diamond }; \\
\partial _{i}\check{\Psi}/\check{\Psi}^{\diamond }; \\
\partial _{i}(\int dy^{3}\ \check{\Im}(\check{\Phi}^{2})^{\diamond })/(%
\check{\Phi})^{\diamond }\check{\Im};%
\end{array}%
\right. \mbox{ and }n_{k}(\tau )=\check{n}_{k}(\tau )=\partial _{k}n(\tau
,x^{i}).$ \\ \hline
$%
\begin{array}{c}
\mbox{polarization functions} \\
\mathbf{\mathring{g}}\rightarrow \widehat{\mathbf{g}}\mathbf{=}[g_{\alpha
}=\eta _{\alpha }\mathring{g}_{\alpha },\ \eta _{i}^{a}\mathring{N}_{i}^{a}]%
\end{array}%
$ &  & $%
\begin{array}{c}
ds^{2}=\eta _{1}[\ _{1}\iota ]\mathring{g}_{1}(r,\theta )[dx^{1}(r,\theta
)]^{2}+\eta _{2}[\ _{2}\iota ]\mathring{g}_{2}(r,\theta )[dx^{2}(r,\theta
)]^{2}+\eta _{3}[\ _{3}\iota ]\mathring{g}_{3}(r,\theta ) \\
\lbrack d\varphi +\eta _{i}^{3}[\ _{5}\iota ]\mathring{N}_{i}^{3}(r,\theta
)dx^{i}(r,\theta )]^{2}+\eta _{4}[\ _{4}\iota ]\mathring{g}_{4}(r,\theta
)[dt+\eta _{i}^{4}[\ _{6}\iota ]\mathring{N}_{i}^{4}(r,\theta
)dx^{i}(r,\theta )]^{2},%
\end{array}%
$ \\ \hline
Prime metric defines a BH &  & $%
\begin{array}{c}
\lbrack \mathring{g}_{i}(r,\theta ),\mathring{g}_{a}=\mathring{h}%
_{a}(r,\theta );\mathring{N}_{k}^{3}=\mathring{w}_{k}(r,\theta ),\mathring{N}%
_{k}^{4}=\mathring{n}_{k}(r,\theta )] \\
\mbox{diagonalizable by frame/ coordinate transforms.} \\
\end{array}%
$ \\
Example of a prime metric &  & $%
\begin{array}{c}
\mathring{g}_{1}=(1-r_{g}/r)^{-1},\mathring{g}_{2}=r^{2},\mathring{h}%
_{3}=r^{2}\sin ^{2}\theta ,\mathring{h}_{4}=(1-r_{g}/r),r_{g}=const \\
\mbox{the Schwarzschild solution, or any BH solution.}%
\end{array}%
$ \\ \hline
Solutions for polarization funct. &  & $%
\begin{array}{c}
\eta _{i}(\tau )=e^{\ \psi (\tau ,x^{k})}/\mathring{g}_{i};\eta _{3}%
\mathring{h}_{3}=-\frac{4[(|\eta _{4}\mathring{h}_{4}|^{1/2})^{\diamond
}]^{2}}{|\int dy^{3}\ _{v}\widehat{\Im }[(\eta _{4}\mathring{h}%
_{4})]^{\diamond }|\ }; \\
\eta _{4}(\tau )=\eta _{4}(\tau ,r,\theta ,\varphi )=\eta _{4}[\ _{4}\iota ]%
\mbox{ as a generating
function}; \\
\ \eta _{i}^{3}(\tau )\ \mathring{N}_{i}^{3}=\frac{\partial _{i}\ \int
dy^{3}\ _{v}\widehat{\Im }(\eta _{4}\ \mathring{h}_{4})^{\diamond }}{\ _{v}%
\widehat{\Im }\ (\eta _{4}\ \mathring{h}_{4})^{\diamond }};\eta _{k}^{4}\
(\tau )\mathring{N}_{k}^{4}=\ _{1}n_{k}+16\ \ _{2}n_{k}\int dy^{3}\frac{%
\left( [(\eta _{4}\mathring{h}_{4})^{-1/4}]^{\diamond }\right) ^{2}}{|\int
dy^{3}\ _{v}\widehat{\Im }[(\eta _{4}\ \mathring{h}_{4})]^{\diamond }|\ }%
\end{array}%
$ \\ \hline
Polariz. funct. with zero torsion &  & $%
\begin{array}{c}
\eta _{i}(\tau )=e^{\ \psi (\tau ,x^{k})}/\mathring{g}_{i};\eta _{4}=\check{%
\eta}_{4}(\tau ,r,\theta ,\varphi )\mbox{ as a generating function}; \\
\eta _{3}(\tau )=-\frac{4[(|\eta _{4}\mathring{h}_{4}|^{1/2})^{\diamond
}]^{2}}{\mathring{g}_{3}|\int dy^{3}\ _{v}\check{\Im}[(\check{\eta}_{4}%
\mathring{h}_{4})]^{\diamond }|\ };\eta _{i}^{3}(\tau )=\frac{\partial _{i}%
\check{A}}{\mathring{w}_{k}},\eta _{k}^{4}(\tau )=\frac{\ \partial _{k}n}{%
\mathring{n}_{k}}%
\end{array}%
$ \\ \hline\hline
\end{tabular}%
\end{eqnarray*}%
} 

\subsection{Nonlinear PDEs for geometric flows with stationary solitonic
hierarchies}

The goal of this subsection is to study explicit examples for constructing
exact and parametric solutions encoding solitonic hierarchies for geometric
flow modified Einstein equations (\ref{solitonhierarcheq}) transformed into
systems of nonlinear PDEs with decoupling (\ref{estatsimpl}).

\subsubsection{Parametric stationary solutions with solitonic sources}

We shall write that in a geometric flow source $~\ _{v}\widehat{\Im }(\tau )$
there is, for instance, a term with left label "0" written $\ _{0}^{int}%
\widehat{\Im }(\tau )=\ _{v}^{int}\widehat{\Im }[\ _{2}\iota ]$ if the
corresponding term in $\ \ ^{eff}\widehat{\mathbf{\Upsilon }}_{\ \nu }^{\mu
}(\tau )\ $ (\ref{dsourcparam}) is defined as a stationary functional on a
solitonic hierarchy $[\ _{2}\iota ].$ If it is written $\ \ _{v}^{int}%
\widehat{\Im }(\tau )$ without a left label "0", such a term correspond to a
general $\ ^{eff}\widehat{\mathbf{\Upsilon }}_{\ \nu }^{\mu }(\tau )$
(without any solitonic specification) encoding contributions from a
distortion tensor $\widehat{\mathbf{Z}}$ (\ref{distr}). In this work, an
effective source term $\ \ _{v}^{fl}\widehat{\Im }$ determined by geometric
flows of the d-metric, $\partial _{\tau }\mathbf{g}_{\alpha ^{\prime }}(\tau
),$ in (\ref{dsourcparam}) is introduced. It is solitonic if the d-metric
coefficients are solitonic. We can consider solitonic hierarchies for Ricci
soliton configurations with $\ _{v}^{fl}\widehat{\Im }=0.$

For this class of solutions, we consider a source (\ref{dsourcparam}) (the
left label $a$ is used for "additive stationary")
\begin{equation}
~\ _{v}^{a}\widehat{\Im }(\tau )=\ _{v}^{a}\widehat{\Im }[\iota ]~=\ _{v}^{a}%
\widehat{\Im }(\tau ,{x}^{i},y^{3})=\ _{v}^{fl}\widehat{\Im }[\ _{1}\iota
]+\ _{v}^{int}\widehat{\Im }[\ _{2}\iota ]+\ _{v}^{int}\widehat{\Im }[\
_{3}\iota ],  \label{adsourcstat}
\end{equation}%
where it is considered that we prescripe an effective solitonic hierarchy
for matter fields even, in general, such gravitational interactions can be of
non-solitonic type. The second equation (\ref{estatsimpl}) with source $\
_{v}\widehat{\Im }[\iota ]$ = $\ _{v}^{a}\widehat{\Im }[\iota ]$ can be
integrated on $y^{3}.$ In result, we construct off-diagonal metrics and
generalized connections encoding solitonic hierarchic determined, , by a
generating function $h_{4}(\tau ,{r,\theta },\varphi )$ with Killing
symmetry on $\partial _{t},$ by effective sources $\ ^{a}\widehat{\Im }%
[\iota ]~=(\ _{h}^{a}\widehat{\Im }[\ _{1}\iota ],\ \ _{v}^{a}\widehat{\Im }%
[\ _{2}\iota ])$ and effective cosmological constant%
\begin{equation}
\ ^{a}\Lambda (\tau )=\ ^{fl}\Lambda (\tau )+\ ^{m}\Lambda (\tau )+\
_{0}^{int}\Lambda (\tau )  \label{adcosmconststat}
\end{equation}%
related to $\ _{v}^{a}\widehat{\Im }[\iota ]~$ (\ref{adsourcstat}) via
nonlinear symmetry transforms (\ref{nsym1a}).

Applying the method summarized in Table 2, we construct such a class of
quadratic elements defining stationary generic off-diagonal solutions
determined by effective sources encoding solitonic hierarchies,
\begin{eqnarray}
ds^{2} &=&e^{\ \psi \lbrack \ _{1}\iota ]}[(dx^{1})^{2}+(dx^{2})^{2}]-\frac{%
[h_{4}{}^{\diamond }(\tau )]^{2}}{|\int dy^{3}\ \ _{v}^{a}\widehat{\Im }%
[\iota ]h_{4}{}^{\diamond }(\tau )|\ h_{4}(\tau )}\left[ dy^{3}+\frac{%
\partial _{i}(\int dy^{3}\ \ \ \ _{v}^{a}\widehat{\Im }[\iota ]\
h_{4}{}^{\diamond }(\tau ))}{\ \ \ _{v}^{a}\widehat{\Im }[\iota ]\
h_{4}{}^{\diamond }(\tau )}dx^{i}\right]  \notag \\
&&+h_{4}(\tau )\left[ dt+\left( \ _{1}n_{k}(\tau )+4\ _{2}n_{k}(\tau )\int
dy^{3}\frac{[h_{4}{}^{\diamond }(\tau )]^{2}}{|\int dy^{3}\ \ \ _{v}^{a}%
\widehat{\Im }[\iota ]h_{4}^{\diamond }(\tau )|\ [h_{4}(\tau )]^{5/2}}%
\right) dx^{k}\right] .  \label{stasdm}
\end{eqnarray}%
Such solutions can be constrained to LC-configurations. The formulas (\ref%
{stasdm}) can be re-defined equivalently in terms of generating functions $%
\Psi (\tau ,{r,\theta },\varphi )$ or $\Phi (\tau ,{r,\theta },\varphi )$
which can be of a general (non--solitonic) character.

\subsubsection{Modified and Einstein gravity with stationary solitonic
generating functions}

We can generate generic off-diagonal stationary solutions using generating
functionals encoding solitonic hierarchies $\Phi (\tau )=\Phi \lbrack i]$
characterized by nonlinear symmetries of type (\ref{nsym1a}) and general
effective sources $\ _{v}\widehat{\Im }(\tau )$ which can be of
non-solitonic character. The second equation into (\ref{estatsimpl})
transforms into
\begin{equation*}
\varpi ^{\diamond }(\tau )[\ \Phi \lbrack i],\Lambda (\tau )]\
h_{4}^{\diamond }(\tau )[\Phi \lbrack i],\Lambda (\tau )]=2h_{3}(\tau )[\Phi
\lbrack i],\Lambda (\tau )]h_{4}(\tau )[\Phi \lbrack i],\Lambda (\tau )]\ \
_{v}\widehat{\Im }(\tau ),
\end{equation*}%
which can be solved together with other equations (\ref{eq1a})-(\ref{eq4a})
following the AFDM, see Table 2.

The solutions for such stationary configurations determined by general
nonlinear functionals for generating functions can be written in all forms (%
\ref{gensolstat}). For simplicity, we present \ here the quadratic line
element only the third type parametrization
\begin{eqnarray}
ds^{2} &=&e^{\ \psi (\tau ,x^{k})}[(dx^{1})^{2}+(dx^{2})^{2}]-\frac{(\Phi
\lbrack i])^{2}[(\Phi \lbrack i])^{2}]^{\diamond }}{|\Lambda (\tau )\int
dy^{3}\ _{v}\widehat{\Im }(\tau )[(\Phi \lbrack i])^{2}]^{\diamond }|\
(h_{4}^{[0]}-\frac{(\Phi \lbrack i])^{2}}{4\Lambda (\tau )})}
\label{stasdmnf} \\
&&[dy^{3}+\frac{\partial _{i}\left( \int dy^{3}\ _{v}\widehat{\Im }(\tau
)[(\Phi \lbrack i])^{2}]^{\diamond }\right) }{\ _{v}\widehat{\Im }(\tau )\
[(\Phi \lbrack i])^{2}]^{\diamond }}dx^{i}]+(h_{4}^{[0]}(\tau ,x^{k})-\frac{%
(\Phi \lbrack i])^{2}}{4\Lambda (\tau )})[dt+(_{1}n_{k}(\tau ,x^{k})+  \notag
\\
&&\ _{2}n_{k}(\tau ,x^{k})\int dy^{3}\frac{\ [(\Phi \lbrack
i])^{2}]^{\diamond }}{|\Lambda (\tau )\int dy^{3}\ \ _{v}\widehat{\Im }(\tau
)\ [(\Phi \lbrack i])^{2}]^{\diamond }|}|h_{4}^{[0]}(\tau ,x^{k})-\frac{\
(\Phi \lbrack i])^{2}}{4\Lambda (\tau )}|^{-5/2})dx^{k}].  \notag
\end{eqnarray}%
For zero torsion constraints, there are extracted LC-configurations,%
\begin{eqnarray}
ds^{2} &=&e^{\psi (\tau )}[(dx^{1})^{2}+(dx^{2})^{2}]-\frac{(\check{\Phi}%
[i])^{2}[(\check{\Phi}[i])^{2}]^{\diamond }}{|\Lambda (\tau )\int dy^{3}\
_{v}\widehat{\Im }(\tau )[(\check{\Phi}[i])^{2}]^{\diamond }|\ (h_{4}^{[0]}-%
\frac{(\check{\Phi}[i])^{2}}{4\Lambda (\tau )})}[dy^{3}+(\partial _{i}\
\check{A}(\tau ))dx^{i}]  \notag \\
&&+(h_{4}^{[0]}(\tau ,x^{k})-\frac{(\check{\Phi}[i])^{2}}{4\Lambda (\tau )})%
\left[ dt+(\partial _{k}n(\tau ))dx^{k}\right] ,  \label{stasdmnflc}
\end{eqnarray}
where $\check{A}(\tau )$ and $n(\tau )$ are also generating functions.

\subsubsection{ Small N-adapted stationary solitonic flow deformations}

\label{assedef}We can study important physical properties of some classes of
solutions if there are considered small parametric deformations from certain
well-known solutions (for instance, from a black hole, BH, configuration of
Kerr or Schwarzschild type).

Let us consider a prime pseudo--Riemannian d-metric $\mathbf{\mathring{g}}=[%
\mathring{g}_{i},\mathring{g}_{a},\mathring{N}_{b}^{j}]$ (\ref{primedm})
when $\partial _{3}\mathring{g}_{4}=\mathring{g}_{4}^{\diamond }\neq 0.$ It
can be diagonalized via coordinate transforms. Our goal is to formulate a
geometric formalism for small generic off--diagonal parametric deformations
of $\mathbf{\mathring{g}}$ into certain target stationary metrics of type $g$
(\ref{dme})
\begin{eqnarray}
ds^{2} &=&\eta _{i}(\varepsilon ,\tau )\mathring{g}_{i}(dx^{i})^{2}+\eta
_{a}(\varepsilon ,\tau )\mathring{g}_{a}(\mathbf{e}^{a})^{2},  \label{targm}
\\
\mathbf{e}^{3} &=&dy^{3}+\ ^{w}\eta _{i}(\varepsilon ,\tau )\mathring{w}%
_{i}dx^{i},\mathbf{e}^{4}=dt+\ ^{n}\eta _{i}(\varepsilon ,\tau )\mathring{n}%
_{i}dx^{i},\   \notag
\end{eqnarray}%
where the coefficients $[g_{\alpha }=\eta _{\alpha }\mathring{g}_{\alpha
},^{w}\eta _{i}\mathring{w}_{i},\ ^{n}\eta _{i}n_{i}]$ depend on a small
parameter $\varepsilon ,$ $0\leq \varepsilon \ll 1,$ and on evolution
parameter $\tau $ (in this work, it is used both for the geometric and curve
flow evolution). We suppose that (\ref{targm}) define a solution of entropic
flow evolution equations reduced to the system of nonlinear PDEs with
decoupling (\ref{estatsimpl}). Some $\varepsilon $ -deformations are
parameterised in the form
\begin{eqnarray}
&&\eta _{i}(\varepsilon ,\tau )=1+\varepsilon \upsilon _{i}(\tau
,x^{k}),\eta _{a}=1+\varepsilon \upsilon _{a}(\tau ,x^{k},y^{3})%
\mbox{  for
the coefficients of d-metrics };  \label{smpolariz} \\
&&^{w}\eta _{i}(\varepsilon ,\tau )=1+\varepsilon \ ^{w}\upsilon _{i}(\tau
,x^{k},y^{3}),\ \ ^{n}\eta _{i}(\tau ,x^{k},y^{3})=1+\varepsilon \ \
^{n}\upsilon _{i}(\tau ,x^{k},y^{3})\mbox{for the coefficients of N-metrics }%
,  \notag
\end{eqnarray}%
where $g_{4}(\tau )=\eta _{4}(\tau )\mathring{g}_{4}=\ \eta _{4}(\tau
,r,\theta ,\varphi )\mathring{g}_{4}(r,\theta ,\varphi )=[1+\varepsilon
\upsilon (\tau ,r,\theta ,\varphi )]\mathring{g}_{4},$ for $\upsilon
=\upsilon _{4}(\tau ,r,\theta ,\varphi )$ and $g_{4}^{\diamond }(\tau )\neq
0 $, as a generating function.

Deformations of $h$-components of a stationary d-metric are written $\
_{\varepsilon }g_{i}=\mathring{g}_{i}(1+\varepsilon \upsilon _{i})=e^{\psi
(\tau ,x^{k})}$ for a solution of the 2-d Laplace equation in (\ref%
{estatsimpl}). For $\ \psi (\tau )=\ \ ^{0}\psi (\tau ,x^{k})+\varepsilon \
^{1}\psi (\tau ,x^{k})$\ and $\ _{h}\widehat{\Im }(\tau )(\tau )=\ _{h}^{0}%
\widehat{\Im }(\tau ,x^{k})+\varepsilon \ _{h}^{1}\widehat{\Im }(\tau
,x^{k}),$ we compute the deformation polarization functions in the form $%
\upsilon _{i}=e^{\ ^{0}\psi }\ ^{1}\psi /\mathring{g}_{i}\ \ _{h}^{0}%
\widehat{\Im }.$ In these formulas, the generating and source functions are
solutions of $\ ^{0}\psi ^{\bullet \bullet }+\ ^{0}\psi ^{\prime \prime }=\
_{h}^{0}\widehat{\Im }$ and $\ ^{1}\psi ^{\bullet \bullet }+\ ^{1}\psi
^{\prime \prime }=\ \ _{h}^{1}\widehat{\Im }.$ Using such $\varepsilon $%
-decomposition of polarization functions of type (\ref{smpolariz}), we
obtain $\varepsilon $-decomposition of the target stationary d-metric and
N-connection coefficients\footnote{%
the vertical components for $\varepsilon $-decompositions of generating
solutions are computed in a similar form} and compute
\begin{eqnarray*}
\mathring{g}_{i}\eta _{i}(\tau ) &=&e^{\ \psi (\tau ,x^{k})}%
\mbox{ as a
solution of 2-d Poisson equations } \\
\ _{\varepsilon }g_{i}(\tau ) &=&[1+\varepsilon e^{\ ^{0}\psi }\ ^{1}\psi /%
\mathring{g}_{i}\ \ _{h}^{0}\widehat{\Im }]\mathring{g}_{i},%
\mbox{ also constructed as a
solution of 2-d Poisson equations for }\ ^{1}\psi \\
\mathring{g}_{3}\eta _{3}(\tau ) &=&-\frac{4[(|\eta _{4}(\tau )\mathring{g}%
_{4}|^{1/2})^{\diamond }]^{2}}{|\int dy^{3}\ \ _{v}\widehat{\Im }(\tau
)[\eta _{4}(\tau )\mathring{g}_{4}]^{\diamond }|\ } \\
&& \mbox{ i.e. } \ _{\varepsilon }g_{3}(\tau ) = [1+\varepsilon \ \upsilon
_{3}]\mathring{g}_{3}\mbox{ for }\upsilon _{3}(\tau ,x^{i},y^{3})=2\frac{%
(\upsilon \mathring{g}_{4})^{\diamond }}{\mathring{g}_{4}^{\diamond }}-\frac{%
\int dy^{3}\ \ _{v}\widehat{\Im }(\tau )(\upsilon \mathring{g}%
_{4})^{\diamond }}{\int dy^{3}\ \ _{v}\widehat{\Im }(\tau )\mathring{g}%
_{4}^{\diamond }}.
\end{eqnarray*}%
In these formulas, a new system of coordinates $[x^{i}(r,\theta ,\varphi
),y^{3}(r,\theta ,\varphi )]$ is used in order to satisfy the condition $(%
\mathring{g}_{4}^{\diamond })^{2}=\mathring{g}_{3}|\int dy^{3}\ _{v}\widehat{%
\Im }(\tau )\mathring{g}_{4}^{\diamond }|,$ which allow to find $\mathring{g}%
_{3}$ for any prescribed values $\mathring{g}_{4}$ and$\ _{v}\widehat{\Im }%
(\tau ).$

The N-connection coefficients are computed
\begin{eqnarray*}
\eta _{i}^{3}(\tau )\mathring{w}_{i} &=&\frac{\partial _{i}\ \int dy^{3}\ \
_{v}\widehat{\Im }(\tau )[\eta _{4}(\tau )\ \mathring{g}_{4}]^{\diamond }}{\
\ \ _{v}\widehat{\Im }(\tau )\ [\eta _{4}(\tau )\ \mathring{g}%
_{4}]^{\diamond }} \\
&& \mbox{ i.e. } \ _{\varepsilon }w_{i}(\tau ) = [1+\varepsilon \
^{w}\upsilon _{i}(\tau )]\mathring{w}_{i}\mbox{ for }\ ^{w}\upsilon
_{i}(\tau ,x^{i},y^{3})=\frac{\partial _{i}\ \int dy^{3}\ \ _{v}\widehat{\Im
}(\tau )(\upsilon \mathring{g}_{4})^{\diamond }}{\partial _{i}\ \int dy^{3}\
\ \ _{v}\widehat{\Im }(\tau )\mathring{g}_{4}^{\diamond }}-\frac{(\upsilon
\mathring{g}_{4})^{\diamond }}{\mathring{g}_{4}^{\diamond }},
\end{eqnarray*}%
when $\mathring{w}_{i}=\partial _{i}\ \int dy^{3}\ _{v}\widehat{\Im }(\tau )%
\mathring{g}_{4}^{\diamond }/\ _{v}\widehat{\Im }(\tau )\mathring{g}%
_{4}^{\diamond }$ is defined for some prescribed $\ _{v}\widehat{\Im }(\tau
) $ and $\mathring{g}_{4}^{\diamond };$%
\begin{eqnarray*}
\eta _{k}^{4}(\tau )\mathring{n}_{k} &=&\ _{1}n_{k}(\tau )+16\
_{2}n_{k}(\tau )\int dy^{3}\frac{\left( [(\eta _{4}(\tau )\mathring{g}%
_{4})^{-1/4}]^{\diamond }\right) ^{2}}{|\int dy^{3}\ _{v}\widehat{\Im }(\tau
)(\eta _{4}\ (\tau )\mathring{g}_{4})^{\diamond }|\ } \\
&& \mbox{ i.e. } \ _{\varepsilon }n_{i}(\tau ) = [1+\varepsilon \
^{n}\upsilon _{i}(\tau )]\mathring{n}_{k}=0\mbox{ for }\ ^{n}\upsilon
_{i}(\tau ,x^{i},y^{3})=0,
\end{eqnarray*}%
if the integration functions are chosen $_{1}n_{k}(\tau )=0$ and $%
_{2}n_{k}(\tau )=0.$

The values with a "circle" are prescribed for a chosen prime solution which
can be a 4-d Kerr metric but subjected to some additional frame and
coordinated transform to satisfy the conditions $\mathring{g}_{4}^{\diamond
}\neq 0$ and relations of $\mathring{g}_{4}^{\diamond }$ to $\mathring{g}%
_{3} $ and $\mathring{w}_{i}$ as we considered above. Fixing a small value $%
\varepsilon $, we can compute such deformations for stationary
configurations and state well-defined conditions of stability if the prime
metric is stable. We conclude that $\varepsilon $--deformed quadratic
elements can be written in a general form
\begin{eqnarray*}
ds_{\varepsilon t}^{2} &=&\ _{\varepsilon }g_{\alpha \beta }(\tau
,x^{k},y^{3})du^{\alpha _{s}}du^{\beta _{s}} \\
&=&\ _{\varepsilon }g_{i}\left( \tau ,x^{k}\right)
[(dx^{1})^{2}+(dx^{2})^{2}]+\ _{\varepsilon }h_{3}(\tau ,x^{k},y^{3})\
[dy^{3}+\ _{\varepsilon }w_{i}(\tau ,x^{k},y^{3})dx^{i}]^{2}+\ _{\varepsilon
}g_{4}(\tau ,x^{k},y^{3})dt^{2}.
\end{eqnarray*}%
We can impose additional constraints in order to extract LC--configurations
with zero torsion.

\subsection{BHs in (off-) diagonal stationary media with solitonic
hierarchies}

We can construct and describe new classes of classes of generic off-diagonal
stationary solutions in terms of $\eta $--polarization functions introduced
in formulas (\ref{dme}) and following the AFDM summarized in Tables 1 and 2.
As a primary metric we consider a primary BH d-metric (for instance, it can
be a Schwarzschild or Kerr metric) defined by geometric data $\mathbf{%
\mathring{g}=}[\mathring{g}_{i}(r,\theta ,\varphi ),\mathring{g}_{a}=%
\mathring{h}_{a}(r,\theta ,\varphi );\mathring{N}_{k}^{3}=\mathring{w}%
_{k}(r,\theta ,\varphi ),\mathring{N}_{k}^{4}=\mathring{n}_{k}(r,\theta
,\varphi )]$ (\ref{primedm}) which can be diagonalized by frame/ coordinate
transforms. The stationary target metrics $\mathbf{g}$ are generated by
nonholonomic $\eta $--deformations, $\mathbf{\mathring{g}}\rightarrow
\mathbf{g}(\tau )\mathbf{=}[g_{i}(\tau ,x^{k})=\eta _{i}(\tau )\mathring{g}%
_{i},g_{b}(\tau ,x^{k},y^{3})=\eta _{b}(\tau )\mathring{g}%
_{b},N_{i}^{a}(\tau ,x^{k},y^{3})=\ \eta _{i}^{a}(\tau )\mathring{N}%
_{i}^{a}],$ and constrained to the conditions to define exact and parametric
solutions of the system of nonlinear PDEs with decoupling (\ref{estatsimpl}%
). The quadratic line elements corresponding to d-metrics $\mathbf{g}$ are
parameterized in some forms similar to (\ref{dme}), {\small
\begin{equation}
ds^{2}=\eta _{i}(\tau ,r,\theta ,\varphi )\mathring{g}_{i}(r,\theta ,\varphi
)[dx^{i}(r,\theta ,\varphi )]^{2}+\eta _{a}(\tau ,r,\theta ,\varphi )%
\mathring{g}_{a}(r,\theta ,\varphi )[d\varphi +\eta _{k}^{a}(\tau ,r,\theta
,\varphi )\mathring{N}_{k}^{a}(r,\theta ,\varphi )dx^{k}(r,\theta ,\varphi
)]^{2},  \label{statsingpf}
\end{equation}%
} with summation on repeating contracted low-up indices. The values $\eta
_{\alpha }(\tau )$ and $\eta _{i}^{a}(\tau )$ are determined by solitonic
flows and nonlinear interactions.

\subsubsection{Stationary solutions generated by solitonic sources}

Considering effective sources determined by solitonic hierarchies $\widehat{%
\mathbf{\Upsilon }}(\tau ,{r,\theta },\varphi )=\ _{v}\widehat{\Im }[\
_{2}\iota ]$ (\ref{dsourcparam}), we compute the coefficients for solutions
of type (\ref{statsingpf}) following formulas \ from Table 2,{\small
\begin{eqnarray}
\eta _{i}(\tau ) &=&\frac{e^{\ \psi (\tau ,x^{k})}}{\mathring{g}}_{i};\eta
_{3}(\tau )=-\frac{4[(|\eta _{4}(\tau )\mathring{h}_{4}|^{1/2})^{\diamond
}]^{2}}{\mathring{h}_{3}|\int dy^{3}\ \ \ _{v}\widehat{\Im }[\ _{2}\iota
](\eta _{4}(\tau )\mathring{h}_{4})^{\diamond }|\ };  \label{statsingpfqpa}
\\
\eta _{4}(\tau ) &=&\eta _{4}(\tau ,r,\theta ,\varphi )%
\mbox{ as a generating
function};  \notag \\
\eta _{i}^{3}(\tau ) &=&\frac{\partial _{i}\ \int dy^{3}\ \ _{v}\widehat{\Im
}[\ _{2}\iota ](\eta _{4}(\tau )\mathring{h}_{4})^{\diamond }}{\mathring{w}%
_{i}\ \ \ _{v}\widehat{\Im }[\ _{2}\iota ]\ (\eta _{4}(\tau )\mathring{h}%
_{4})^{\diamond }};\ \eta _{k}^{4}\ (\tau )=\frac{\ _{1}n_{k}}{\mathring{n}%
_{k}}+16\ \ \frac{\ _{2}n_{k}}{\mathring{n}_{k}}\int dy^{3}\frac{\left(
[(\eta _{4}(\tau )\mathring{h}_{4})^{-1/4}]^{\diamond }\right) ^{2}}{|\int
dy^{3}\ \ _{v}\widehat{\Im }[\ _{2}\iota ](\eta _{4}\ (\tau )\mathring{h}%
_{4})^{\diamond }|},  \notag
\end{eqnarray}%
} for integration functions $\ _{1}n_{k}(\tau ,{r,\theta )}$ \ and $\
_{2}n_{k}(\tau ,{r,\theta ).}$

In (\ref{statsingpfqpa}), the gravitational polarization $\eta _{4}(r,\theta
,\varphi )$ is taken as a (non) singular generating function which following
nonlinear symmetries (\ref{nsym1a}) can be related to other type generating
functions,
\begin{eqnarray*}
\Phi ^{2}(\tau ) &=&-4\ \Lambda (\tau )h_{4}(\tau )=-4\ \Lambda \eta
_{4}(\tau ,{r,\theta },\varphi )\mathring{h}_{4}(\tau ,{r,\theta },\varphi ),
\\
\ (\Psi ^{2})^{\diamond }(\tau ) &=&-\int d\varphi \ \ \ _{v}\widehat{\Im }%
[\ _{2}\iota ][\eta _{4}(\tau ,{r,\theta },\varphi )\mathring{h}_{4}(\tau ,{%
r,\theta },\varphi )]^{\diamond }.
\end{eqnarray*}%
It should be noted that the values $\Phi ,h_{4}$ and $\eta _{4}$ may not
encode soltionic hierarchies but $\Psi $ and other coefficients of such
d-metric are solitonic ones if they are computed using $\ _{v}\widehat{\Im }%
[\ _{2}\iota ].$ We can constrain the coefficients (\ref{statsingpfqpa}) to
a subclass of data generating target stationary off-diagonal metrics of type
(\ref{lcsolstat}) with zero torsion.

The nonlinear functionals for the soliton v-source and (effective)
cosmological constant considered above can be changed into additive
functionals $\ _{v}\widehat{\Im }\ \rightarrow $ $\ _{v}^{a}\widehat{\Im }$
and $\Lambda \rightarrow $ $\ ^{a}\Lambda $ as $\ _{v}^{a}\widehat{\Im
}[\iota ]$ (\ref{adsourcstat}) and $\ ^{a}\Lambda $ (\ref{adcosmconststat}). The singular behaviour of such solutions is generated by some prime BH data $\mathbf{\mathring{g}}=[\mathring{g}_{i},\mathring{g}_{a},\mathring{N}%
_{b}^{j}]$ (\ref{primedm}) which can be preserved or changed for different
classes of generating and integration functions. For certain classes of
generating functions and sources and small nonholonomic deformations, the
same type of singularity is preserved. Similar stationary configurations can
be computed for general solitonic hierarchies. The constructions depend on
the type of explicit geometric evolution or dynamical model we construct
(for instance, with one type solitonic wave, nonlinear superposition of
solitonic waves on $\tau $, solitonic stationary distributions etc.). Such
generic off-diagonal stationary entropic solutions can be considered as
certain conventional nonholonomically deformed BH configurations imbedded
into some aether (non) singular media with flows and off-diagonal
interactions determined by stationary solitonic and non-solitonic fields
modeling dark, usual matter quasiperiodic distributions and patern forming
structures.

\subsubsection{BH solutions deformed by solitonic generating functions}

Solutions with entropic $\eta $--polarizations (\ref{statsingpf}) can be
constructed with coefficients of the d-metrics determined by nonlinear
generating functionals $\Phi \lbrack i],$ or any additive functionals $\
^{a}\Phi \lbrack i],$ including terms with integration functions$\
h_{4}^{[0]}(\tau ,{r,\theta })$ for $h_{4}[i].$ Such configurations are
defined also by some prescribed data $\ \ _{v}\widehat{\Im }\ (\tau ,{%
r,\theta },\varphi )$ and $\Lambda (\tau ),$ which are not obligatory of
solitonic nature. Using nonlinear symmetries (\ref{nsym1a}), we can compute
(recurrently) corresponding nonlinear functionals, $\ \eta _{4}(\tau ,{%
r,\theta },\varphi )$ (for simplicity, we omit here similar formulas for
additive functionals $\ ^{a}\eta _{4}(\tau ,{r,\theta },\varphi ))$ and
related polarization functions,
\begin{eqnarray*}
\ \eta _{4}[i] &=&-\Phi ^{2}[i]/4\Lambda (\tau )\mathring{h}_{4}({r,\theta }%
,\varphi ), \\
\ [\Psi ^{2}(\tau )]^{\diamond } &=&-\int d\varphi \ \ \ _{v}\widehat{\Im }%
(\tau ,{r,\theta },\varphi )h_{4}^{\ \diamond }(\tau )=-\int d\varphi \ \
_{v}\widehat{\Im }(\tau ,{r,\theta },\varphi )[\ \eta _{4}[i]\mathring{h}%
_{4}({r,\theta },\varphi )]^{\diamond }.
\end{eqnarray*}

Using such formulas for Table 2, the coefficients of d-metric (\ref%
{statsingpf}) are computed {\small
\begin{eqnarray}
\eta _{i}(\tau ) &=&\frac{e^{\ \psi (\tau ,x^{k})}}{\mathring{g}_{i}};\eta
_{3}=-\frac{4[(|\ \eta _{4}[i]\mathring{h}_{4}|^{1/2})^{\diamond }]^{2}}{%
\mathring{h}_{3}|\int dy^{3}\ \ _{v}\widehat{\Im }(\tau )\ \eta _{4}[i]%
\mathring{h}_{4})^{\diamond }|\ };  \label{statsingpfqp1} \\
\eta _{4}(\tau ) &=&\eta _{4}(\tau ,r,\theta ,\varphi )=\eta _{4}[i]%
\mbox{ as a generating
function};  \notag \\
\eta _{i}^{3}(\tau ) &=&\frac{\partial _{i}\ \int d\varphi \ \ _{v}\widehat{%
\Im }(\tau )(\eta _{4}[i]\ \mathring{h}_{4})^{\diamond }}{\mathring{w}_{i}\
\ _{v}\widehat{\Im }(\tau )\ (\eta _{4}[i]\ \mathring{h}_{4})^{\diamond }}%
;\eta _{k}^{4}(\tau )=\frac{\ _{1}n_{k}(\tau )}{\mathring{n}_{k}}+16\ \
\frac{\ _{2}n_{k}(\tau )}{\mathring{n}_{k}}\int d\varphi \frac{\left(
\lbrack (\eta _{4}[i]\mathring{h}_{4})^{-1/4}]^{\diamond }\right) ^{2}}{%
|\int dy^{3}\ _{v}\widehat{\Im }(\tau )(\eta _{4}[i]\mathring{h}%
_{4})^{\diamond }|\ },  \notag
\end{eqnarray}%
} for integrating functions $\ _{1}n_{k}\tau ,{r,\theta )}$ and $\
_{2}n_{k}\tau ,{r,\theta ).}$

Using (\ref{statsingpfqp1}), target stationary off-diagonal metrics (\ref%
{lcsolstat}) with zero torsion can be generated by polarization functions
subjected to additional nonholonomic constraints and integrability
conditions,
\begin{eqnarray*}
\eta _{i}(\tau ) &=&\frac{e^{\ \psi (\tau ,x^{k})}}{\mathring{g}_{i}};\ \eta
_{3}(\tau )=-\frac{4[(|\ \check{\eta}_{4}[i]\mathring{h}_{4}|^{1/2})^{%
\diamond }]^{2}}{\mathring{h}_{3}|\int dy^{3}\ \ _{v}\check{\Im}(\tau )(%
\check{\eta}_{4}[i]\mathring{h}_{4})^{\diamond }|\ }; \\
\eta _{4}(\tau ) &=&\check{\eta}_{4}(\tau ,r,\theta ,\varphi )=\check{\eta}%
_{4}[i]\mbox{ as a
generating function };\eta _{i}^{3}(\tau )=\frac{\partial _{i}\ \check{A}%
(\tau )}{\mathring{w}_{k}},\eta _{k}^{4}(\tau )=\frac{\ \partial _{k}n(\tau )%
}{\mathring{n}_{k}},
\end{eqnarray*}%
for an integrating functions $n(\tau ,{r,\theta })$ and a generating
function $\ \check{A}(\tau ,{r,\theta },\varphi ).$

The solutions construced in this subsection describe certain
nonholonomically deformed BH configurations self-consistently imbedded into
a solitonic gravitational evolution media modeling certain aether properties
for nonholonomic dark energy distributions.

\subsubsection{Stationary BH deformations by solitonic sources \& solitonic
generating functions}

More general classes of stationary solitonic deformations of BHs can be constructed using nonlinear functionals both for the generating functions and sources. Nonlinear superpositions of solutions of type (\ref{statsingpf}) and (\ref{statsingpfqp1}) can be performed if the coefficients of d-metric are computed {\small
\begin{eqnarray}
\eta _{i}(\tau ) &=&\frac{e^{\ \psi (\tau ,x^{k})}}{\mathring{g}_{i}};\eta
_{3}(\tau )=-\frac{4[(|\eta _{4}[\ _{4}\iota ]\mathring{h}%
_{4}|^{1/2})^{\diamond }]^{2}}{\mathring{h}_{3}|\int d\varphi \ \ _{v}%
\widehat{\Im }[\iota ](\eta _{4}[\ _{4}\iota ]\mathring{h}_{4})^{\diamond
}|\ };  \label{statsingpfqp12} \\
\eta _{4}(\tau ) &=&\eta _{4}(\tau ,r,\theta ,\varphi )=\ \eta _{4}[\
_{4}\iota ]\mbox{ as a generating
function};\eta _{i}^{3}(\tau )=\frac{\partial _{i}\ \int d\varphi \ \ _{v}%
\widehat{\Im }[\iota ](\eta _{4}[\ _{4}\iota ]\ \mathring{h}_{4})^{\diamond }%
}{\mathring{w}_{i}\ \ \ _{v}\widehat{\Im }[\iota ]\ (\eta _{4}[\ _{4}\iota
]\ \mathring{h}_{4})^{\diamond }};\   \notag \\
\eta _{k}^{4}(\tau ) &=&\frac{\ _{1}n_{k}(\tau )}{\mathring{n}_{k}}+16\ \
\frac{\ _{2}n_{k}(\tau )}{\mathring{n}_{k}}\int d\varphi \frac{\left(
\lbrack (\eta _{4}[\ _{4}\iota ]\mathring{h}_{4})^{-1/4}]^{\diamond }\right)
^{2}}{|\int dy^{3}\ \ _{v}\widehat{\Im }[\iota ](\eta _{4}[\ _{4}\iota ]\
\mathring{h}_{4})^{\diamond }|\ },  \notag
\end{eqnarray}%
} where $\ _{1}n_{k}(\tau ,x^{k})$ and $\ _{2}n_{k}(\tau ,x^{k})$ are
integration functions.

In (\ref{statsingpfqp12}), we consider a nonlinear generating functional $%
\Phi \lbrack \ _{4}\iota ]$ and prescribed nonlinear functional $\ \ _{v}%
\widehat{\Im }[\iota ]$ and running constant $\Lambda (\tau )$ related via
nonlinear symmetries generalizing (\ref{nsym1a}). This allows us to compute
corresponding nonlinear functionals $\ \eta _{3}(\tau ,{r,\theta },\varphi
)=\ \eta _{3}[\iota ,\ _{4}\iota ,...]$ and polarization functions,
\begin{equation}
\ \eta _{4}(\tau ) =-\ \Phi ^{2}[\ _{4}\iota ]/4\ \Lambda (\tau )\mathring{h%
}_{4}({r,\theta },\varphi ),\    (\ \Psi ^{2}(\tau ))^{\diamond } =-\int d\varphi \ \ _{v}\widehat{\Im } [\iota ]h_{4}^{\ \diamond }[\ _{4}\iota ]=-\int d\varphi
 \ \ _{v}\widehat{\Im }[\iota ][\ \eta _{4}(\tau )\mathring{h}_{4}({r,\theta },\varphi
)]^{\diamond }.  \label{nsym1b}
\end{equation}%
Imposing additional conditions for a zero torsion, target stationary metrics
(\ref{lcsolstat}) are generated.

A corresponding quadratic line element can be written for generating data $%
\left( \Phi \lbrack \ _{4}\iota ],\Lambda (\tau )\right) :$
\begin{eqnarray}
ds^{2} &=&e^{\ \psi (\ _{4}\iota )}[(dx^{1})^{2}+(dx^{2})^{2}]-\frac{(\Phi
\lbrack \ _{4}\iota ])^{2}[(\Phi \lbrack \ _{4}\iota ])^{2}]^{\diamond }}{%
|\Lambda (\tau )\int dy^{3}\ _{v}\widehat{\Im }[\iota ][(\Phi \lbrack \
_{4}\iota ])^{2}]^{\diamond }|\ (h_{4}^{[0]}-\frac{(\Phi \lbrack \ _{4}\iota
])^{2}}{4\Lambda (\tau )})}  \label{statsingpfqp12a} \\
&&[dy^{3}+\frac{\partial _{i}\left( \int dy^{3}\ _{v}\widehat{\Im }[\iota
][(\Phi \lbrack \ _{4}\iota ])^{2}]^{\diamond }\right) }{\ _{v}\widehat{\Im }%
[\iota ]\ [(\Phi \lbrack \ _{4}\iota ])^{2}]^{\diamond }}%
dx^{i}]+(h_{4}^{[0]}(\tau ,x^{k})-\frac{(\Phi \lbrack \ _{4}\iota ])^{2}}{%
4\Lambda (\tau )})[dt+(_{1}n_{k}(\tau ,x^{k})+  \notag \\
&&\ _{2}n_{k}(\tau ,x^{k})\int dy^{3}\frac{\ [(\Phi \lbrack \ _{4}\iota
])^{2}]^{\diamond }}{|\Lambda (\tau )\int dy^{3}\ \ _{v}\widehat{\Im }[\iota
]\ [(\Phi \lbrack \ _{4}\iota ])^{2}]^{\diamond }|}|h_{4}^{[0]}(\tau ,x^{k})-%
\frac{\ (\Phi \lbrack \ _{4}\iota ])^{2}}{4\Lambda (\tau )}|^{-5/2})dx^{k}].
\notag
\end{eqnarray}%
The data for a primary BH can be extracted using nonlinear symmetries (\ref%
{nsym1b}), when $\mathring{g}_{i}=e^{\ \psi (\tau ,x^{k})}/\eta _{i}(\tau )$
and $\mathring{h}_{4}({r,\theta },\varphi )=-\ \Phi ^{2}[\ _{4}\iota ]/4\
\Lambda (\tau )\eta _{4}(\tau )$ certain values for a $\tau _{0}$ are such
way prescribed that the integration functions $h_{4}^{[0]}(\tau
,x^{k}),_{1}n_{k}(\tau ,x^{k})$ and $_{2}n_{k}(\tau ,x^{k})$ encode a prime
d-metric $\mathbf{\mathring{g}}=[\mathring{g}_{i},\mathring{g}_{a},\mathring{%
N}_{b}^{j}]$ (\ref{primedm}) and describes certain evoluton for $\tau >$ $%
\tau _{0}$ parameterized in the form (\ref{statsingpfqp12a}). This class of
stationary solutions with gravitational polarizations (\ref{statsingpfqp12})
describes nonholonomic solitonic hierarchies deformations of a BH
self-consistently imbedded into solitonic gravitational (dark energy)
backgrounds and solitonic dark and/or standard matter.

\subsection{Off--diagonal deformations of Kerr metrics by solitonic flow
sources}

In this section, we study how effective sources for geometric flows with
solitonic hierarchies flows result in generic off--diagonal deformations and
generalizations of the 4-d Kerr metric and construct such new classes of
exact solutions of systems of nonlinear PDEs (\ref{estatsimpl}).

\subsubsection{The Kerr BH solution in nonholonomic variables}

\label{asskerr} To apply the AFDM\ is necessary to define \ some special
classes of nonholonomic variables which allow decoupling and integration of
certain systems of equations describing N-adapted nonholonomic deformations,
for instance, of a Kerr metric black hole, BH, solution as prime d-metric $%
\mathbf{\mathring{g}=}[\mathring{g}_{i},\mathring{g}_{a}=\mathring{h}_{a};%
\mathring{N}_{k}^{3}=\mathring{w}_{k},\mathring{N}_{k}^{4}=\mathring{n}_{k}]$
(\ref{primedm}). We cite here \cite{misner} as a standard monograph on GR
with necessary details on geometry of BHs and \cite%
{gheorghiu14,bubuianu18,ruchin13,gheorghiu16} for examples of nonholonomic
deformations of BH solutions in geometric flows and MGTs.

Let consider a 4-d ansatz for a prime metric,
\begin{equation*}
ds_{[0]}^{2}=\mathring{g}_{\alpha \beta }du^{\alpha }du^{\beta
}=Y^{-1}e^{2h}(d\rho ^{2}+dz^{2})+Y(d\varphi +Adt)^{2}-\rho ^{2}Y^{-1}dt^{2}.
\end{equation*}
This nonlinear quadratic line element is determined by three functions $%
(h,Y,A)$ on coordinates $x^{i}=(\rho ,z).$ It defines the Kerr solution of
the vacuum Einstein equations (for rotating BHs) if the coefficients are
chosen
\begin{eqnarray*}
Y &=&\frac{1-(p\widehat{x}_{1})^{2}-(q\widehat{x}_{2})^{2}}{(1+p\widehat{x}%
_{1})^{2}+(q\widehat{x}_{2})^{2}},\ A=2M\frac{q}{p}\frac{(1-\widehat{x}%
_{2})(1+p\widehat{x}_{1})}{1-(p\widehat{x}_{1})-(q\widehat{x}_{2})}, \\
e^{2h} &=&\frac{1-(p\widehat{x}_{1})^{2}-(q\widehat{x}_{2})^{2}}{p^{2}[(%
\widehat{x}_{1})^{2}+(\widehat{x}_{2})^{2}]},\ \rho ^{2}=M^{2}(\widehat{x}%
_{1}^{2}-1)(1-\widehat{x}_{2}^{2}),\ z=M\widehat{x}_{1}\widehat{x}_{2}.
\end{eqnarray*}%
For $M=const$ and $\rho =0,$ we obtain result a horizon $\widehat{x}_{1}=0$
and the "north / south" segments of the rotation axis, $\widehat{x}%
_{2}=+1/-1.$ For our purposes, it is convenient to write this Kerr metric in
the form
\begin{equation}
ds_{[0]}^{2}=(dx^{1})^{2}+(dx^{2})^{2}+Y(\mathbf{e}^{3})^{2}-\rho ^{2}Y^{-1}(%
\mathbf{e}^{4})^{2},  \label{kerr1}
\end{equation}%
where the coordinates $x^{1}(\widehat{x}_{1},\widehat{x}_{2})$ and $x^{2}(%
\widehat{x}_{1},\widehat{x}_{2})$ are defined for any%
\begin{equation*}
(dx^{1})^{2}+(dx^{2})^{2}=M^{2}e^{2h}(\widehat{x}_{1}^{2}-\widehat{x}%
_{2}^{2})Y^{-1}\left( \frac{d\widehat{x}_{1}^{2}}{\widehat{x}_{1}^{2}-1}+%
\frac{d\widehat{x}_{2}^{2}}{1-\widehat{x}_{2}^{2}}\right)
\end{equation*}%
and the v-coordinates are changed $y^{3}=\varphi +\widehat{y}%
^{3}(x^{1},x^{2},t),$ $y^{4}=t+\widehat{y}^{4}(x^{1},x^{2}).$ We can
consider an N-adapted basis $\mathbf{e}^{3}=dy^{3}+(\partial _{i}\widehat{y}%
^{3})dx^{i}$ and $\mathbf{e}^{4}=dt+(\partial _{i}\widehat{y}^{4})dx^{i},$
for some functions $\widehat{y}^{a},$ $a=3,4,$ with $\partial _{t}\widehat{y}%
^{3}=-A(x^{k}).$

We can use the Kerr metric in the so--called Boyer--Linquist coordinates $%
(r,\vartheta ,\varphi ,t),$ for $r=m_{0}(1+p\widehat{x}_{1}),\widehat{x}%
_{2}=\cos \vartheta ,$ which are more convenient for applying of the AFDM.
Such coordinates are be related to parameters $p,q$ involving the total BH
mass, $m_{0}$ and the total angular momentum, $am_{0},$ for the
asymptotically flat, stationary and anti-symmetric Kerr spacetime.
Considering $m_{0}=Mp^{-1}$ and $a=Mqp^{-1}$ with $p^{2}+q^{2}=1$ and $%
m_{0}^{2}-a^{2}=M^{2},$ we write the metric (\ref{kerr1}) as a d-metric
\begin{eqnarray}
ds_{[0]}^{2} &=&(dx^{1^{\prime }})^{2}+(dx^{2^{\prime }})^{2}+(\overline{C}-%
\overline{B}^{2}/\overline{A})(\mathbf{e}^{3^{\prime }})^{2}+\overline{A}(%
\mathbf{e}^{4^{\prime }})^{2},  \label{kerrbl} \\
\mathbf{e}^{3^{\prime }} &=&dy^{3^{\prime }}=d\varphi ,\mathbf{e}^{4^{\prime
}}=dt+d\varphi \overline{B}/\overline{A}=dy^{4^{\prime }}-\partial
_{i^{\prime }}(\widehat{y}^{4^{\prime }}+\varphi \overline{B}/\overline{A}%
)dx^{i^{\prime }},  \notag
\end{eqnarray}%
with coordinate functions $\ x^{1^{\prime }}(r,\vartheta ),\ x^{2^{\prime
}}(r,\vartheta ),\ y^{3^{\prime }}=\varphi ,\ y^{4^{\prime }}=t+\widehat{y}%
^{4^{\prime }}(r,\vartheta ,\varphi )+\varphi \overline{B}/\overline{A}%
,\partial _{\varphi }\widehat{y}^{4^{\prime }}=-\overline{B}/\overline{A}.$
In formulas (\ref{kerrbl}), $(dx^{1^{\prime }})^{2}+(dx^{2^{\prime
}})^{2}=\Xi \left( \Delta ^{-1}dr^{2}+d\vartheta ^{2}\right) ,$ and the
coefficients are defined in the form%
\begin{eqnarray*}
\overline{A} &=&-\Xi ^{-1}(\Delta -a^{2}\sin ^{2}\vartheta ),\overline{B}%
=\Xi ^{-1}a\sin ^{2}\vartheta \left[ \Delta -(r^{2}+a^{2})\right] , \\
\overline{C} &=&\Xi ^{-1}\sin ^{2}\vartheta \left[ (r^{2}+a^{2})^{2}-\Delta
a^{2}\sin ^{2}\vartheta \right] ,\mbox{ and }\Delta =r^{2}-2m_{0}+a^{2},\
\Xi =r^{2}+a^{2}\cos ^{2}\vartheta .
\end{eqnarray*}

The quadratic linear elements (\ref{kerr1}) and/or (\ref{kerrbl}) can be
written as a stationary prime metric (\ref{primedm}) with coefficients
\begin{eqnarray}
\mathring{g}_{1} &=&1,\mathring{g}_{2}=1,\mathring{g}_{3}=Y,\mathring{g}%
_{4}=-\rho ^{2}Y^{-1},\mathring{N}_{i}^{a}=\partial _{i}\widehat{y}^{a},%
\mbox{
or \ }  \label{dkerr} \\
\mathring{g}_{1^{\prime }} &=&1,\mathring{g}_{2^{\prime }}=1,\mathring{g}%
_{3^{\prime }}=\overline{C}-\overline{B}^{2}/\overline{A},\ \mathring{g}%
_{4^{\prime }}=\overline{A},\mathring{N}_{i^{\prime }}^{3}=\mathring{w}%
_{i^{\prime }}=0,\mathring{N}_{i^{\prime }}^{4}=\mathring{n}_{i^{\prime
}}=-\partial _{i^{\prime }}(\widehat{y}^{3^{\prime }}+\varphi \overline{B}/%
\overline{A}),  \notag
\end{eqnarray}%
Such d-metrics define BH solutions of the vacuum Einstein equations with
zero sources.

\subsubsection{Nonholonomic evolution of Kerr metrics with induced (or zero)
torsion}

We consider the coefficients (\ref{dkerr}) as a prime metric $\mathbf{%
\mathring{g}}$ when $\mathring{g}_{1^{\prime }}=1,\mathring{g}_{2^{\prime
}}=1,\mathring{g}_{3^{\prime }}=\overline{C}-\overline{B}^{2}/\overline{A}\ $%
together with some coordinate transforms $g_{4^{\prime }}=\widehat{A}(%
\overline{A},y^{3})\rightarrow \overline{A}$ $,\mathring{N}_{i^{\prime
}}^{3}=\mathring{w}_{i^{\prime }}({r,\theta },\varphi )\rightarrow 0$ and $%
\mathring{N}_{i^{\prime }}^{4}=\mathring{n}_{i^{\prime }}=-\partial
_{i^{\prime }}(\widehat{y}^{3^{\prime }}+\varphi \overline{B}/\overline{A}%
)\rightarrow 0,$ to a local coordinate system when $\mathring{g}%
_{4}^{\diamond }\neq 0.$\footnote{%
we can define nonholonomic deformations with $\mathring{g}_{4}^{\diamond }=0$
and/or $g_{4}^{\diamond }\neq 0$ when the solutions are constructed on
certain hypersurfaces and certain models are with singular geometric
evolution as we considered in our previous works \cite%
{gheorghiu14,bubuianu18}} This allows us to construct nonholonomic
deformations following the geometric formalism outlined in section \ref%
{assedef} and Table 2.

For general $\eta $--deformations (\ref{targm}) and constraints $n_{i}=0,$
the solitonic flow modifications of the Kerr metric are computed
\begin{eqnarray}
ds^{2} &=&e^{\ \psi (\tau ,x^{k^{\prime }})}[(dx^{1^{\prime
}})^{2}+(dx^{2^{\prime }})^{2}]-\frac{4[(|\eta _{4}[\ _{4}\iota ]\overline{A}%
|^{1/2})^{\diamond }]^{2}}{|\int dy^{3}\ _{v}\widehat{\Im }[\iota ][\eta
_{4}[\ _{4}\iota ]\overline{A}]^{\diamond }|\ }(\overline{C}-\frac{\overline{%
B}^{2}}{\overline{A}})(\mathbf{e}^{3^{\prime }})^{2}+\eta _{4^{\prime }}[\
_{4}\iota ]\overline{A}(\mathbf{e}^{4^{\prime }})^{2},  \notag \\
\mathbf{e}^{3^{\prime }} &=&dy^{3^{\prime }}+\frac{\partial _{i^{\prime }}\
\int dy^{3}\ _{v}\widehat{\Im }[\iota ][\eta _{4^{\prime }}[\ _{4}\iota ]%
\overline{A}]^{\diamond }}{\ _{v}\widehat{\Im }[\iota ]\ [\eta _{4^{\prime
}}[\ _{4}\iota ]\overline{A}]^{\diamond }}dx^{i^{\prime }},\mathbf{e}%
^{4^{\prime }}=dt,\   \label{ofindtmga}
\end{eqnarray}%
where $\eta _{4^{\prime }}(\tau )=\eta _{4^{\prime }}(\tau ,x^{k^{\prime
}},y^{3^{\prime }})=\eta _{4}[\ _{4}\iota ]$ is a generating function and $\
_{v}\widehat{\Im }(\tau )=\ \ _{v}\widehat{\Im }[\iota ]$ is a flow
generating source as in (\ref{statsingpfqp12}) and $\psi (\tau ,x^{k^{\prime
}})$ is a solution of a 2-d Poisson equation (\ref{offdstat}).

\subsubsection{Small parametric modifications of BHs and effective entropic
flow sources}

We study models of geometric and courve flows for nonholonomic distributions
describing $\varepsilon $-deformations described by formulas (\ref{smpolariz}%
). Such deformations of a prime Kerr metric (\ref{dkerr}) with $\overline{A}%
(r,\theta )\rightarrow $ $\widehat{A}[x^{i}(r,\theta ),y^{3}]$ for $%
\mathring{g}_{4}^{\diamond }\neq 0$ result in stationary target metrics of
type (\ref{targm}). The corresponding quadratic line elements are written in
the form%
\begin{eqnarray}
ds^{2} &=&[1+\varepsilon e^{\ ^{0}\psi }\frac{\ ^{1}\psi }{\mathring{g}_{i}\
}\ _{h}^{0}\widehat{\Im }]\mathring{g}_{i}(dx^{i})^{2}+\left[ 1+\varepsilon
\ (2\frac{[\upsilon \mathring{g}_{4}]^{\diamond }}{\mathring{g}%
_{4}^{\diamond }}-\frac{\int dy^{3}\ _{v}\widehat{\Im }([\upsilon \mathring{g%
}_{4}])^{\diamond }}{\int dy^{3}\ _{v}\widehat{\Im }\mathring{g}%
_{4}^{\diamond }})\right] \mathring{g}_{3}(\mathbf{e}^{3})^{2}+[1+%
\varepsilon \upsilon ]\mathring{g}_{4}(\mathbf{e}^{4})^{2},  \notag \\
\mathbf{e}^{3} &=&dy^{3}+[1+\varepsilon (\frac{\partial _{i}\ \int dy^{3}\
_{v}\widehat{\Im }(\upsilon \mathring{g}_{4})^{\diamond }}{\partial _{i}\
\int dy^{3}\ _{v}\widehat{\Im }\mathring{g}_{4}^{\diamond }}-\frac{(\upsilon
\mathring{g}_{4})^{\diamond }}{\mathring{g}_{4}^{\diamond }})]\ \mathring{w}%
_{i}dx^{i},\mathbf{e}^{4}=dx^{4}=dt,  \label{smalparstat}
\end{eqnarray}%
where $\ ^{0}\psi (\tau )=\ ^{0}\psi (\tau ,x^{k})$ and $\ ^{1}\psi (\tau
)=\ ^{1}\psi (\tau ,x^{k})$ are solutions of 2-d Poisson equations with a
generating h-source $\ _{h}\widehat{\Im }(\tau )=\ _{h}\widehat{\Im }(\tau
,x^{k})=\ _{h}^{0}\widehat{\Im }(\tau ,x^{k})+\varepsilon \ _{h}^{1}\widehat{%
\Im }(\tau ,x^{k})$ as described in section \ref{assedef}. In this formula, $%
\ _{v}\widehat{\Im }(\tau )=\ \ _{v}\widehat{\Im }[\iota ]$ is a generating
v-source for which a $\varepsilon $-decomposition is possible and $\upsilon
_{4}=\upsilon (\tau )=\upsilon (\tau ,x^{k},y^{3})=\upsilon \lbrack \
_{4}\iota ]$ is a generating function. The formula (\ref{smalparstat}) is
for a N-adapted system of references and space coordinates $[x^{i}(r,\theta
,\varphi ),y^{3}(r,\theta ,\varphi )]$ for which the condition $(\mathring{g}%
_{4}^{\diamond })^{2}=\mathring{g}_{3}|\int dy^{3}\ _{v}\widehat{\Im }%
\mathring{g}_{4}^{\diamond }|$ allows to compute $\mathring{g}_{3}$ and $%
\mathring{w}_{i}=\partial _{i}\ [dy^{3}\ \ _{v}\widehat{\Im }\mathring{g}%
_{4}^{\diamond }]/\ _{v}\widehat{\Im }\mathring{g}_{4}^{\diamond }$ when
there are prescribed some $\ _{v}\widehat{\Im }$ and $\mathring{g}%
_{4}^{\diamond }\neq 0.$ For simplicity, we fix the conditions $%
_{1}n_{k}(\tau )=0$ and $_{2}n_{k}(\tau )=0$ for which $N_{i}^{4}=n_{i}=0$
but a non-zero $N_{i}^{3}=w_{i}(\varepsilon ,\tau ,x^{k},y^{3})$ results in
trivial nonholonomic torsion and anholonomy coefficients. We can impose
additional constraints on $\upsilon (\tau )$ and sources which allow us to
extract LC--configurations as described in footnote \ref{lcconda}.

Using nonlinear symmetries of type (\ref{nsym1b}) with $\eta _{4}(\tau )=-\
\Phi ^{2}[\ _{4}\iota ]/4\ \Lambda (\tau )\mathring{g}_{4}$ for $\ $(\ref%
{smalparstat}), we can use as a generating function determined by solitonic
hierarchies the value
\begin{equation}
\varepsilon \upsilon \lbrack \ _{4}\iota ]=-\left( 1+\Phi ^{2}[\ _{4}\iota
]/4\ \Lambda (\tau )\overline{A}\right) \mbox{ or }\Phi \lbrack \ _{4}\iota
]\simeq 2\sqrt{|\Lambda (\tau )\overline{A}|}(1-\frac{\varepsilon }{2}%
\upsilon \lbrack \ _{4}\iota ]).  \label{infgenerf}
\end{equation}%
Other types solitonic hierarchies can be encoded into generated sources $\
_{h}\widehat{\Im }(\tau ,x^{k})=\ _{h}^{0}\widehat{\Im }(\tau
,x^{k})+\varepsilon \ _{h}^{1}\widehat{\Im }(\tau ,x^{k})$ and $\ _{v}%
\widehat{\Im }(\tau )=\ \ _{v}\widehat{\Im }[\iota ].$ The geometric
solitonic data $(\Lambda (\tau ),\upsilon \lbrack \ _{4}\iota ],\ _{h}^{0}%
\widehat{\Im }+\varepsilon \ _{h}^{1}\widehat{\Im },_{v}\widehat{\Im }[\iota
])$ determine this class of parametric solutions for geometric flows and
respective Perelman's thermodynamic for GIFs and QGIFs.

\section{Computing Perelman's thermodynamic values for stationary geometric
flows and QGIFs}

\label{s6}

We show how G. Pereman's W-entropy and related thermodynamic values can be
computed for nonholonomic Einstein systems, NES, describing stationary
solitonic and nonholonomically deformed black hole, BH, solutions under
geometric flow evolution. There are provided formulas for extensions of main
concepts and physical values to GIFs and QGIFs with entanglement elaborated
in sections \ref{s3}-\ref{s5}. In this work, there are studied only
stationary configurations (see \cite{vacaru19e} as a partner work on QGIFs
and applications in modern cosmology and \cite{ruchin13,rajpoot17} for
locally anisotropic cosmological solutions and related geometric
thermodynamic models). Similar constructions can be performed for any type
of stationary exact and parametric solitonic solutions parameterized in
Table 2 and/or (in nonholonomic geometric dual form) for solutions with
Killing symmetry on a time like vector.

\subsection{Fixed values for normalization and integration functions}

To study geometric evolution of NES when $\widehat{\mathbf{R}}_{\alpha \beta }=%
\widehat{\mathbf{\Upsilon }}_{\alpha \beta }$ and $\ \ _{s}\widehat{R}=%
\widehat{\mathbf{\Upsilon }}_{a}^{a},$ we can chose a constant value for the
normalizing function, $\ \widehat{f}(\tau )=\ \widehat{f}_{0}=const=0,$ in (\ref{normalizfunct}). This prescribes a geometric vertical scale for flow
evolution determined by data $\left( \Phi (\tau ),\ \Lambda (\tau )\right) $
of such physical models and related via nonlinear symmetries (\ref{nsym1a})
to a generating source $\ _{v}\widehat{\Im }(\tau )$ (such a h-scale is
determined by a 2-d Poisson equation as described in section \ref{assedef}).
Fixing additionally certain constants for integration functions, we can
simplify substantially the formulas for G.\ Perelman's thermodynamic values.%
\footnote{Computing such values in a convenient system of reference/coordinates, we
can consider changing to any system of reference and curved (co)tangent
Lorentz manifolds and other type normalizations for their geometric
evolution.} In result, the formulas for the F- and W-functionals in
canonical geometric variables (see respectively (\ref{ffcand})) are written
\begin{equation}
\widehat{\mathcal{F}}=\frac{1}{8\pi ^{2}}\int \tau ^{-2}\sqrt{|\mathbf{g}%
[\Phi (\tau )]|}\delta ^{4}u [\ _{h}\Lambda (\tau )+\Lambda (\tau )],\
\widehat{\mathcal{W}}=\frac{1}{4\pi ^{2}}\int \tau ^{-2}\sqrt{|\mathbf{g}%
[\Phi (\tau)]|}\delta ^{4}u (\tau \left[\ _{h}\Lambda (\tau)+\Lambda (\tau)%
\right] ^{2}-1),  \label{wn}
\end{equation}%
where $\sqrt{|\mathbf{g}[\Phi (\tau )\mathbf{]}|}=\sqrt{|q_{1}q_{2}\mathbf{q}%
_{3}(_{q}N)|}=2e^{\ \psi (\tau )}\left\vert \Phi (\tau )\right\vert \sqrt{%
\frac{\left\vert [\Phi ^{2}(\tau )]^{\diamond }\right\vert }{|\Lambda (\tau
)\int dy^{3}\ _{v}\widehat{\Im }(\tau )[\Phi ^{2}(\tau )]^{\diamond }|\ }}$
is computed for d-metrics parameterized in the form (\ref{decomp31}) with
\begin{equation*}
q_{1}(\tau )=q_{2}(\tau )=e^{\ \psi (\tau )},\mathbf{q}_{3}(\tau )=-\frac{%
4[\Phi ^{2}(\tau )]^{\diamond }}{|\int dy^{3}\ _{v}\widehat{\Im }(\tau
)[\Phi ^{2}(\tau )]^{\diamond }|\ },[\ _{q}N(\tau )]^{2}=h_{4}(\tau
,x^{k},y^{3})=-\frac{\Phi ^{2}(\tau )}{4\Lambda (\tau )}
\end{equation*}%
for $h_{4}^{[0]}=0.$ The N-adapted differential $\delta ^{4}u=dx^{1}dx^{2}%
\mathbf{e}^{3}\mathbf{e}^{4}=dx^{1}dx^{2}[dy^{3}+w_{i}(\tau
)dx^{i}][dt+n_{i}(\tau )dx^{i}]$ is taken for respective values of
N-connection coefficients when
 $N_{i}^{a}=[w_{i}(\tau )=\frac{\partial _{i}\left( \int dy^{3}\ _{v}\widehat{%
\Im }(\tau )[\Phi ^{2}(\tau )]^{\diamond }\right) }{\ _{v}\widehat{\Im }%
(\tau )[\Phi ^{2}(\tau )]^{\diamond }},n_{i}(\tau )=0]$
for fixed integration functions $_{1}n_{k}(\tau )=0$ and $_{2}n_{k}(\tau
)=0. $

The thermodynamic generating function (\ref{genfcanv}) corresponding to $%
\widehat{\mathcal{W}}$ (\ref{wn}) and fixed $\ \widehat{f}$--normalization
is
\begin{equation}
\widehat{\mathcal{Z}}[\mathbf{g}(\tau )]=\frac{1}{4\pi ^{2}}\int \tau ^{-2}d%
\mathcal{V}(\tau ),  \label{genfn}
\end{equation}%
where the effective integration volume functional $d\mathcal{V}(\tau )=d%
\mathcal{V(}\psi (\tau ),\Phi (\tau ),\ _{v}\widehat{\Im }(\tau ),\Lambda
(\tau )),$%
\begin{equation}
d\mathcal{V}(\tau )=e^{\ \psi (\tau )}\left\vert \Phi (\tau )\right\vert
\sqrt{\frac{\left\vert [\Phi ^{2}(\tau )]^{\diamond }\right\vert }{|\Lambda
(\tau )\int dy^{3}\ _{v}\widehat{\Im }(\tau )[\Phi ^{2}(\tau )]^{\diamond
}|\ }}dx^{1}dx^{2}\left[ dy^{3}+\frac{\partial _{i}\left( \int dy^{3}\ _{v}%
\widehat{\Im }(\tau )[\Phi ^{2}(\tau )]^{\diamond }\right) }{\ _{v}\widehat{%
\Im }(\tau )\ [\Phi ^{2}(\tau )]^{\diamond }}dx^{i}\right] dt  \label{eiv}
\end{equation}%
is completely determined by data $(\psi (\tau ),\Phi (\tau ),\ _{v}\widehat{%
\Im }(\tau ),\Lambda (\tau )).$ These formulas allow us to compute analogous
thermodynamic values for stationary configurations,%
\begin{equation}
\widehat{\mathcal{E}}\ (\tau )=-\frac{\tau ^{2}}{4\pi ^{2}}\int \left( \left[
\ _{h}\Lambda (\tau )+\Lambda (\tau )\right] -\frac{2}{\tau }\right) \tau
^{-2}d\mathcal{V}(\tau ),\ \widehat{\mathcal{S}}(\tau )\ =-\frac{1}{4\pi ^{2}%
}\int \left( \tau \left[\ _{h}\Lambda (\tau )+\Lambda (\tau )\right]
-2\right) \tau ^{-2}d\mathcal{V}(\tau ).  \label{thvcann}
\end{equation}%
In this work, we omit and do not provide applications of cumbersome formulas
for computing flow fluctuations $\widehat{\eta }$ (\ref{thvcanon}).

Using geometric thermodynamic values (\ref{thvcann}) for $\widehat{f}_{0}=0,$
we can compute in canonical variables the respective free energy and
relative entropy (\ref{telativenerg}). If two d-metrics are defined by
different classes of solutions defined by respective geometric data $\ _{1}%
\mathbf{g}(\ _{1}\psi ,\ _{1}\Phi ,\ _{v}^{1}\widehat{\Im },\ _{1}\Lambda )$
and $\mathbf{g}(\psi ,\Phi ,\ _{v}\widehat{\Im },\Lambda ),$ we obtain
\begin{eqnarray}
\widehat{\mathcal{F}\ }(\ _{1}\psi ,\ _{1}\Phi ,\ _{v}^{1}\widehat{\Im },\
_{1}\Lambda ) &=&\widehat{\mathcal{S}}(\ _{1}\psi ,\ _{1}\Phi ,\ _{v}^{1}%
\widehat{\Im },\ _{1}\Lambda )-\beta ^{-1}\widehat{\mathcal{S}}(\ _{1}\psi
,\ _{1}\Phi ,\ _{v}^{1}\widehat{\Im },\ _{1}\Lambda )\mbox{ and }
\label{relatn} \\
\widehat{\mathcal{S}}(\ \ _{1}\psi ,\ _{1}\Phi ,\ _{v}^{1}\widehat{\Im },\
_{1}\Lambda )\shortparallel \psi ,\Phi ,\ _{v}\widehat{\Im },\Lambda )
&=&\beta \lbrack \widehat{\mathcal{F}}((\ _{1}\psi ,\ _{1}\Phi ,\ _{v}^{1}%
\widehat{\Im },\ _{1}\Lambda ))-\widehat{\mathcal{F}}(\psi ,\Phi ,\ _{v}%
\widehat{\Im },\Lambda )],  \notag \\
\mbox{ where }\widehat{\mathcal{E}}(\ _{1}\psi ,\ _{1}\Phi ,\ _{v}^{1}%
\widehat{\Im },\ _{1}\Lambda ) &=&-\frac{\tau ^{2}}{4\pi ^{2}}\int \left( %
\left[ \ _{h}^{1}\Lambda (\tau )+\ _{1}\Lambda (\tau )\right] -\frac{2}{\tau
}\right) \tau ^{-2}d\ _{1}\mathcal{V}(\tau ),  \notag \\
\widehat{\mathcal{S}}(\ _{1}\psi ,\ _{1}\Phi ,\ _{v}^{1}\widehat{\Im },\
_{1}\Lambda ) &=&-\frac{1}{4\pi ^{2}}\int \left( \tau \left[ \
_{h}^{1}\Lambda (\tau )+\ _{1}\Lambda (\tau )\right] -2\right) \tau ^{-2}d\
_{1}\mathcal{V}(\tau )  \notag
\end{eqnarray}%
for 
$d\ _{1}\mathcal{V}(\tau )=e^{\ _{1}\psi (\tau )}\left\vert \ _{1}\Phi (\tau
)\right\vert \sqrt{\frac{\left\vert [\ _{1}\Phi ^{2}(\tau )]^{\diamond
}\right\vert }{|\ _{1}\Lambda (\tau )\int dy^{3}\ _{v}^{1}\widehat{\Im }%
(\tau )[\ _{1}\Phi ^{2}(\tau )]^{\diamond }|\ }}dx^{1}dx^{2}[dy^{3}+\frac{%
\partial _{i}\left( \int dy^{3}\ _{v}^{1}\widehat{\Im }(\tau )[\ _{1}\Phi
^{2}(\tau )]^{\diamond }\right) }{\ _{v}^{1}\widehat{\Im }(\tau )\ [\
_{1}\Phi ^{2}(\tau )]^{\diamond }}dx^{i}]dt$,  Similar values for $%
\mathbf{g}(\psi ,\Phi ,\ _{v}\widehat{\Im },\Lambda )$ are given by formulas
(\ref{eiv}) and (\ref{thvcann}). Here we emphasize that the free energy and
relative entropy values (\ref{relatn}) can be defined and computed for the
same class of NES but subjected to different types geometric flows.

A density state (see definition in footnote \ref{fndensst}) is a functional $%
\widehat{\rho }[\mathbf{g}(\tau )]=\widehat{\mathcal{Z}}^{-1}(\tau
)e^{-\beta E}$ where $\widehat{\mathcal{Z}}$ for stationary solitonic
solutions are computed following formula (\ref{genfn}) (we use equivalently
two symbols $\rho $ and/or $\sigma$). In QGIF theory, there are considered
also the geometric evolution densities $\widehat{\rho }[\ _{1}\mathbf{g}]$
and $\widehat{\rho }^{\prime }[\ _{1}\mathbf{g}],$ where the left label 1 is
used in order to distinguish two d-metrics $\mathbf{g}$ and $\ _{1}\mathbf{g}
$ which in this work may define two different geometric flows NES (in
principle, such systems can be similar ones). Conventionally, we consider
two stationary configurations for GIFs and NES systems, $\widehat{A}(\tau )=%
\widehat{A}(\psi (\tau ),\Phi (\tau ),\ _{v}\widehat{\Im }(\tau ),\Lambda
(\tau ))$ and $\widehat{B}(\tau )=\widehat{B}(\ _{1}\psi (\tau ),\ _{1}\Phi
(\tau ),\ _{v}^{1}\widehat{\Im }(\tau ),\ _{1}\Lambda (\tau )).$ Such two
systems thermodynamic geometric flow models are elaborated on $\mathbf{V}%
\otimes \mathbf{V}$ when the normalizing function is fixed $\ _{AB}\widehat{f%
}(u,\ _{1}u)=0.$ The respective generating function (\ref{twogenf}) and
entropy (\ref{twoentr}) are computed
\begin{eqnarray}
\ _{AB}\widehat{\mathcal{Z}}(\tau ) &=&\frac{1}{16\pi ^{4}}\int \ _{1}\int
\tau ^{-4}d\mathcal{V}(\tau )d\ _{1}\mathcal{V}(\tau ),\mbox{ for }\mathbf{%
V\otimes \mathbf{\mathbf{V}}}  \label{twogenfn} \\
\ _{AB}\widehat{\mathcal{S}}(\tau ) &=&\widehat{\mathcal{S}}\ [\widehat{A},%
\widehat{B}] =-\frac{1}{16\pi ^{4}}\int \left( \tau \left[\ _{h}\Lambda
(\tau )+\Lambda (\tau )\right]-2\right) \left(\tau \lbrack \ _{h}^{1}\Lambda
(\tau )+\ _{1}\Lambda (\tau )]-2\right) \tau ^{-4} d\mathcal{V}(\tau )d\ _{1}%
\mathcal{V}(\tau ).  \label{twoentrn}
\end{eqnarray}

In similar forms, the three partite thermodynamic generation function and
entropy (see formulas (\ref{threegenf}) and entropy (\ref{threeentr})) of
GIF and NES stationary systems $\widehat{A}(\tau )=\widehat{A}(\psi (\tau
),\Phi (\tau ),\ _{v}\widehat{\Im }(\tau ),\Lambda (\tau )),$\newline
$\widehat{B}(\tau )=\widehat{B}(\ _{1}\psi (\tau ),\ _{1}\Phi (\tau ),\
_{v}^{1}\widehat{\Im }(\tau ),\ _{1}\Lambda (\tau ))$ and $\widehat{C}(\tau
)=\widehat{C}(\ _{2}\psi (\tau ),\ _{2}\Phi (\tau ),\ _{v}^{2}\widehat{\Im }%
(\tau ),\ _{2}\Lambda (\tau ))$ are considered for a fixed normalizing function
$\ _{ABC}\widehat{f}(u,\ _{1}u,\ _{2}u)=0.$ They are characterized by
respective formulas,
\begin{eqnarray}
\ _{ABC}\widehat{\mathcal{Z}}(\tau ) &=&\frac{1}{64\pi ^{6}}\int \ \ \
_{1}\int \tau ^{-4}d\mathcal{V}(\tau )d\ _{1}\mathcal{V}(\tau )d\ _{2}%
\mathcal{V}(\tau ),\mbox{ for }\mathbf{V\otimes \mathbf{\mathbf{V}}\otimes
\mathbf{\mathbf{V}}}  \label{threeegenfn} \\
\ _{ABC}\widehat{\mathcal{S}}(\tau ) &=&\widehat{\mathcal{S}}\ [\widehat{A},%
\widehat{B},\widehat{C}]\ =-\frac{1}{64\pi ^{6}}\int \left( \tau \left[\
_{h}\Lambda (\tau )+\Lambda (\tau )\right] -2\right) \left( \tau \lbrack \
_{h}^{1}\Lambda (\tau )+\ _{1}\Lambda (\tau )] -2\right)  \notag \\
&&\left(\tau \lbrack \ _{h}^{2}\Lambda (\tau )+\ _{2}\Lambda (\tau )]
-2\right) \tau ^{-6}d\mathcal{V}(\tau )d\ _{1}\mathcal{V}(\tau )d\ _{2}%
\mathcal{V}(\tau ).  \label{threeentrn}
\end{eqnarray}

Using formulas (\ref{twogenfn}), (\ref{twoentrn}) and (\ref{threeegenfn}), (%
\ref{threeentrn}) we can elaborate on GIF models for stationary d-metrics.
For such a solution $\widehat{A}(\tau )=\widehat{A}[\mathbf{g}(\psi (\tau
),\Phi (\tau ),\ _{v}\widehat{\Im }(\tau ),\Lambda (\tau ))],$ we can
associate a quantum system $\mathcal{A}(\tau )$ when the density matrix%
\begin{equation}
\widehat{\rho }_{\mathcal{A}}(\tau ):=\widehat{\mathcal{Z}}^{-1}(\tau
)e^{-\tau ^{-1}\widehat{\mathcal{E}}\ }  \label{statdmstat}
\end{equation}%
is determined by values $\widehat{\mathcal{Z}}(\psi (\tau ),\Phi (\tau ),\
_{v}\widehat{\Im }(\tau ),\Lambda (\tau ))$ (\ref{genfn}) and $\widehat{%
\mathcal{E}}(\psi (\tau ),\Phi (\tau ),\ _{v}\widehat{\Im }(\tau ),\Lambda
(\tau ))$ (\ref{thvcann}). In result, we can compute the entanglement
entropy (\ref{entangentr}) for stationary configurations
\begin{equation}
\ _{q}\widehat{\mathcal{S}}[\ \widehat{\rho }_{\mathcal{A}}(\psi (\tau
),\Phi (\tau ),\ _{v}\widehat{\Im }(\tau ),\Lambda (\tau ))]:=Tr[\ \widehat{%
\rho }_{\mathcal{A}}\ (\psi (\tau ),\Phi (\tau ),\ _{v}\widehat{\Im }(\tau
),\Lambda (\tau ))\log \widehat{\rho }_{\mathcal{A}}(\psi (\tau ),\Phi (\tau
),\ _{v}\widehat{\Im }(\tau ),\Lambda (\tau ))],  \label{entangentrn}
\end{equation}%
when $\widehat{\rho }_{\mathcal{A}}$ is computed using formulas (\ref{statdmstat}). This entanglement entropy is a QGIF version of the G. Perelman thermodynamic entropy $\widehat{\mathcal{S}}(\tau )$ (\ref{thvcann}).

For stationary QGIF systems, we can compute the R\'{e}nyi entropy (\ref%
{renentr}) using the replica method with $\ \widehat{\rho }_{\mathcal{A}%
}(\psi (\tau ),\Phi (\tau ),\ _{v}\widehat{\Im }(\tau ),\Lambda (\tau ))$ (%
\ref{statdmstat}). Considering an integer replica parameter $r,$ the R\'{e}%
nyi entropy for the mentioned class of stationary flow configurations with
3+1 splitting (\ref{decomp31}),
\begin{equation}
\ _{r}\ \widehat{\mathcal{S}}_{\mathcal{A}}(\tau )=\ _{r}\ \widehat{\mathcal{%
S}}(\widehat{\mathcal{A}}):=\frac{1}{1-r}\log [tr_{\mathcal{A}}(\widehat{%
\rho }_{\mathcal{A}}(\tau ))^{r}]  \label{renyistat}
\end{equation}%
for stationary solitonic QGIF\ system determined by the matrix $\ \widehat{%
\rho }_{\mathcal{A}}$ $(\tau )$ (\ref{statdmstat}) and associated
thermodynamic model $\widehat{\mathcal{A}}(\tau )=\left[ \widehat{\mathcal{Z}%
}(\tau ),\ \widehat{\mathcal{E}}(\tau ),\widehat{\mathcal{S}}(\tau )\right]
. $ Applying a standard computational formalism elaborated for an analytic
continuation of $r$ to a real number with a well defined limit $\ _{q}%
\widehat{\mathcal{S}}(\widehat{\rho }_{\mathcal{A}}(\tau))=\lim_{r%
\rightarrow 1}\ \ _{r}\ \widehat{\mathcal{S}}(\widehat{\mathcal{A}}(\tau ))$
and normalization $tr_{\mathcal{A}}(\widehat{\rho }_{\mathcal{A}})$ for $%
r\rightarrow 1.$ For such limits and stationary solitonic flow d-metrics,
the R\'{e}nyi entropy reduces to the entanglement entropy (\ref{entangentrn}%
). In result, we can formulate QGIFs models from section \ref{s3} for
stationary configurations.

\subsection{Thermodynamic values for stationary
solitonic generating functions and generating sources}

Such values are computed for a 3+1 spitting (\ref{decomp31}) determined by
stationary solitonic d-metric (\ref{stasdmnf}) (for LC-configurations, we
can consider (\ref{stasdmnflc})), when
\begin{eqnarray*}
q_{1}&=&q_{2}[\ _{h}i]=e^{\ \psi \lbrack \ _{h}i]},\mathbf{q}_{3}[\ _{h}i]=-%
\frac{4[(\Phi \lbrack \ _{4}i])^{2}]^{\diamond }}{|\int dy^{3}\ _{v}\widehat{%
\Im }[i][(\Phi \lbrack \ _{4}i])^{2}]^{\diamond }|\ },\ \lbrack \ _{q}N(\tau
)]^{2} = h_{4}[\ _{4}i]=-\frac{(\Phi \lbrack \ _{4}i])^{2}}{4\Lambda (\tau )}
\\
\mbox{ and }N_{i}^{a} &=&[w_{i}(\tau )=\frac{\partial _{i}\left( \int
dy^{3}\ _{v}\widehat{\Im }[i][\Phi ^{2}[\ _{4}i]]^{\diamond }\right) }{\ _{v}%
\widehat{\Im }[i][\Phi ^{2}[\ _{4}i]]^{\diamond }},n_{i}(\tau )=0].
\end{eqnarray*}%
Using these coefficients and prescribing solitonic hierarchies for the
effective volume (\ref{eiv}), we obtain%
\begin{eqnarray*}
d\mathcal{V}[\ _{h}i,\ _{4}i,i] &=&e^{\ \psi \lbrack \ _{h}i]}\left\vert
\Phi \lbrack \ _{4}i]\right\vert \sqrt{\frac{\left\vert [\Phi ^{2}[\
_{4}i]]^{\diamond }\right\vert }{|\Lambda (\tau )\int dy^{3}\ _{v}\widehat{%
\Im }[i][\Phi ^{2}[\ _{4}i]]^{\diamond }|\ }}dx^{1}dx^{2} \\
&&\left[ dy^{3}+\frac{\partial _{i}\left( \int dy^{3}\ _{v}\widehat{\Im }%
[i][\Phi ^{2}[\ _{4}i]]^{\diamond }\right) }{\ _{v}\widehat{\Im }[i]\ [\Phi
^{2}[\ _{4}i]]^{\diamond }}dx^{i}\right] dt
\end{eqnarray*}
and respective thermodynamic generating function (\ref{genfn}) $\widehat{%
\mathcal{Z}}[\ _{h}i,\ _{4}i,i]=\frac{1}{4\pi ^{2}}\int \tau ^{-2}d\mathcal{V%
}[\ _{h}i,\ _{4}i,i].$

The value $\widehat{\mathcal{Z}}[\ _{h}i,\ _{4}i,i]$ determine the
thermodynamic values (\ref{thvcann}) for geometric flows of such solitonic
hierarchies,
\begin{eqnarray*}
\widehat{\mathcal{E}}\ [\ _{h}i,\ _{4}i,i] &=&-\frac{\tau ^{2}}{4\pi ^{2}}%
\int \left( \left[ \ _{h}\Lambda (\tau )+\Lambda (\tau )\right] -\frac{2}{%
\tau }\right) \tau ^{-2}d\mathcal{V}[\ _{h}i,\ _{4}i,i], \\
\ \widehat{\mathcal{S}}[\ _{h}i,\ _{4}i,i]\ &=&-\frac{1}{4\pi ^{2}}\int
\left( \tau \left[ \ _{h}\Lambda (\tau )+\Lambda (\tau )\right] -2\right)
\tau ^{-2}d\mathcal{V}[\ _{h}i,\ _{4}i,i].
\end{eqnarray*}%
In a similar form, using and $d\mathcal{V}[\ _{h}i,\ _{4}i,i],$ we compute
two partite and three partite thermodynamic values (\ref{twogenfn}), (\ref%
{twoentrn}) and (\ref{threeegenfn}), (\ref{threeentrn}).

To any GIF solitonic stationary system $\widehat{A}[\ _{h}i,\ _{4}i,i]=%
\widehat{A}[\mathbf{g}(\psi \lbrack \ _{h}i],\Phi \lbrack \ _{4}i],\ _{v}%
\widehat{\Im }[i],\Lambda (\tau ))],$ we can associate a quantum system $%
\mathcal{A}[\ _{h}i,\ _{4}i,i]$ when the density matrix $\widehat{\rho }_{%
\mathcal{A}}[\ _{h}i,\ _{4}i,i]:=\widehat{\mathcal{Z}}^{-1}[\ _{h}i,\
_{4}i,i]e^{-\tau ^{-1}\widehat{\mathcal{E}}[\ _{h}i,\ _{4}i,i]\ }$ allows us
to compute the entanglement entropy (\ref{entangentr}) for stationary
solitonic configurations
\begin{equation*}
\ _{q}\widehat{\mathcal{S}}[\ \widehat{\rho }_{\mathcal{A}}(\psi \lbrack \
_{h}i],\Phi \lbrack \ _{4}i],\ _{v}\widehat{\Im }[i],\Lambda (\tau ))]:=Tr[\
\widehat{\rho }_{\mathcal{A}}(\psi \lbrack \ _{h}i],\Phi \lbrack \ _{4}i],\
_{v}\widehat{\Im }[i],\Lambda (\tau ))\log \widehat{\rho }_{\mathcal{A}%
}(\psi \lbrack \ _{h}i],\Phi \lbrack \ _{4}i],\ _{v}\widehat{\Im }%
[i],\Lambda (\tau ))].
\end{equation*}%
For such stationary solitonic QGIF systems, the R\'{e}nyi entropy (\ref%
{renentr}) is computed using the replica method for the density matric $%
\widehat{\rho }_{\mathcal{A}}[\ _{h}i,\ _{4}i,i],$
\begin{equation*}
\ _{r}\ \widehat{\mathcal{S}}_{\mathcal{A}}[\ _{h}i,\ _{4}i,i]=\ _{r}\
\widehat{\mathcal{S}}(\widehat{\mathcal{A}}[\ _{h}i,\ _{4}i,i]):=\frac{1}{1-r%
}\log [tr_{\mathcal{A}}(\widehat{\rho }_{\mathcal{A}}([\ _{h}i,\
_{4}i,i]))^{r}],
\end{equation*}%
characterizing an associated thermodynamic model $\widehat{\mathcal{A}}[\
_{h}i,\ _{4}i,i])=\left[ \widehat{\mathcal{Z}}[\ _{h}i,\ _{4}i,i]),\
\widehat{\mathcal{E}}[\ _{h}i,\ _{4}i,i],\widehat{\mathcal{S}}[\ _{h}i,\
_{4}i,i]\right] .$

\subsection{Perelman's thermodynamics for BHs deformed by stationary
solitonic hierarchies}

Such GIF and QGIF systems encode stationary solitonic d-metrics (\ref%
{statsingpfqp12a}), or LC-configurations (\ref{ofindtmga}), when the
d-metric $\mathbf{\mathring{g}}=[\mathring{g}_{i},\mathring{g}_{a},\mathring{%
N}_{b}^{j}]$ (\ref{primedm}) defines a Kerr BH solution (\ref{dkerr}). The
solitonic generating function (\ref{nsym1b}) can be written in the form $\
\Phi \lbrack \ _{4}\iota ]=2\sqrt{|\Lambda (\tau )\ \eta _{4}[\ _{4}\iota ]%
\overline{A}|}$ which results in such an effective volume form%
\begin{equation*}
d\mathcal{V}[\ _{h}i,\ \eta _{4}[\ _{4}\iota ],i,\overline{A}]=2e^{\ \psi
\lbrack \ _{h}i]}\sqrt{\frac{\left\vert \eta _{4}[\ _{4}\iota ]\overline{A}%
|\ \eta _{4}[\ _{4}\iota ]|^{\diamond }\right\vert }{|\int dy^{3}\ _{v}%
\widehat{\Im }[i]|\ \eta _{4}[\ _{4}\iota ]|^{\diamond }|\ }}dx^{1}dx^{2}%
\left[ dy^{3}+\frac{\partial _{i}\left( \int dy^{3}\ _{v}\widehat{\Im }[i]|\
\eta _{4}[\ _{4}\iota ]|^{\diamond }\right) }{\ _{v}\widehat{\Im }[i]\ |\
\eta _{4}[\ _{4}\iota ]|^{\diamond }}dx^{i}\right] dt
\end{equation*}%
and corresponding GIF thermodynamic generating function (\ref{genfn})
\begin{equation*}
\widehat{\mathcal{Z}}[\ _{h}i,\ \eta _{4}[\ _{4}\iota ],i,\overline{A}]=%
\frac{1}{4\pi ^{2}}\int \tau ^{-2}d\mathcal{V}[\ _{h}i,\ \eta _{4}[\
_{4}\iota ],i,\overline{A}].
\end{equation*}

Both the primary BH and solitonic data are encoded also in the thermodynamic
values (\ref{thvcann}),
\begin{eqnarray*}
\widehat{\mathcal{E}}[\ _{h}i,\ \eta _{4}[\ _{4}\iota ],i,\overline{A}] &=&-%
\frac{\tau ^{2}}{4\pi ^{2}}\int \left( \left[ \ _{h}\Lambda (\tau )+\Lambda
(\tau )\right] -\frac{2}{\tau }\right) \tau ^{-2}d\mathcal{V}[\ _{h}i,\ \eta
_{4}[\ _{4}\iota ],i,\overline{A}], \\
\ \widehat{\mathcal{S}}[\ _{h}i,\ \eta _{4}[\ _{4}\iota ],i,\overline{A}]
&=&-\frac{1}{4\pi ^{2}}\int \left( \tau \left[ \ _{h}\Lambda (\tau )+\Lambda
(\tau )\right] -2\right) \tau ^{-2}d\mathcal{V}[\ _{h}i,\ \eta _{4}[\
_{4}\iota ],i,\overline{A}]
\end{eqnarray*}%
and (in similar forms via $d\mathcal{V}[\ _{h}i,\ \eta _{4}[\ _{4}\iota ],i,%
\overline{A}])$ in two partite and three partite thermodynamic values (\ref%
{twogenfn}), (\ref{twoentrn}) and (\ref{threeegenfn}), (\ref{threeentrn}).

Such a BH deformed GIF solitonic stationary system $\widehat{A}[\ _{h}i,\
\eta _{4}[\ _{4}\iota ],i,\overline{A}]=\widehat{A}[\mathbf{g}(\psi \lbrack
\ _{h}i],\eta _{4}[\ _{4}\iota ],\ _{v}\widehat{\Im }[i],\Lambda (\tau ))]$
can be associated to a quantum system $\mathcal{A}[\ _{h}i,\ \eta _{4}[\
_{4}\iota ],i,\overline{A}]$ characterized by a respective density matrix%
\begin{equation*}
\widehat{\rho }_{\mathcal{A}}[\ _{h}i,\ \eta _{4}[\ _{4}\iota ],i,\overline{A%
}]:=\widehat{\mathcal{Z}}^{-1}[\ _{h}i,\ \eta _{4}[\ _{4}\iota ],i,\overline{%
A}]e^{-\tau ^{-1}\widehat{\mathcal{E}}[\ _{h}i,\ \eta _{4}[\ _{4}\iota ],i,%
\overline{A}]\ }.
\end{equation*}%
This value can be used constructing a QGIF system and computing the
entanglement entropy (\ref{entangentr}) for stationary BH solitonic
deformations
\begin{equation*}
\ _{q}\widehat{\mathcal{S}}[\ \widehat{\rho }_{\mathcal{A}}(\psi \lbrack \
_{h}i],\eta _{4}[\ _{4}\iota ],i],\ _{v}\widehat{\Im }[i],\Lambda (\tau
))]:=Tr[\ \widehat{\rho }_{\mathcal{A}}(\psi \lbrack \ _{h}i],\eta _{4}[\
_{4}\iota ],\ _{v}\widehat{\Im }[i],\Lambda (\tau ))\log \widehat{\rho }_{%
\mathcal{A}}(\psi \lbrack \ _{h}i],\eta _{4}[\ _{4}\iota ],\ _{v}\widehat{%
\Im }[i],\Lambda (\tau ))]
\end{equation*}%
and the R\'{e}nyi entropy (\ref{renentr}), $\ \ _{r}\ \widehat{\mathcal{S}}_{%
\mathcal{A}}[\ _{h}i,\eta _{4}[\ _{4}\iota ],i]=\ _{r}\ \widehat{\mathcal{S}}%
(\widehat{\mathcal{A}}[\ _{h}i,\eta _{4}[\ _{4}\iota ],i]):=\frac{1}{1-r}%
\log [tr_{\mathcal{A}}(\widehat{\rho }_{\mathcal{A}}([\ _{h}i,\eta _{4}[\
_{4}\iota ],i]))^{r}].$

\subsection{Small parametric stationary solitonic BH deformations and
geometric flow thermodynamics}

The d-metrics for such parametric solutions are described by quadratic
elements  (\ref{smalparstat}) \ and generating functions $\Phi \lbrack \
_{4}\iota ,\overline{A}]\simeq 2\sqrt{|\Lambda (\tau )\overline{A}|}(1-\frac{%
\varepsilon }{2}\upsilon \lbrack \ _{4}\iota ])$  (\ref{infgenerf}) and a
primary BH metric  (\ref{dkerr}). Using the respective effective volume form%
\begin{eqnarray*}
d\mathcal{V}[\ _{h}i,\ \varepsilon \upsilon \lbrack \ _{4}\iota ],i,%
\overline{A}] &=&2e^{\ \psi \lbrack \ _{h}i]}\left\vert (1-\frac{\varepsilon
}{2}\upsilon \lbrack \ _{4}\iota ])\right\vert \sqrt{\frac{\left\vert \
\overline{A}|\ \upsilon \lbrack \ _{4}\iota ]|^{\diamond }\right\vert }{%
|\int dy^{3}\ _{v}\widehat{\Im }[i]|\ \upsilon \lbrack \ _{4}\iota
]|^{\diamond }|\ }}dx^{1}dx^{2} \\
&&\left[ dy^{3}+\frac{\partial _{i}\left( \int dy^{3}\ _{v}\widehat{\Im }%
[i]|\ \upsilon \lbrack \ _{4}\iota ]|^{\diamond }\right) }{\ _{v}\widehat{%
\Im }[i]\ |\ \upsilon \lbrack \ _{4}\iota ]|^{\diamond }}dx^{i}\right] dt,
\end{eqnarray*}%
we compute corresponding thermodynamic generating function (\ref{genfn}) and
canonical energy and entropy (\ref{thvcann}) for stationary geometric
solitonic flow parametric deformations
\begin{eqnarray*}
\widehat{\mathcal{Z}}[\ _{h}i,\ \varepsilon \upsilon \lbrack \ _{4}\iota ],i,%
\overline{A}] &=&\frac{1}{4\pi ^{2}}\int \tau ^{-2}d\mathcal{V}[\ _{h}i,\
\varepsilon \upsilon \lbrack \ _{4}\iota ],i,\overline{A}]\mbox{ and } \\
\widehat{\mathcal{E}}\ [\ _{h}i,\ \varepsilon \upsilon \lbrack \ _{4}\iota
],i,\overline{A}] &=&-\frac{\tau ^{2}}{4\pi ^{2}}\int \left( \left[ \
_{h}\Lambda (\tau )+\Lambda (\tau )\right] -\frac{2}{\tau }\right) \tau
^{-2}d\mathcal{V}, \\
\ \widehat{\mathcal{S}}[\ _{h}i,\ \varepsilon \upsilon \lbrack \ _{4}\iota
],i,\overline{A}]\  &=&-\frac{1}{4\pi ^{2}}\int \left( \tau \left[ \
_{h}\Lambda (\tau )+\Lambda (\tau )\right] -2\right) \tau ^{-2}d\mathcal{V}%
[\ _{h}i,\ \varepsilon \upsilon \lbrack \ _{4}\iota ],i,\overline{A}]
\end{eqnarray*}%
and (in similar form but for $d\mathcal{V}[\ _{h}i,\ \varepsilon \upsilon
\lbrack \ _{4}\iota ],i,\overline{A}]$) two partite and three partite
thermodynamic values (\ref{twogenfn}), (\ref{twoentrn}) and (\ref%
{threeegenfn}), (\ref{threeentrn}).

For any  BH parametric deformed GIF solitonic stationary system
$$\widehat{A}%
[\ _{h}i,\ \varepsilon \upsilon \lbrack \ _{4}\iota ],i,\overline{A}]=%
\widehat{A}[\mathbf{g}(\psi \lbrack \ _{h}i],\varepsilon \upsilon \lbrack \
_{4}\iota ],\ _{v}\widehat{\Im }[i],\Lambda (\tau ),\overline{A})],$$ we can
associate to a quantum system $\mathcal{A}[\ _{h}i,\ \varepsilon \upsilon
\lbrack \ _{4}\iota ],i,\overline{A}]$ and compute the density matrix%
\begin{equation*}
\widehat{\rho }_{\mathcal{A}}[\ _{h}i,\ \varepsilon \upsilon \lbrack \
_{4}\iota ],i,\overline{A}]:=\widehat{\mathcal{Z}}^{-1}[\ _{h}i,\
\varepsilon \upsilon \lbrack \ _{4}\iota ],i,\overline{A}]e^{-\tau ^{-1}%
\widehat{\mathcal{E}}[\ _{h}i,\ \varepsilon \upsilon \lbrack \ _{4}\iota ],i,%
\overline{A}]\ }.
\end{equation*}%
A corresponding  QGIF stationary solitonic system with small parameter is
characterized by  the entanglement entropy (\ref{entangentr})
\begin{eqnarray*}
&&\ _{q}\widehat{\mathcal{S}}[\ \widehat{\rho }_{\mathcal{A}}(\psi \lbrack \
_{h}i],\ \varepsilon \upsilon \lbrack \ _{4}\iota ],i],\ _{v}\widehat{\Im }%
[i],\Lambda (\tau )),\overline{A}]:= \\
&& Tr[\ \widehat{\rho }_{\mathcal{A}}(\psi
\lbrack \ _{h}i],\ \varepsilon \upsilon \lbrack \ _{4}\iota ],\ _{v}\widehat{%
\Im }[i],\Lambda (\tau ),\overline{A})\log \widehat{\rho }_{\mathcal{A}%
}(\psi \lbrack \ _{h}i],\ \varepsilon \upsilon \lbrack \ _{4}\iota ],\ _{v}%
\widehat{\Im }[i],\Lambda (\tau ),\overline{A})],
\end{eqnarray*}%
when, for instance, the R\'{e}nyi entropy (\ref{renentr}),
\begin{equation*}
\ \ _{r}\ \widehat{\mathcal{S}}_{\mathcal{A}}[\ _{h}i,\ \varepsilon \upsilon
\lbrack \ _{4}\iota ],i,\overline{A}]=\ _{r}\ \widehat{\mathcal{S}}(\widehat{%
\mathcal{A}}[\ _{h}i,\ \varepsilon \upsilon \lbrack \ _{4}\iota ],i,%
\overline{A}]):=\frac{1}{1-r}\log [tr_{\mathcal{A}}(\widehat{\rho }_{%
\mathcal{A}}([\ _{h}i,\ \varepsilon \upsilon \lbrack \ _{4}\iota ],i,%
\overline{A}]))^{r}].
\end{equation*}%
All formulas and inequalities for entropies of NES QGIFs can be computed for
such small parametric stationary deformations. For instance, prescribing a solitonic wave, all values and equations can be computed in explicit form for such a BH solitonic stationary configuration, see a number of examples in our previous works \cite{vacaru01,vacaru10,anco06,vacaru15,vacaru01a,vacaru02,vacaru09c,rajpoot15,
vacaru06,vacaru07b,vacaru08,bubuianu16,gheorghiu14,vacaru13}.

Finally, we emphasize that it is not possible to define and compute the Bekenstein--Hawking entropy for the exact and parametric stationary solitonic and/or BH solutions constructed in section \ref{s5}. Such geometric flows of NES systems are characterized by respective thermodynamic values, GIF and QGIF models computed in section \ref{s6}.

\section{Outlook, conclusions, and discussion}

\label{sconcl}

This is the third our work on the theory of classical and quantum geometric
information flows (respectively, GIFs and QGIFs), see \cite%
{vacaru19b,vacaru19c} on QGIF of relativistic classical and quantum
mechanical systems. On entanglement and QGIF of Einstein-Maxwell and
Kaluza--Klein gravity theories, we cite the forth partner work \cite%
{vacaru19e}. In a more general context, such papers belong to a series of
articles \cite{vacaru07,vacaru09b,vacaru09,ruchin13,gheorghiu16,rajpoot17,
bubuianu19,vacaru19,vacaru19a} on generalized (relativistic, modified, or noncommutative and
nonassociative, supersymmetric etc.) Ricci flows and applications to modified gravity theories, MGTs, and general relativity, GR. The key idea of this and partner articles is that all such theories and fundamental physical equations,  their symmetries and solutions, and associated thermodynamic and information models  can be derived from certain types of nonholonomically deformed Perelman--Lyapunov type F- and W- functionals \cite{perelman1}. In our research, we do not attempt to use such results for formulating and providing proofs for certain relativistically generalized Thurston--Poincar\'{e} conjectures (as it was performed due to G. Perelman
\cite{perelman1} and R. Hamilton \cite{hamilt1}, see reviews in \cite%
{monogrrf1,monogrrf2,monogrrf3}). The goal of this work is to elaborate on
applications in quantum information theory of the concept W-entropy
(Perelman's W-functional can be treated as a "minus" entropy) and associated
thermodynamic models of geometric flows containing MGTs and GR, as certain
particular nonholonomic Ricci soliton configurations.

In this paper, relativistic versions of Perelman's functionals and
associated thermodynamic models are formulated in canonical nonholonomic
variables (with "hats" on geometric objects). In result, any exact and
parametric solution in gravity theories and geometric flow models\footnote{%
in principle, being generic off-diagonal, with generalized nonlinear and
linear connections, various types of effective and matter field sources, and
depending on all modified spacetime/ phase space coordinates, see various
examples and applications in modern cosmology and astrophysics \cite%
{vacaru11,ruchin13,gheorghiu16,rajpoot17,vacaru19,vacaru18tc,bubuianu18,
bubuianu19,vacaru19a}} can be derived  and characterized thermodynamically considering a respective W-entropy. Such constructions are more general than those for gravitational
thermodynamics and black hole / (anti) de Sitter physics \cite%
{bekenstein72,bekenstein73,bardeen73,hawking75} (with a conventional
hypersurface-area entropy) intensively elaborated during last two decades
with new concepts of entanglement, entropic and holographic gravity \cite%
{preskill,witten18,ryu06,raamsdonk10,bub21,faulkner14,swingle12,jacobson15, taylorm18,pastawski15,casini11,aolita14,nishioka18}.

In the framework of the theory of relativistic geometric flows of metrics on
Lorentz spacetime manifolds, various MGTs and GR are modelled as certain
nonholonomic Einstein structures, NES, running on a temperature like parameter and on a time like parameter. Such theories are described equivalently as some self-similar systems (i.e. nonholonomic Ricci solitons) which are characterized by a corresponding
W-entropy and other type associated relativistic thermodynamic parameters. In this
article, we formulate an approach to the theory of GIFs and QGIFs and NES with a temperature like parameter. We apply and develop for such gravitational flow evolution models and
geometric thermodynamical systems, the standard concepts and methods of
information theory and quantum physics and gravity \cite%
{cover,nielsen,nielsen10,preskill,weedbrook11,hayashi17,watrous18,
witten18,aolita14,nishioka18}. The constructions are generalized for the
Shannon/von Neumann/conditional/relative entropy determined by thermodynamic
generating functions and density matrices encoding geometric data for GIFs
and NES, and characterized by respective Perelman W-entropy. The concept of
quantum geometric flow and gravitational entanglement and main properties
(inequalities) are formulated and studied for new classes of theories of
QGIFs for NES.

In sections \ref{s4}-\ref{s6}, we shown how to construct in explicit form
stationary generic off-diagonal solutions for relativistic geometric flows,
nonholonomic Ricci solitons and generalized gravitational field equations.
 Such configurations are not characterized, in general, by certain entropy-area,
holographic or duality conditions and can not described as some modified
 Bekenstein-Hawking BH thermodynamic systems. In our works, it is developed an alternative and
more general way when stationary and cosmological solutions in geometric  flow evolution theories, MGTs and GR, GIFs and QGIFs, can be defined and
characterized by nonholonomic deformations of Perelman's W-entropy and associated statistical thermodynamic models.

Finally, we note that further developments of our approach will involve
explicit examples with computations of the W-entropy and QGIF and NES
entanglement, R\'{e}nye and other type entropies for various classes of
stationary and cosmological type solutions in MGTs and quantum gravity
models, spinor and noncommutative variables, see some previous our results in \cite%
{vacaru09,vacaru18tc,bubuianu18}.

\vskip3pt

\textbf{Acknowledgments:} This research develops former programs partially supported by IDEI, PN-II-ID-PCE-2011-3-0256, CERN and DAAD and extended to adjunct positions and  collaborations with California State University at Fresno, the USA, and Yu. Fedkovych Chernivtsi National University,  Ukraine.

\end{document}